\documentclass[preprint]{aastex} 
\shorttitle{Analysis of BEST data.}
\shortauthors{C. Karoff et al.}
\begin{document}
\title{Identification of Variable Stars in CoRoT's first main observing field (LRc1)}
\author{C. Karoff\altaffilmark{1}, H. Rauer\altaffilmark{2}, A. Erikson, H. Voss\altaffilmark{3}, P. Kabath and T. Wiese}
\affil{Institut f\"ur Planetenforschung, Deutsches Zentrum f\"ur Luft- und Raumfahrt,
  Rutherfordstrasse 2, D-12489 Berlin, Germany}
\email{heike.rauer@dlr.de}
\author{M. Deleuil, C. Moutou and J.C. Meunier}
\affil{Laboratoire d'Astrophysique de Marseille, CNRS - Universit\'e de
  Provence, BP 8, 13376 Marseille Cedex 12, France}
 \and
\author{H. Deeg}
\affil{Instituto de Astrofisica de Canarias, E-38200 La Laguna, Spain}
\altaffiltext{1}{Visiting Astronomer, Department of Physics and Astronomy, University of Aarhus, Denmark}
\altaffiltext{2}{Professor at Zentrum f\"ur Astronomie und Astrophysik, Technische Universit\"at Berlin, Germany}
\altaffiltext{3}{Present address: Th\"uringer Landessternwarte Tautenburg, Tautenburg, Germany}
\begin{abstract}
The \it COROT \rm space mission will monitor several target fields for up to 150 days to perform asteroseismology and to search for extrasolar planets by photometric transits. Variable stars in the target fields are important objects for additional scientific studies but can also disturb the search for planetary transits. A variability characterization of the target fields prior to \it COROT \rm observations is therefore important for two reasons: to find interesting variable stars to monitor further and to make an analysis of the impact of the variable stars on detecting extrasolar planet transits with \it COROT\rm. 

The Berlin Exoplanet Search Telescope (BEST) is a small wide-angle telescope dedicated to high-precision photometry. It has observed a 9 deg$^2$ field of view centered at ($\alpha$,$\delta$)=(19$^{\mathrm{h}}$00$^{\mathrm{m}}$00$\fs$0,$+$00$^{\circ}$01$^{\prime}$55$\farcs$2)
(J2000.0) over 98 nights to search for variable stars in the surroundings of the first long-run target field (LRc1) of the \it COROT \rm space mission. In this data set we identified 92 periodic variable stars, 86 of which are new discoveries and 6 of which are known from the General Catalogue of Variable Stars (GCVS). For five of the GCVS stars, variability could not be confirmed. Forty-three of the 92 detected periodic variable stars are identified as eclipsing binaries. We have evaluated the completeness of our survey for eclipsing binaries by comparing it to the expected fraction of eclipsing binaries based on \it HIPPARCOS \rm observations. From this evaluation we show that the BEST data set presented here has a completeness of 20\% -- 30\% for periods longer than 1 day and is complete relative to \it HIPPARCOS \rm for short-period binaries. 
\end{abstract}

\section{Introduction}
Small wide-angle telescopes such as the Berlin Exoplanet Search Telescope \citep[BEST;][]{2004PASP..116...38R} offer a unique possibility of identifying stellar variability in a large number of stars. Apart from giving us important information on the individual targets, these kinds of wide-field data also contain information on the statistical distribution of variable stars, which can be used not only to understand the structure and evolution of the Galaxy, but also to analyze the expected false-alarm rates for transiting extrasolar planet searches \citep{2003ApJ...593L.125B}. For space missions searching for extrasolar planets by photometric transits, knowledge of the variable stars in the target fields helps reduce the number of false alarms, thereby reducing the amount of follow-up observation needed. In addition, the variable stars provide interesting targets for additional science. 

The \it COROT \rm space mission  \citep{2002sshp.conf...17B, 2003A&A...405.1137B}, launched on 2006 December 27, will observe a number of target fields for up to 150 days in order to perform asteroseismic studies of the target stars and to search for extrasolar planets by photometric transits. The observation of the first 150 day "long-run" field (LRc1) started in spring 2007. This field has been observed with BEST in order to identify variable stars. \it COROT \rm is expected to obtain a precision of 0.66 ppm in 5 days of observation in the asteroseismic channel and 700 ppm in 1 hr of observation in the exoplanet channel \citep{2002sshp.conf...17B}. This is an order of magnitude better than what can be obtained with BEST, because \it COROT \rm benefits from not having to observe through the atmosphere. Therefore, many of the transits 
that are expected to be detected with \it COROT \rm will not be visible in the BEST data, but despite this order-of-magnitude gap the BEST data will be highly valuable to the \it COROT \rm mission.
 
In the exoplanet channel of the \it COROT \rm satellite, stars have a large point-spread function (PSF) due to a prism installed in front of the detectors to obtain the colors of each star. The color information will help to distinguish stellar activity from exoplanet transits. In dense target fields, however, the resulting relatively large PSFs of the stellar spectra can overlap. Prior knowledge of variables overlapping the PSFs of interesting target stars will therefore be helpful to avoid misinterpretation. 

In addition to searching exoplanetary transits, additional science programs will be performed in the \it COROT \rm fields on selected targets. Some of these additional programs are interested in variable stars, e.g., binaries. To be able to place integration windows in the satellite field on such targets, their position has 
to be known before observing the field with \it COROT\rm. A search for variable stars in the COROT fields is therefore also helpful for the additional science program of the mission by detecting previously unknown variables in the field and confirming or rejecting the variability of the stars mentioned in variable star 
catalogs. 

A variability characterization like the one presented here of the target fields prior to \it COROT \rm observations is therefore important for two reasons: to find interesting variable stars to monitor further by preselecting windows and to make an analysis of the impact of the variable stars on detecting extrasolar planet transits with \it COROT \rm.
 
The present paper is arranged as follows: Section~4.2 describes the observations, Section~4.3 summarizes the data reduction procedure, Section~4.4 describes the data analysis, including search and detection of periodic variable stars, Section~4.5 contains the variable star catalog of the observed field, and Section~4.6 describes the analysis of the completeness of the data set. Concluding remarks are found in Section~4.7. 

\section{Observation}
BEST is a small wide-angle telescope operated by the Institut f{\"u}r Planetenforschung of the Deutsches Zentrum f{\"u}r Luft- und Raumfahrt (DLR) and is located at the Observatoire de Haute-Provence (OHP), France. It consists of a Schmidt telescope with a 20 cm effective aperture, an $f / 2.7$ focal ratio, and an AP-10 CCD camera, with 2048 $\times$ 2048 pixels of 14 $\mu$m size \citep{2004PASP..116...38R}. The result is a 3.1$^{\circ}$ $\times$ 3.1$^{\circ}$ field of view (FOV) with a pixel scale of 5$\farcs$5.  This is comparable to \it COROT\rm, which has a pixel scale of 2$\farcs$3 and a FOV for each of the four CCDs of 1.3$^{\circ}$ $\times$ 1.3$^{\circ}$.

We present the results from a monitoring program of the \it COROT \rm field LRc1 in the Galactic center direction centered at ($\alpha$,$\delta$)=(19$^{\mathrm{h}}$00$^{\mathrm{m}}$00$\fs$0,$+$00$^{\circ}$01$^{\prime}$55$\farcs$2) (J2000.0) obtained from 2005 June 6 to September 7 . The data set consists of a 
total of 35 nights of observations spread out over 98 nights in summer 2005. No data were obtained on nights with bad weather, around midsummer, or around full Moon. The observations were made without a filter in front of the CCD (the spectral response of the CCD corresponds to a wide $R$ bandpass) in order to detect as many photons as possible and to provide a maximum sampling rate. An equal number of frames 
were obtained with 40, 120, and 240 s integration times in order to observe stars in a large magnitude range and to analyze the effect of crowding in the frames. We only present data from the 240 s integration frames here, corresponding to 727 frames in total. Fig.~4.1 shows the 3.1$^{\circ}$ $\times$ 3.1$^{\circ}$ FOV observed. 

\section{Data Reduction}
The reduction of the data was performed as described by \citet{2006ASPC..349..261K}. Based on a comparative test of three different photometry algorithms (SExtractor \citep{1996A&AS..117..393B}, MOMF
\citep{1992PASP..104..413K} and ISIS \citep{2000A&AS..144..363A}) a data reduction pipeline was built \citep{2006ASPC..349..261K} in order to reduce noise originating from the stellar background, extending 
the data pipeline of BEST data, which was optimized for less crowded fields observed at its previous location at the Th{\"u}ringer Landessternwarte, Germany (Rauer et al. 2004). We outline the different steps in the pipeline below. 

The frames are calibrated from night to night by standard procedures as described in \citep{2004PASP..116...38R}. Hereafter, the reduction generally follows the method given in the ISIS package \citep{2000A&AS..144..363A}.

\subsection{Image Subtraction}
Image subtraction is performed in order to remove both the sky and the stellar background. This is done by subtracting a reference frame from all the frames, which removes the sky background and all the constant stars. When image subtraction is performed in the right way it will produce uncrowded frames where 
only the variable stars are present. The noise will then be dominated by the photon noise from the target stars and from the stellar background. 

Before image subtraction can be made all the frames need to be transformed into the same ($x, y$)-coordinate grid. The frame with the best seeing from the middle of the observation run was selected as the template frame for this transformation. Then ISIS produced a reference frame by combining a number of best-seeing frames. The reference frame was then subtracted from all the other frames with the ISIS parameters listed in Table~4.1. This means that the reference frame is convolved with a convolution kernel in order to match all the stars in the frames. 

Having subtracted the reference frame from all the other frames, ISIS constructs a frame that is the mean of the absolute normalized deviations of all the subtracted frames. This frame only contains nonconstant stars, as all the constant stars have been removed by the subtraction. Unfortunately, this only works for nonconstant stars that have a large amplitude, as is discussed below.
 
The ISIS package also includes a photometry routine that performs PSF-weighted photometry on the variable stars detected, thereby producing light curves of the differential fluxes of the stars. For different reasons, discussed below, we did not use this routine. We used the output of ISIS to distinguish between true periodic variable stars and stars that showed variability because of contamination from another nearby variable star. 

\subsection{Photometry}
There are three reasons why we do not use the photometry algorithm in ISIS. The first is that ISIS only produces light curves of differential fluxes; in order to convert these into magnitudes, photometry needs to also be performed on the reference frame in the same way as on the subtracted frames. The second reason, 
which has been noted by \citet{2002AJ....123.3460M} and \citet{2004AJ....128.1761H}, is that the detection of variables in ISIS works best for variables with an amplitude of at least 0.1 mag, since small-amplitude variables tend to get washed out in the overall noise in the subtracted frames. The last reason is that in order to do PSF-weighted photometry, the stars need to have a well-defined PSF; this is not always true if the stars are undersampled. We therefore preferred to perform simple unit-weight aperture photometry on the subtracted frames. \citet{2005AJ....130.2241H} analyzed differences in performing PSF-weighted and unit-weighted aperture photometry on the subtracted frames and concluded that PSF-weighted photometry works best for the relatively faint stars, while unit-weighted aperture photometry works best for the relatively bright 
stars, but also that PSF-weighted photometry gives rise to a constant error term in the light curves. In addition, it is important that the transformation from differential fluxes to magnitudes is correct when evaluating the rms noise level obtained by different photometry algorithms and when comparing the noise 
level for ground-based transit surveys like BEST to that of space-based transit surveys like \it COROT\rm. By using the same unit-weighted aperture photometry algorithm on the subtracted frames, as well as on the reference frame, we are confident that this transformation is correct. 

\citet{2004AJ....128.1761H} note that if they use an aperture radius below 7 pixels for stars with a FWHM of the order of 3 pixels, the amplitude of the light curves is artificially reduced. As the FWHMs of the stars we analyze here are in the range from 1 to 3 pixels, depending on the seeing on different nights, we use 
an aperture radius of 7 pixels. We did try to reduce the aperture radius, which resulted not only in reduced amplitudes of the light curves but also reduced rms noise level in the light curves, as expected because of a decrease in the amount of background noise. In other words it might be possible to increase the precision by using smaller or nonconstant apertures, but then the amplitude and noise determination would not be as reliable. It is also noted here that although the PSFs in the nonsubtracted frames may be small, there may be large non-Gaussian residuals in the subtracted frames that require large apertures in order to ensure flux preservation. It is important to note here that the aperture radius of 7 pixels is only used on the subtracted frames where all the constant stars are removed. Therefore, using an aperture radius of 7 pixels does not mean that we are not able to resolve stars separated by less than 7 pixels (as long as they are not variables). 
Figure 4.2 shows the rms scatter in a single night (June 29) of all the light curves as a function of stellar luminosity, with the identified periodic variable stars marked. The figure shows that periodic variable stars are found roughly equally distributed over stellar luminosity. The obtained precision is comparable to what has been obtained by other small wide-angle telescopes \citep{2004AJ....128.1761H, 2004MNRAS.353..689K, 2005MNRAS.362..117K, 2005MNRAS.364.1091K}.

\subsection{Astrometry and Calibration}
As noted by \citet{2005AJ....130.2241H} the light curves need to be calibrated for systematic changes in the zero points. This is done by selecting a number of nonsaturated stars with low rms noise in the light curves and then calculating the mean change of the light curves. We selected 2182 stars for calculating these zero-point changes and found that the corrections from frame to frame were below 0.02 mag, except for a few bad frames which contained clouds or airplane tracks. 

Even after the zero-point correction, residuals of systematic trends in the light curves caused by, e.g., air mass, cloud cover, and temperature variations were still present in the data (these residuals are sometimes referred to as "red noise"). We therefore applied the algorithm introduced by \cite{2005MNRAS.356.1466T} in order to correct for systematic effects. This clearly improved the rms noise level, in particular in the brightest stars. A detailed discussion of the photometric limitations of the BEST survey will be presented in a forthcoming paper. 

The astrometric calibration was performed by matching the 1500 brightest stars in the stellar catalog obtained from the reference frame with the 2000 brightest stars (magnitude limit 11.4) of the USNO-A2.0 catalog \citep{1998USNO2.C......0M}. The problem of matching the observed stars to stellar catalogs is not trivial. The main reason for this is the large image scale of BEST (5$\farcs$5
pixel$^{-1}$. This means that stars in a crowded part of the frame will have multiple matches in the stellar catalogs. The astrometric calibration was performed using MATCH \citep{1995PASP..107.1119V}. MATCH finds a linear, quadratic, or cubic transformation between two stellar catalogs. We used a cubic transformation to calculate the transformation from the ($x, y$)-coordinates of the CCD to right ascension and declination. We note that a linear transformation does not work because of the large FOV. 

Out of the brightest 1500 stars in the stellar catalog obtained from the reference frame it was possible to match 902 with a matching radius of 2$\cdot10^{-4}$ degree. From these 902 stars we calculated an absolute magnitude difference between the observations and the USNO-A2.0 catalog $R$ band. We note that we  do not intend to perform a real absolute photometric calibration of the stars, since only relative magnitudes are needed to identify variables. 

\section{Data Analysis}
\subsection{Identification of Suspected Variable Stars}
We used the variability index $J$ defined by \citet{1996PASP..108..851S} to identify stars as suspected variable stars. The index $J$ is defined as the normalized sum of the deviation of each data point (or pair of data points with small separations in time) in the light curve compared to the expected noise level for that star (the expected noise level is calculated based on the signal-to-noise ratio of the star). We considered data points to be pairs if they were separated by less than 0.03 day. Fig.~4.3 shows the index $J$ as 
a function of magnitude. 

We classified all 9112 stars with an index $J$ greater than 2 as suspected variables, as all these stars show some kind of variability. Most of these stars are clearly variable stars, but either the period cannot be properly determined because of an imperfect duty cycle or the stars are nonperiodic or multiperiodic variable stars. In this paper we concentrate on the periodic variable stars.
\subsection{Identification of Periodic Variable Stars and Period Determination}
All the stars classified as suspected variables were searched for periodicity with the method introduced by \citet{1996ApJ...460L.107S}. This method fits a set of periodic orthogonal polynomials to the observations and evaluates the quality of the fit with the use of an analysis-of-variance statistic. All the suspected variable stars were fitted with two polynomials, and all objects with a fit quality better than 0.9 were examined visually. 

In the visual examination we selected stars as periodic variable stars if they showed a clear periodicity between 0.286 and 120 days. We rejected light curves with periods close to 1 day, or harmonics of 1 day, because the periodicities in these light curves are likely to be caused by periodic changes in background 
level, temperature, air mass, etc. 

We have identified 92 periodic variable stars in the data set (Tables~4.2 and 4.3). It is seen in Fig.~4.1 that the periodic variable stars are detected all over the CCD, although the density depends on the stellar density in that part of the field. 

\section{Variable Star Catalog}
The GCVS \citep{1998GCVS4.C......0K} contains 60 variable stars in the field, 11 of which have periods less than 120 days. We have identified all 11 stars in the BEST data set. The phased light curves are shown in Fig.~4.4, and the corresponding parameters are given in Table 4.3. By looking at the light curves and comparing them with the index $J$ we were able to confirm variability in six cases (G1, G2, G8, G9, G10, and G11). None of the 11 stars were detected by the algorithm we had set up to search for periodicity. G8, G9, and G11 are RR Lyrae stars with such a characteristic light curve that they should have been detected by the algorithm. By going back in the analysis to see why these stars were not detected, we see that they were all excluded during the visual inspection. There are three reasons for this exclusion. 
First, the method introduced by \citet{1996ApJ...460L.107S} provides a period that is equal to half the true period. This means that the phased light curves mainly show a falling trend. Second, the method introduced by Schwarzenberg-Czerny (1996) provides a period offset from the true period (the period of G8 is off by 5\%, G9 by 0.2\%, and G11 by 33\%), thereby making the phased light curve look a lot more noisy. Third, G8 and G11 have periods very close to 1 day. G1, G2, and G10 do not show strongly periodic signals, and detection was not expected. 

We were not able to confirm variability in five cases (G3, G4, G5, G6, and G7). There are three possible reasons why we were not able to confirm variability in these three stars: (1) the identification as variable stars in the GCVS is wrong, (2) the stars are not variable anymore, and (3) the variability of the stars is below the noise level in the BEST data. In Table~4.3 it is seen that all three stars should have amplitudes larger than 1 mag, and in Fig.~4.4 it is seen that the observed variability in the light curves is well below 1 mag. This rules out the third possibility. 

We have divided the periodic variable stars identified by our search into two groups: periodic pulsating variables and eclipsing binaries. The two groups each contain 43 stars. For most of the stars it is easy to make this classification, but for some stars the classification may not be unique. The light curves of the 
periodic variable stars are shown in Figs 4.5 and 4.6, and the parameters are listed in Table 4.2. 

The detected periodic variable stars were cross-identified with the USNO-A2.0 catalog. For each star we checked that the cross-identification was unique. If there were multiple or no identifications we did not assign a USNO-A2.0 ID to the star. The USNO-A2.0 IDs are given in Table 4.2. We also looked up all the detected periodic variable stars in the SIMBAD catalog but were only able to identify one of the periodic variable stars, i.e., V66 = V1225 Aql. 

Some of the stars show problems with blending. This means that the stars next to the variable stars also show variability. For each case where we had two or more stars showing the same kind of variability within the radius of the aperture (7 pixels) we have only identified one of the stars as a periodic variable. It is difficult to verify from the light curves alone which star is the true variable, because a faint star next to a bright variable star can show larger relative variability than the true bright variable star. In order to overcome this problem we identified all the variable stars that were affected by blending in the variability frame of our ISIS pipeline (the frame that was the mean of the absolute normalized deviations of all the subtracted frames), because this frame will only show truly variable stars. This approach only works well if the relative 
amplitudes in the variable stars are high. 

A few stars, such as V20, V31, and V50, seem to show some additional variability apart from the periodic signal. By inspecting the light curves we found that the additional variability originated from a few nights where the noise level was unusually high. It is therefore most unlikely that this additional variability is of stellar origin. 

Forty of the 92 identified periodic variable stars are in the LRc1 FOV of one of the two \it COROT \rm CCDs dedicated to the planet program. These are V3, V4, V6, V9, V11, V12, V13, V15, V21, V23, V25, V26, V27, V32, V38, V39, V40, V41, V42, V43, V44, V45, V47, V49, V50, V53, V54, V58, V59, V60, V61, V65, V69, V78, V83, V84, V85, V86, G2, and G9. Some of the remaining identified periodic variable stars are expected to be in the FOV of some of the future pointings. 

We have identified all the periodic variable stars in the ExoDat catalog \citep{exodat}. The ExoDat catalog is part of the preparation of \it COROT \rm and has been made in order to select which stars to observe with $COROT$, taking into account spectral type, luminosity class, variability, and contamination (overlapping PSF on the focal plane). It contains, among other things, $BVRI$ photometry of the field observed with the 2.5 m Isaac Newton Telescope at La Palma. 

We cross-identify the periodic variable stars from BEST with the ExoDat catalog using a matching radius of 1$\cdot10^{-3}$ deg. In this way we obtain 81 matches out of the 86 periodic variable stars. More identifications could have been obtained by increasing the matching radius, but that would have also increased the number of mismatches. 

Fig.~4.7 shows a color-magnitude diagram of the ExoDat catalog, with the discovered periodic variable stars marked. The stars in the $V$ -- $I$ versus $I$ diagram follow two branches: to the left the main-sequence stars and to the right the giant stars. For the faint stars there is a mismatch between stars on the main sequence and stars on the giant branch, since we do not know anything about the distance to the stars and therefore cannot provide their absolute luminosity. There is also a mismatch for nearby late main-sequence stars, which are bright with red colors and fall within or to the right of the giant branch. The diamonds in the figure represent eclipsing binaries, and the crosses represent periodic pulsating stars. A vague tendency for eclipsing binaries to fall mainly on the main sequence can be seen (15 out of 43; 35\%), compared to the number of periodic pulsating stars that fall on the main sequence (7 out of 49; 14\%). This is in agreement with what we expect, since most of the periodic pulsating stars we have found are Cepheids, Miras, or RR Lyrae stars, and none of these stars fall on the main-sequence branch in the $V$ -- $I$ versus $I$ diagram. 

Due to their short periodicity, the identified eclipsing binaries cannot be comprised of giant stars. Hence, from their location in Fig.~4.7, they range from solar-type stars (at the typical main-sequence field star distance of \it COROT \rm fields) to later but more nearby main-sequence stars. 

\section{Data Completeness}
A way to evaluate the capability of ground-based surveys to detect variable stars is to compare the number of eclipsing binaries detected with a certain system with the expected number of eclipsing binaries in general. Evaluating the capability of ground-based surveys this way has the advantage that it includes not only the photometric precision of the data in the evaluation but also the phase coverage and the detection algorithms. 

Fig.~4.8 shows the number of eclipsing binaries in the BEST data compared with the \it HIPPARCOS \rm data \citep{2005A&A...442.1003S}. A description of how the eclipsing binaries are detected in the \it HIPPARCOS \rm data can be found in \citep{2005A&A...442.1003S}. 

The \it HIPPARCOS \rm stars have been separated into two different absolute magnitude bins ($M_v$ = --2.2 to 0.9) and ($M_v$ = 0.3 to 4.0). This separation corresponds to a separation into two different parallax bins \citep[($\pi > 4.5$ and $\pi$ = 1.0 -- 4.5)][]{2005A&A...442.1003S}. As we do not know the parallaxes for the stars observed with BEST, we do not know the absolute magnitudes. We have tried to compare the BEST data with the \it HIPPARCOS \rm data from both bins with more or less the same result, although the close \it HIPPARCOS \rm sample seems to fit the BEST data slightly better than the distant sample.
 
Fig.~4.8 shows that BEST detected more eclipsing binaries at low periods, which is caused by the higher sampling rate of BEST compared to \it HIPPARCOS\rm. The fact that BEST detected more eclipsing binaries than HIPPARCOS at low periods indicates that \it HIPPARCOS \rm may not be perfect at low periods. At higher 
periods (log $P$ = 0.0 to 0.5) BEST only detected 20\% -- 30\% of the number of eclipsing binaries detected by \it HIPPARCOS\rm.  \citet{2005A&A...442.1003S} have calculated the relative number of eclipsing binaries as a function of period in three different amplitude bins ($\Delta m >$ 0.1, 0.4 and 0.7 mag). As almost all the eclipsing binaries we have detected have amplitudes between 0.1 and 0.4 mag, we only discuss the results from the first bin here, i.e., the bin with the largest eclipse probabilities. This statistical analysis shows that the BEST data set presented here has a completeness of 20\% -- 30\% compared with the \it HIPPARCOS \rm data set for periods longer than 1 day and is complete compared with the \it HIPPARCOS \rm data set for short-period binaries.

As there were six periodic variable stars in the GCVS that we did not detect with our periodic variable star detection algorithm, there is reason to believe that our algorithm is not perfect. This means that by improving the algorithm more periodic variable stars would be detected and the completeness of the data set would increase. In particular, it is seen that we detect a lower number of eclipsing binaries with periods around 1 day than what is expected. This might mean that we have been too conservative with excluding periodic variable stars that have a period close to 1 day.

\section{Conclusions}
BEST has observed the first long-run field (LRc1) of the \it COROT \rm mission for 35 nights. We have found 92 periodic variable stars in the data: 43 eclipsing binaries and 49 periodic pulsating stars. 

Data reduction has been performed using image subtraction. Astrometric and photometric calibration were made by matching stars to the USNO-A2.0 catalog. We used the index $J$ defined by \citet{1996PASP..108..851S}  to identify suspected variable stars and searched these stars for periodicity with the method introduced by \citet{1996ApJ...460L.107S}. In this way we have identified 92 periodic variable stars, compared to the already known 11 variable stars in the GCVS with periods less than 120 days. 

The periodic variable stars have been identified in the ExoDat catalog, and the color-magnitude diagram of the stars in the field shows that the eclipsing binaries have a slight tendency to lie on the main sequence compared to the periodic pulsating stars. This reflects that most of the periodic pulsating stars detected in the 
data are Cepheids, Miras, or RR Lyrae stars. 

We also present a new tool to evaluate the completeness of a survey data set by comparing the number of eclipsing binaries detected in the data set with statistical estimates of the relative number of eclipsing binaries. The analysis shows that this BEST data set has a completeness of 20\% -- 30\% compared with the 
\it HIPPARCOS \rm data set for periods greater than 1 day and is complete relative to \it HIPPARCOS \rm for shorter periods.

\section*{Acknowledgments}
We would like to thank T. Arentoft for helpful discussions and careful reading of the manuscript and the referee, B. S. Gaudi, for many useful suggestions and comments. We thank the Observatoire de Haute-Provence for great support of the BEST survey. C. K. gratefully acknowledges financial support by the Danish National Science Research Council through the Danish AsteroSeismology Centre (DASC) and the Instrument Center for Danish Astrophysics (IDA). This research has made use of the SIMBAD database, operated at CDS, Strasbourg, France.

\begin{figure}
\centering
\includegraphics[width=\columnwidth]{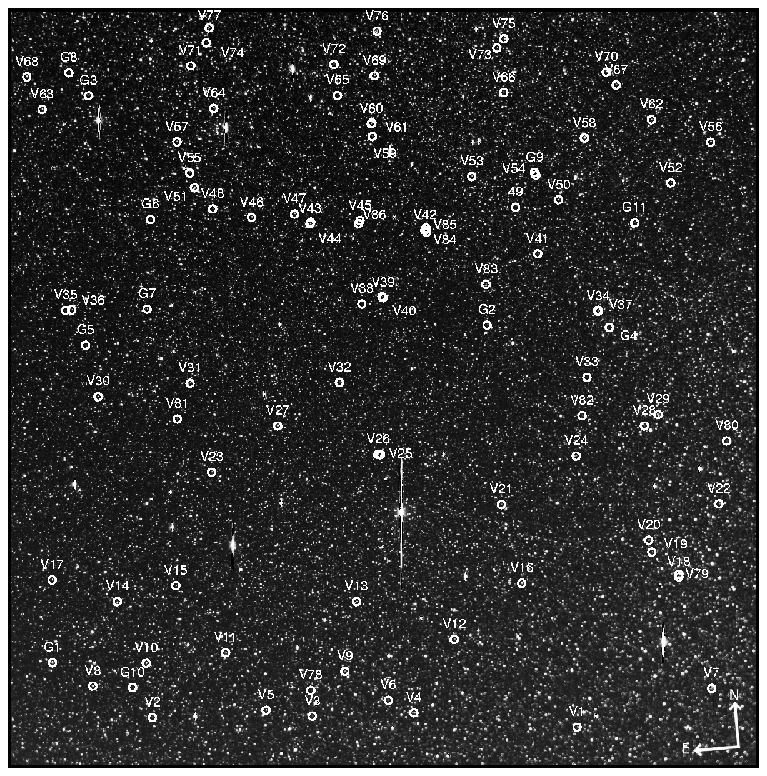}
\caption{The reference frame. The newly identified periodic variable
stars are marked (V) as well as the stars from the GCVS (G).}
\label{fig:fie}
\end{figure}

\begin{figure}
\centering
\includegraphics[width=\columnwidth]{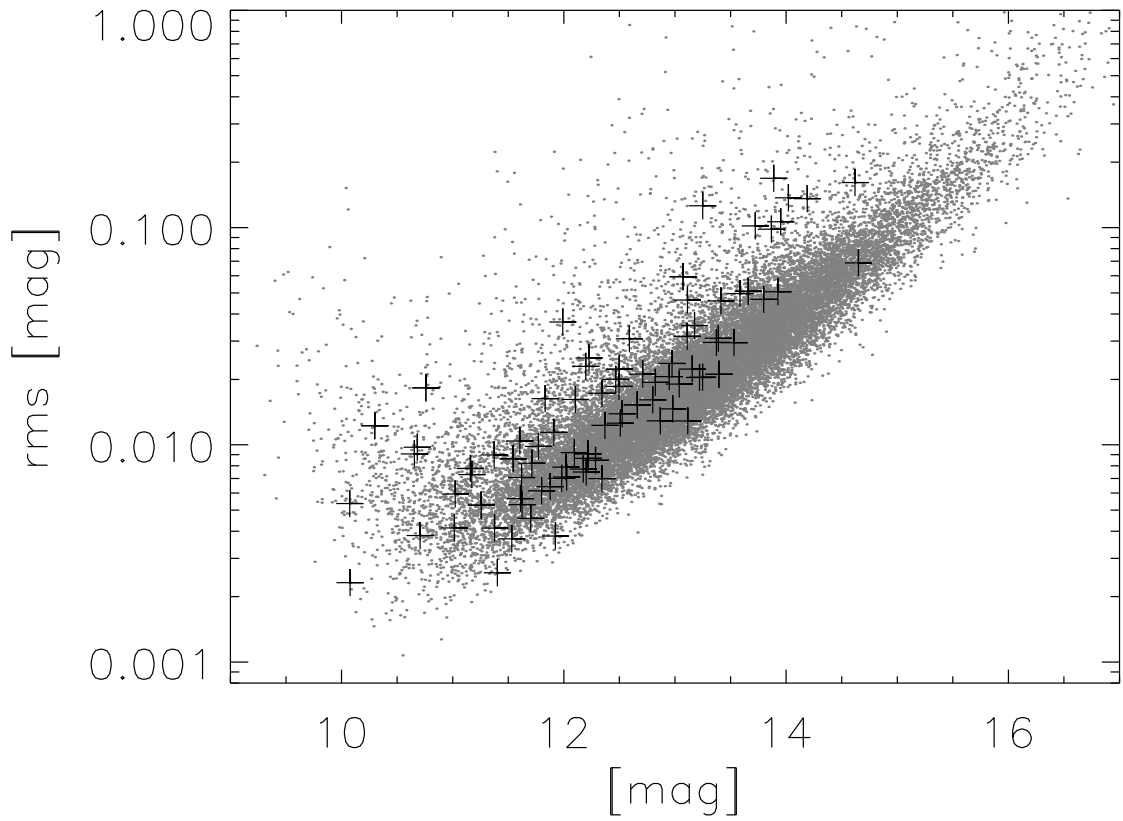}
\caption{The rms scatter of all the light curves (\it gray dots\rm) and of
  the identified periodic variable stars (\it crosses\rm). }
\label{fig:noise}
\end{figure}

\begin{figure}
\centering
\includegraphics[width=\columnwidth]{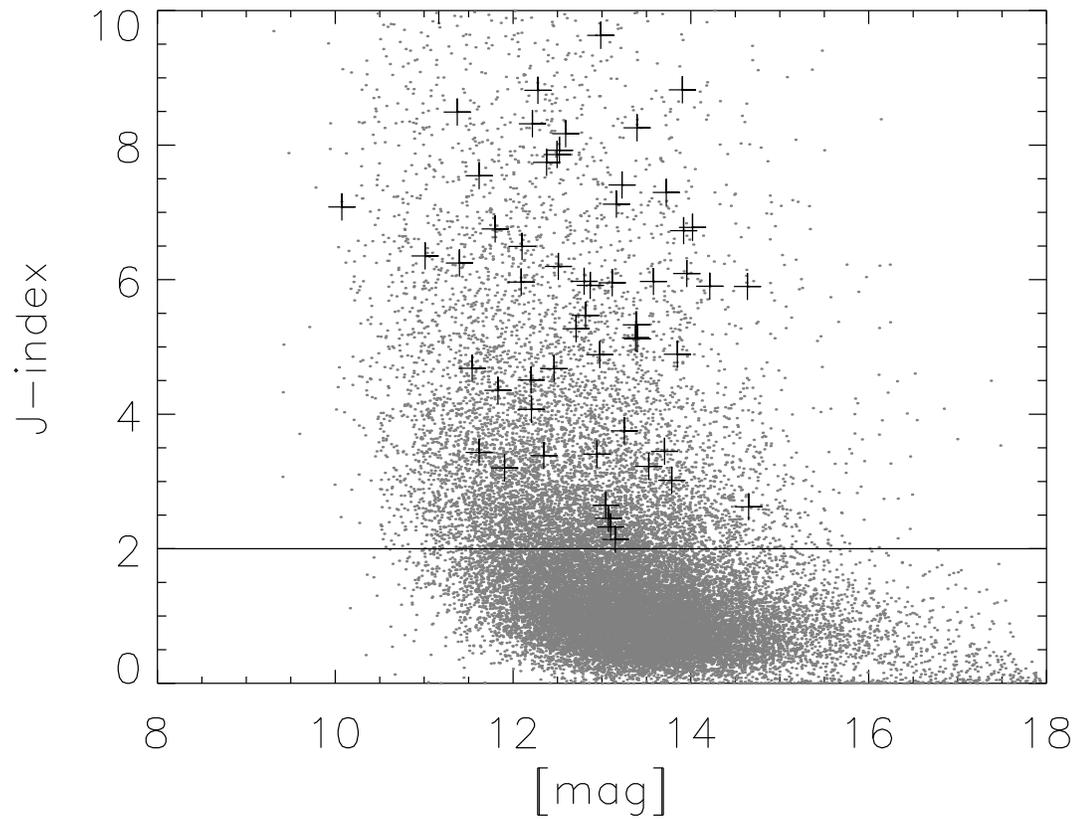}
\caption{The variability index $J$ vs. $R$ magnitude. The gray dots
  represents all the observed stars while the crosses represents the
  periodic variable stars that we have identified. The 9112 stars
  above the line with a index $J$ larger then 2.0 were selected as
  suspected variable stars.}
\label{fig:j}
\end{figure}

\begin{figure*}[!ht]
\centering
\hbox{\includegraphics[width=4.65cm]{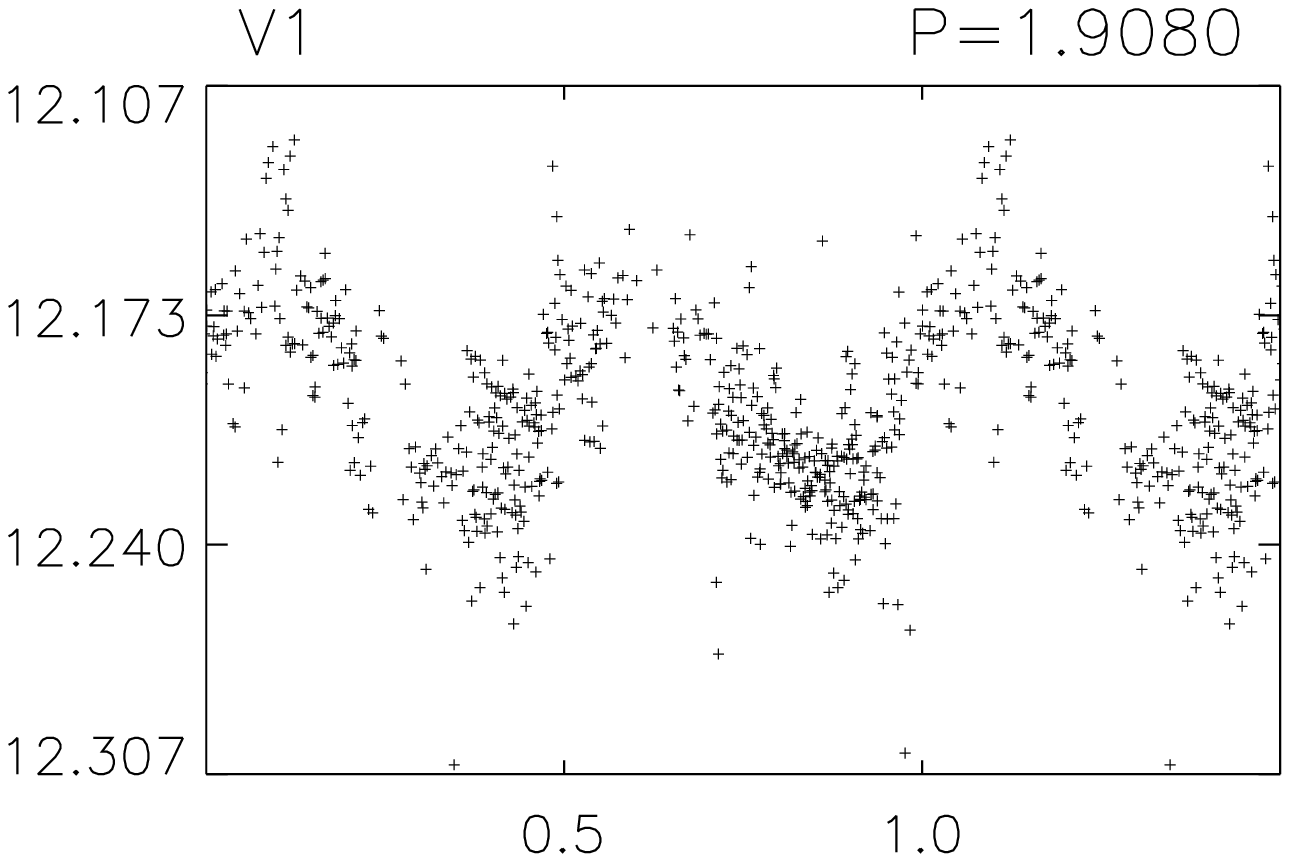}
   \hspace{-0.7cm}\includegraphics[width=4.65cm]{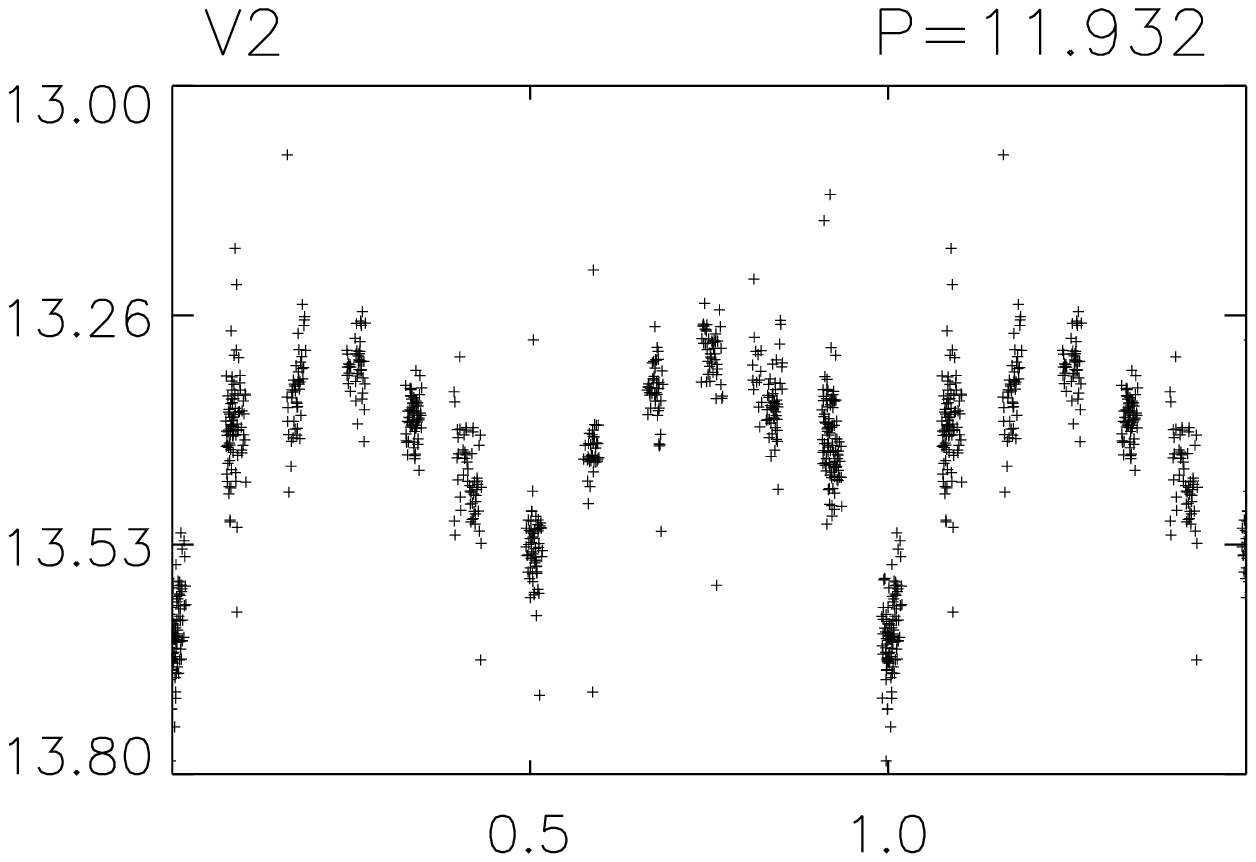}
   \hspace{-0.7cm}\includegraphics[width=4.65cm]{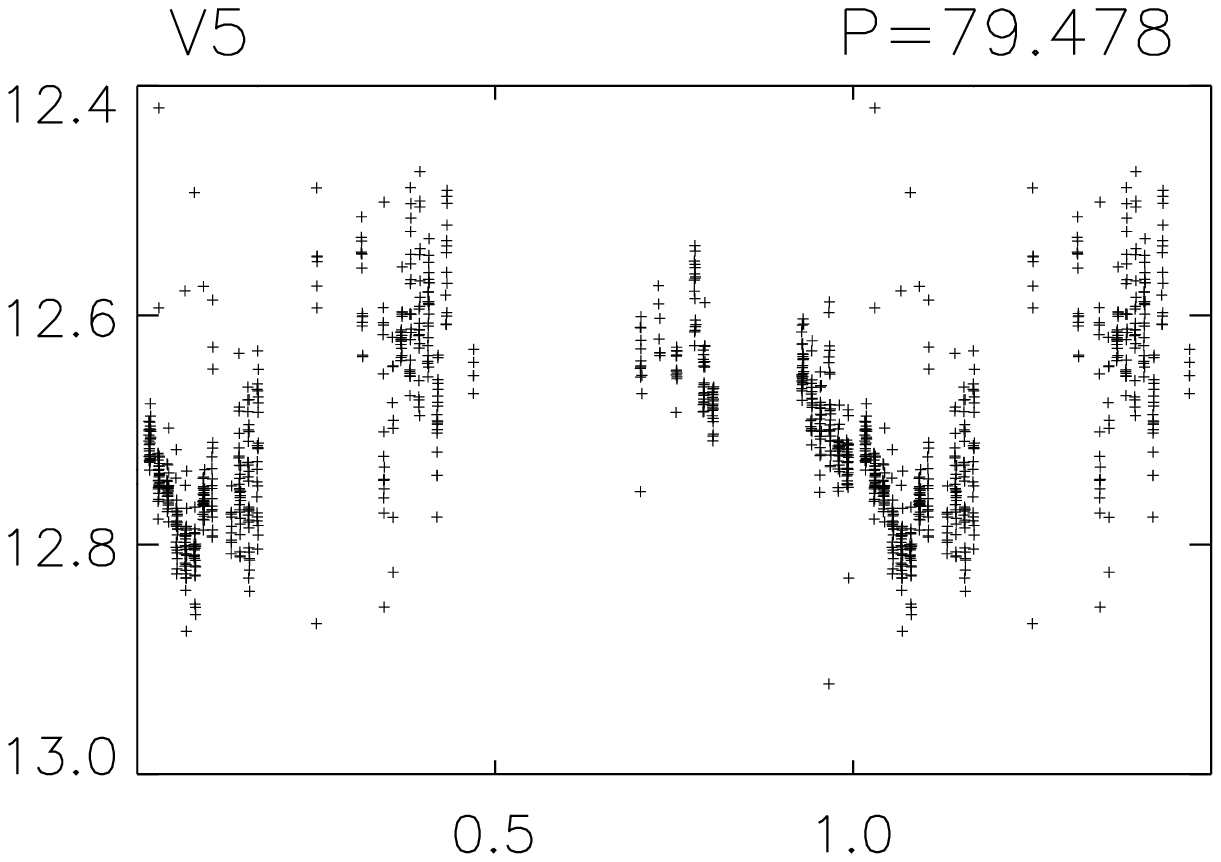}}
\vspace{-0.3cm}
   \hbox{\includegraphics[width=4.65cm]{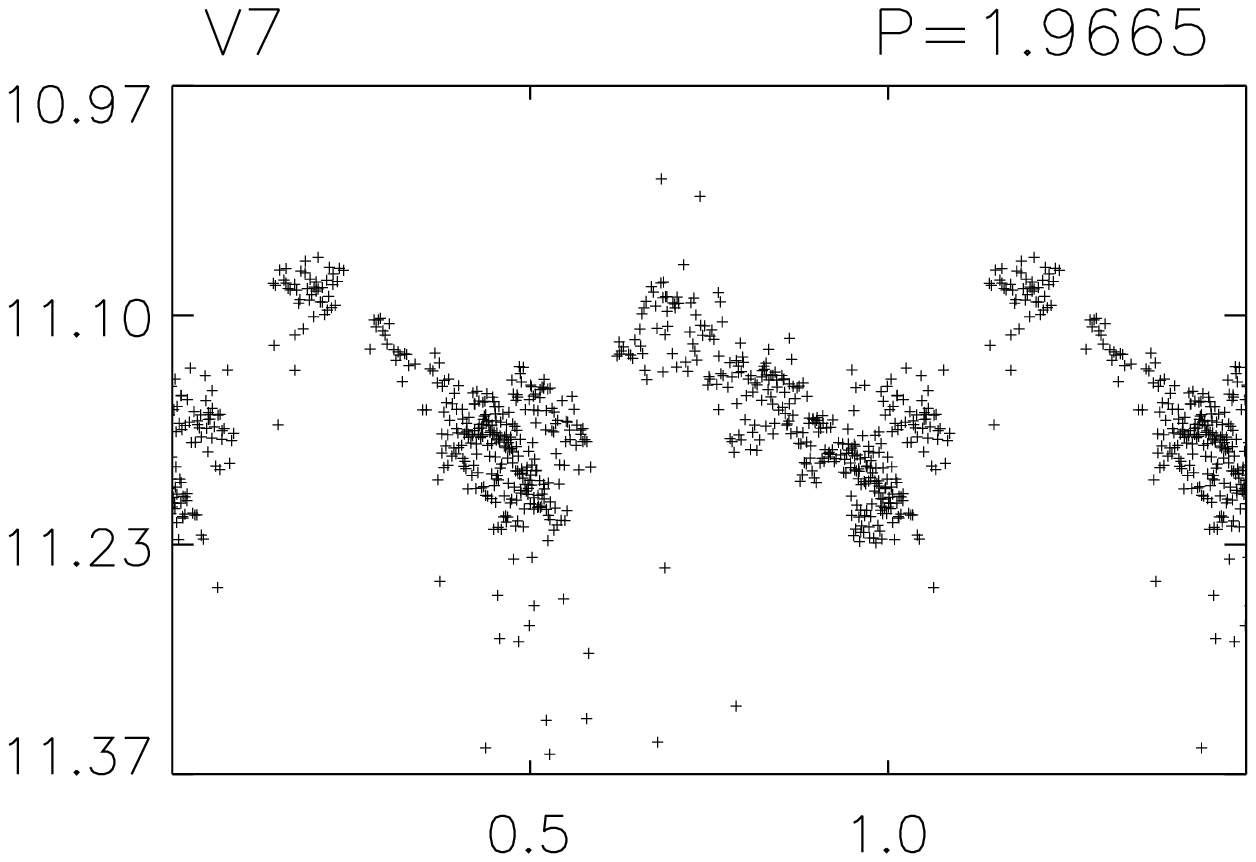}
   \hspace{-0.7cm}\includegraphics[width=4.65cm]{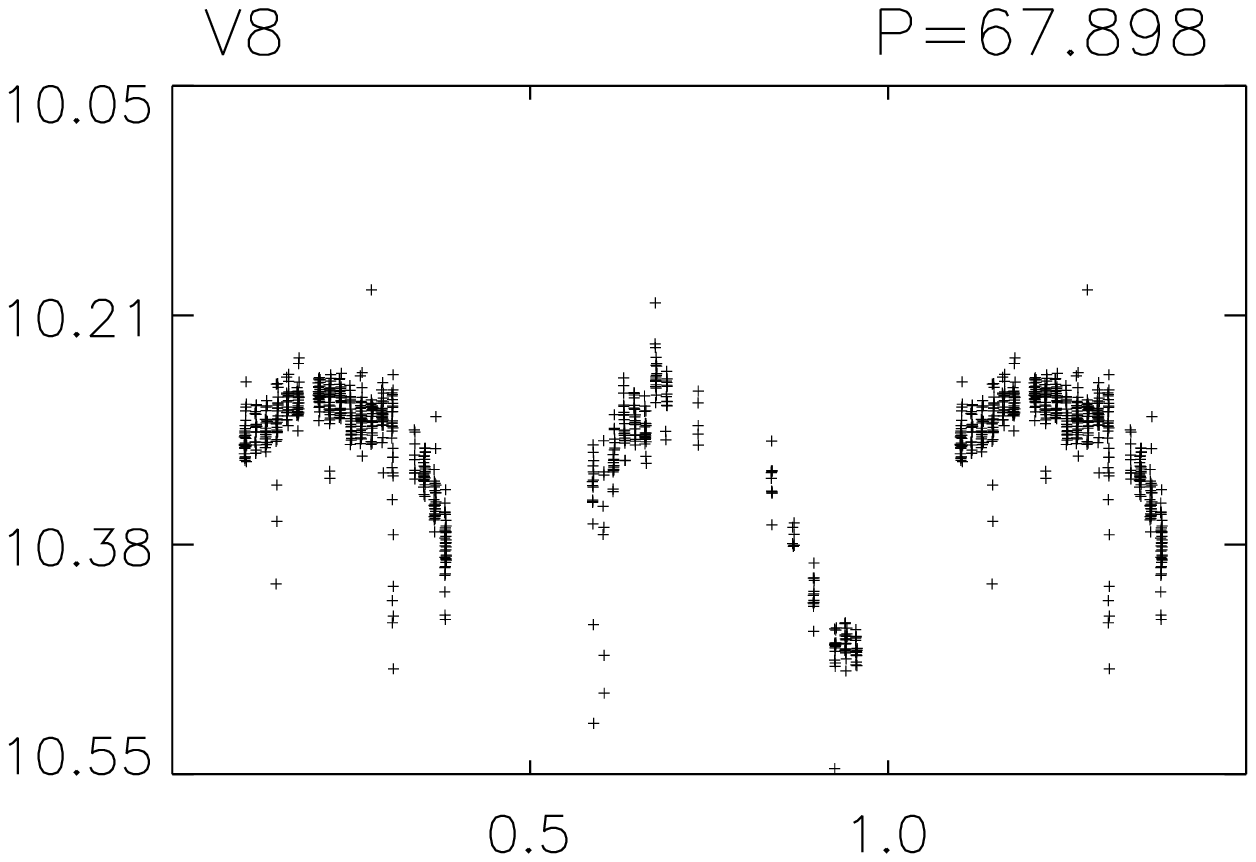}
   \hspace{-0.7cm}\includegraphics[width=4.65cm]{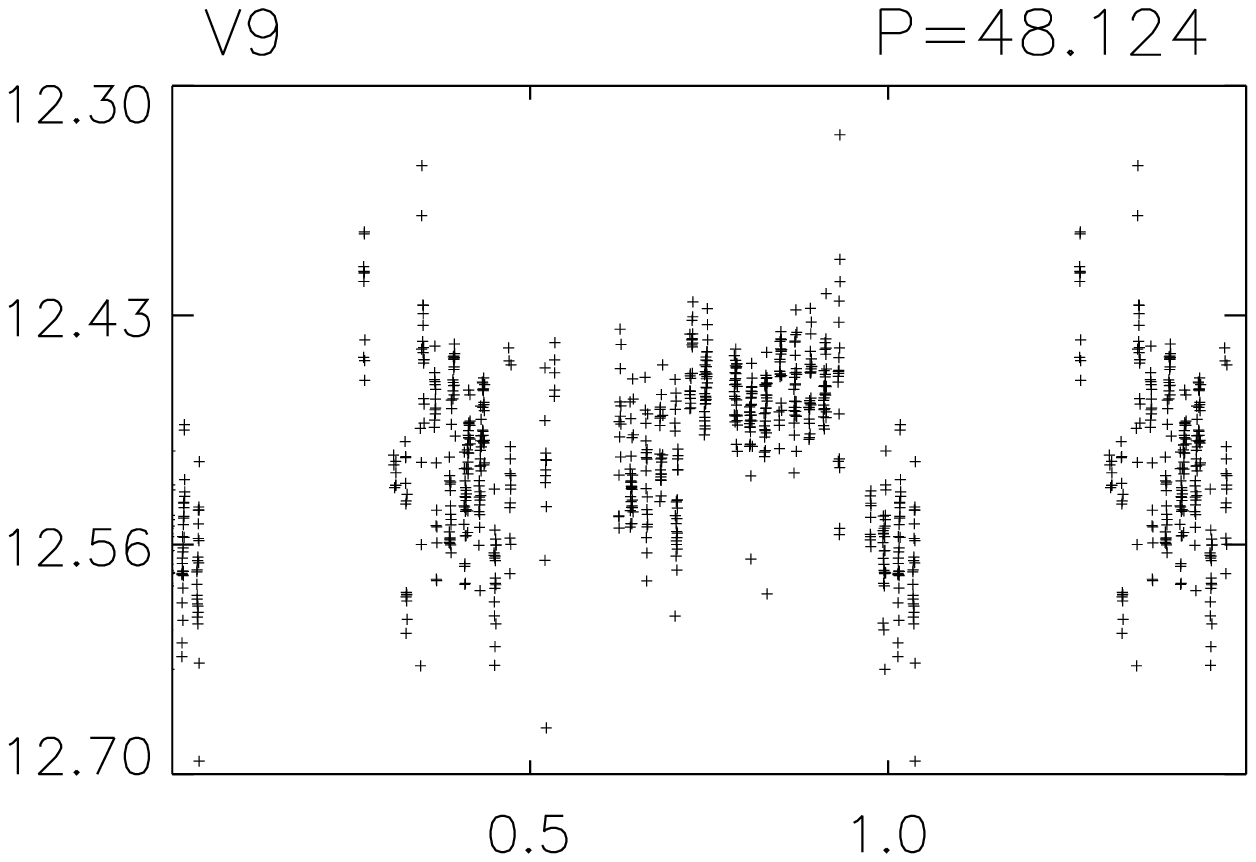}}
\vspace{-0.3cm}
   \hbox{\includegraphics[width=4.65cm]{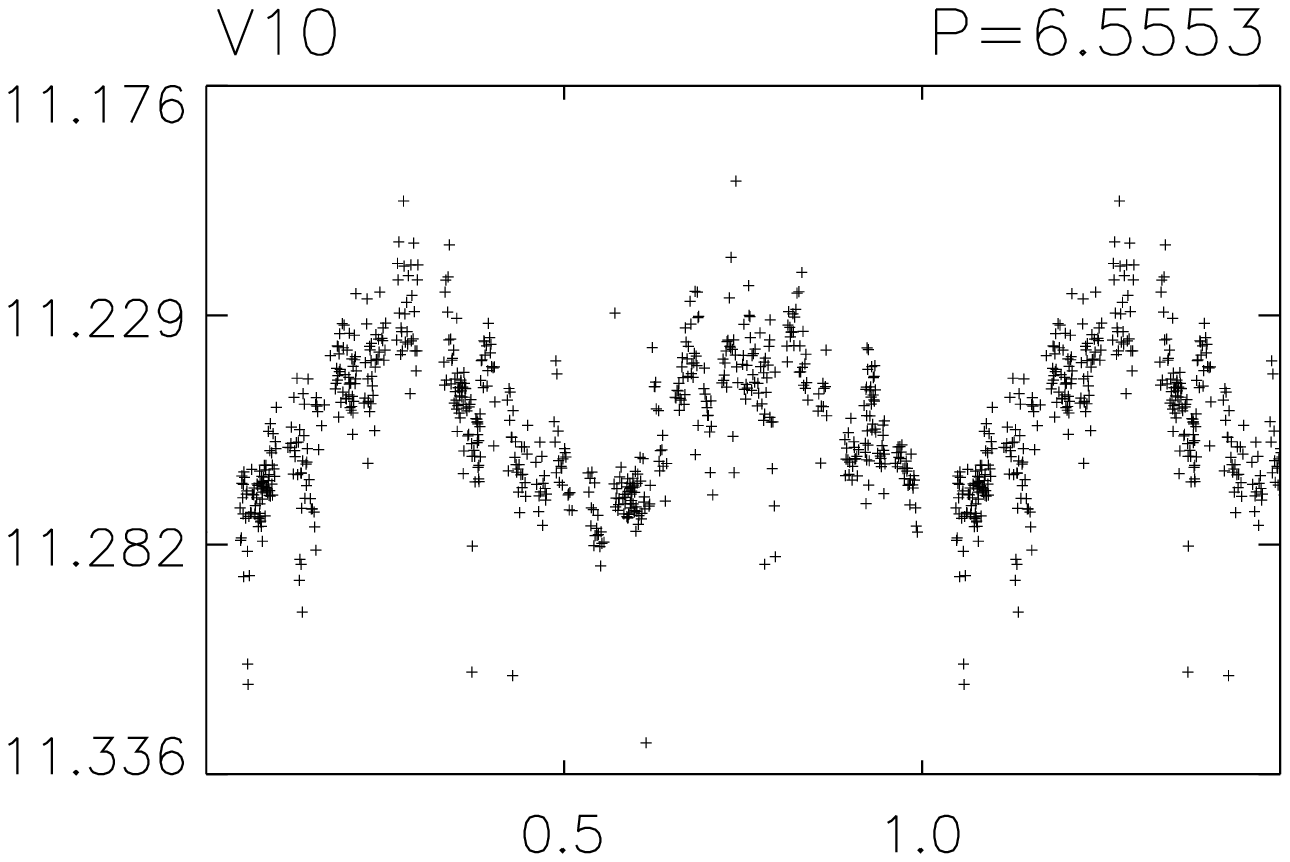}   
   \hspace{-0.7cm}\includegraphics[width=4.65cm]{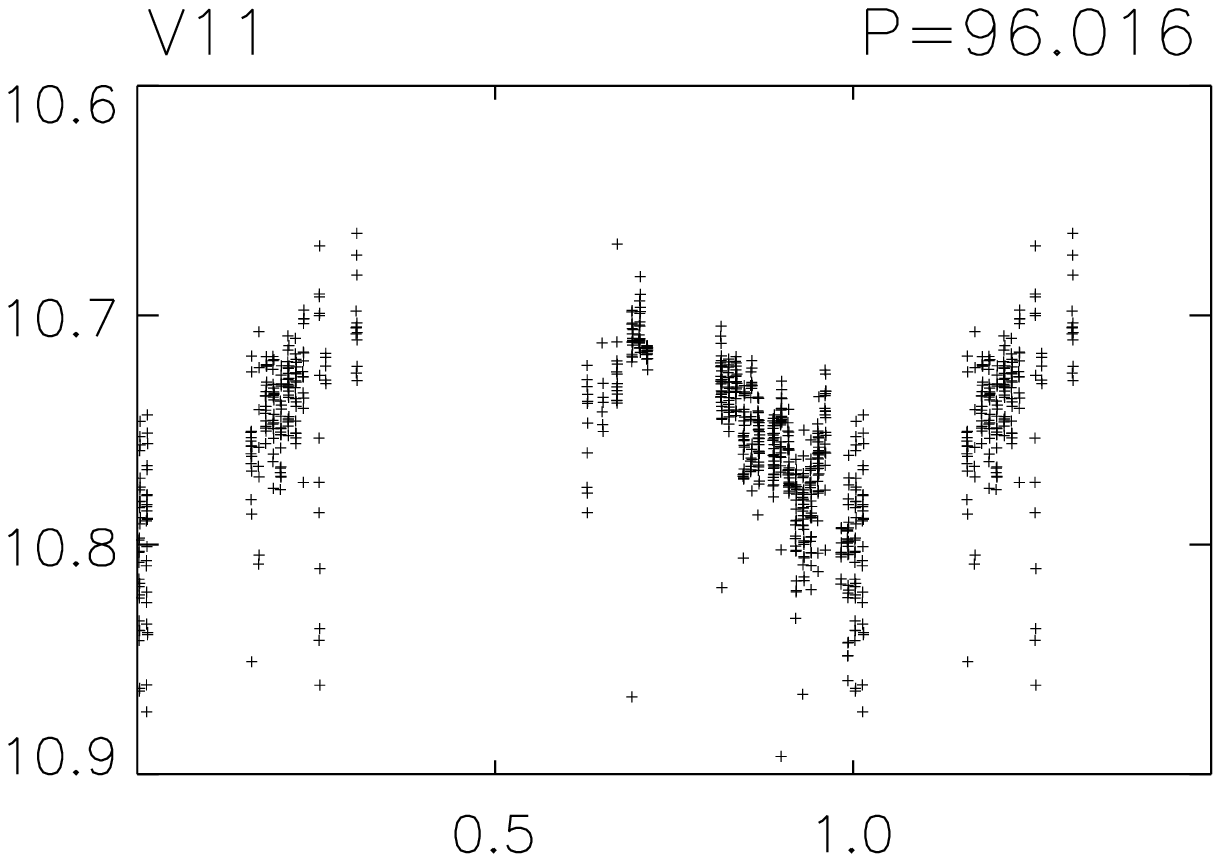}
   \hspace{-0.7cm}\includegraphics[width=4.65cm]{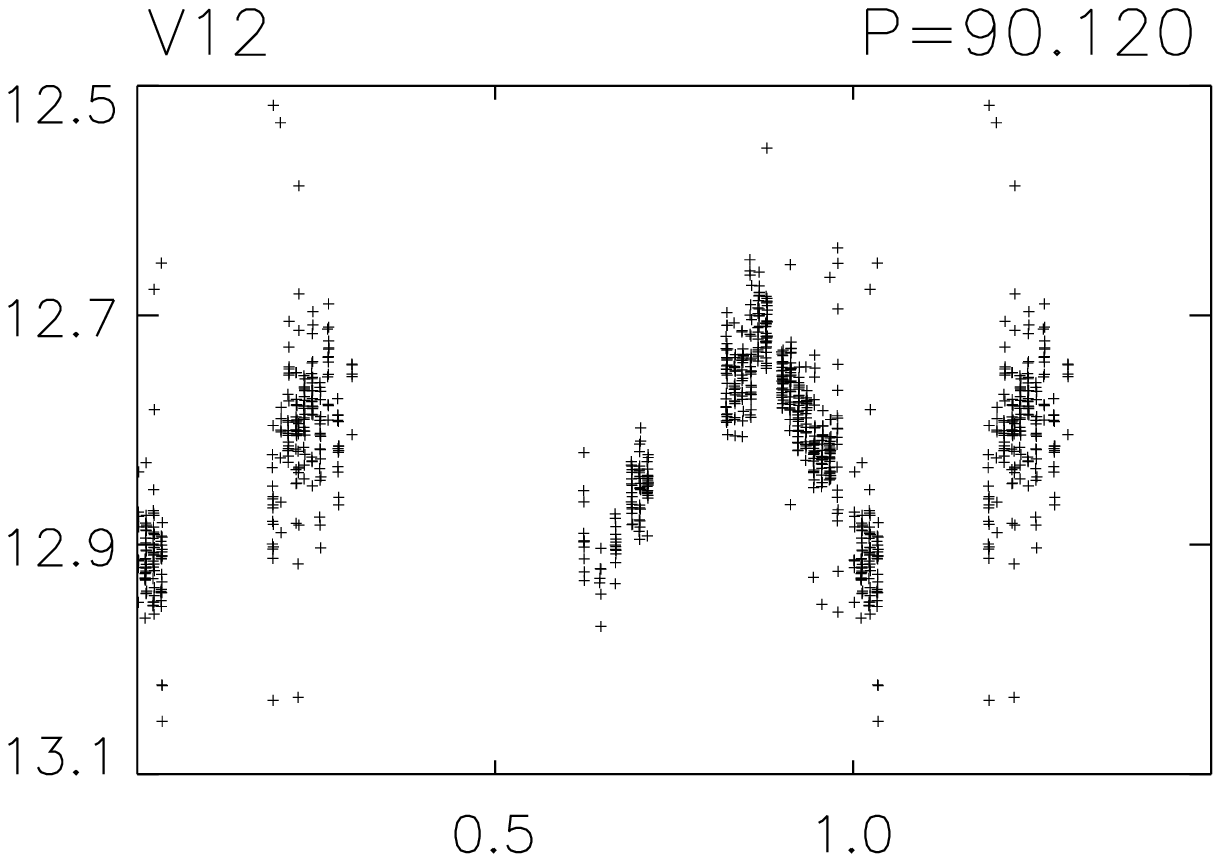}}
\vspace{-0.3cm}
\hbox{\includegraphics[width=4.65cm]{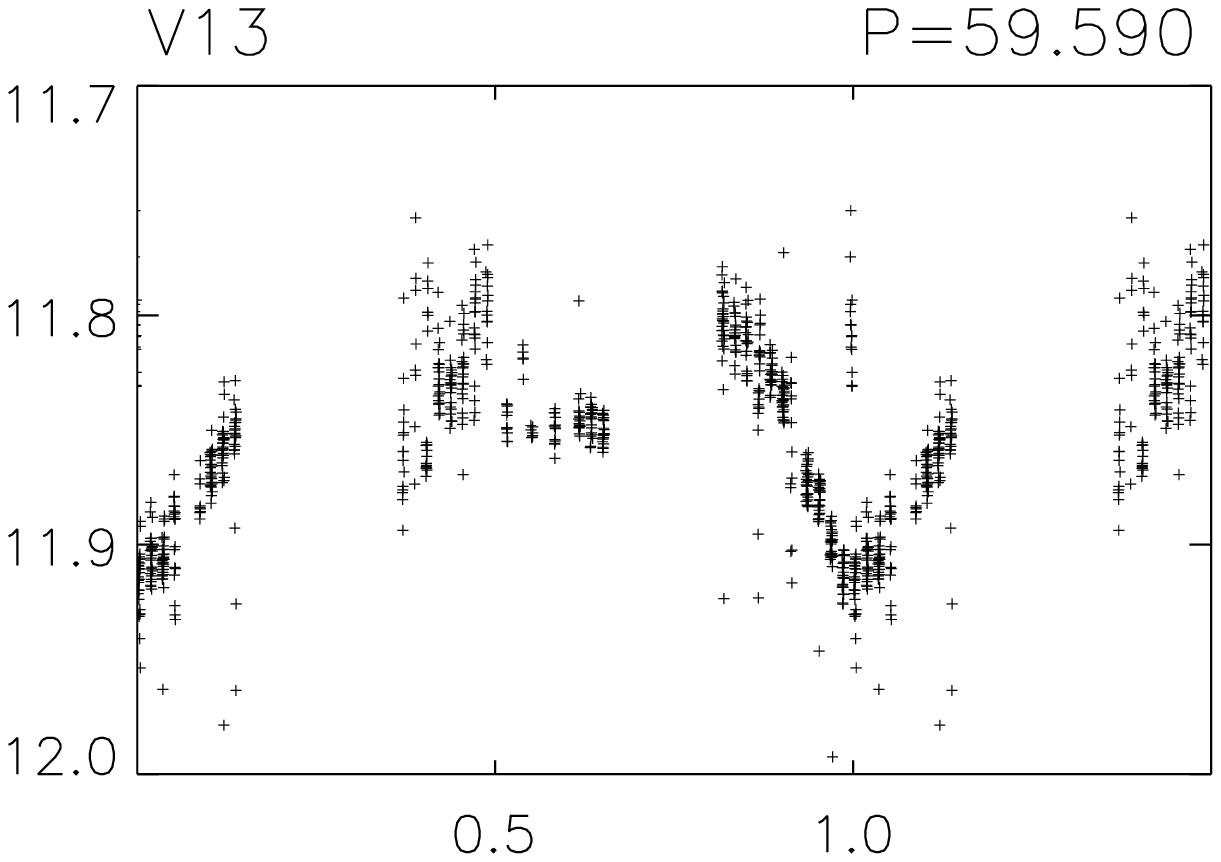}
   \hspace{-0.7cm}\includegraphics[width=4.65cm]{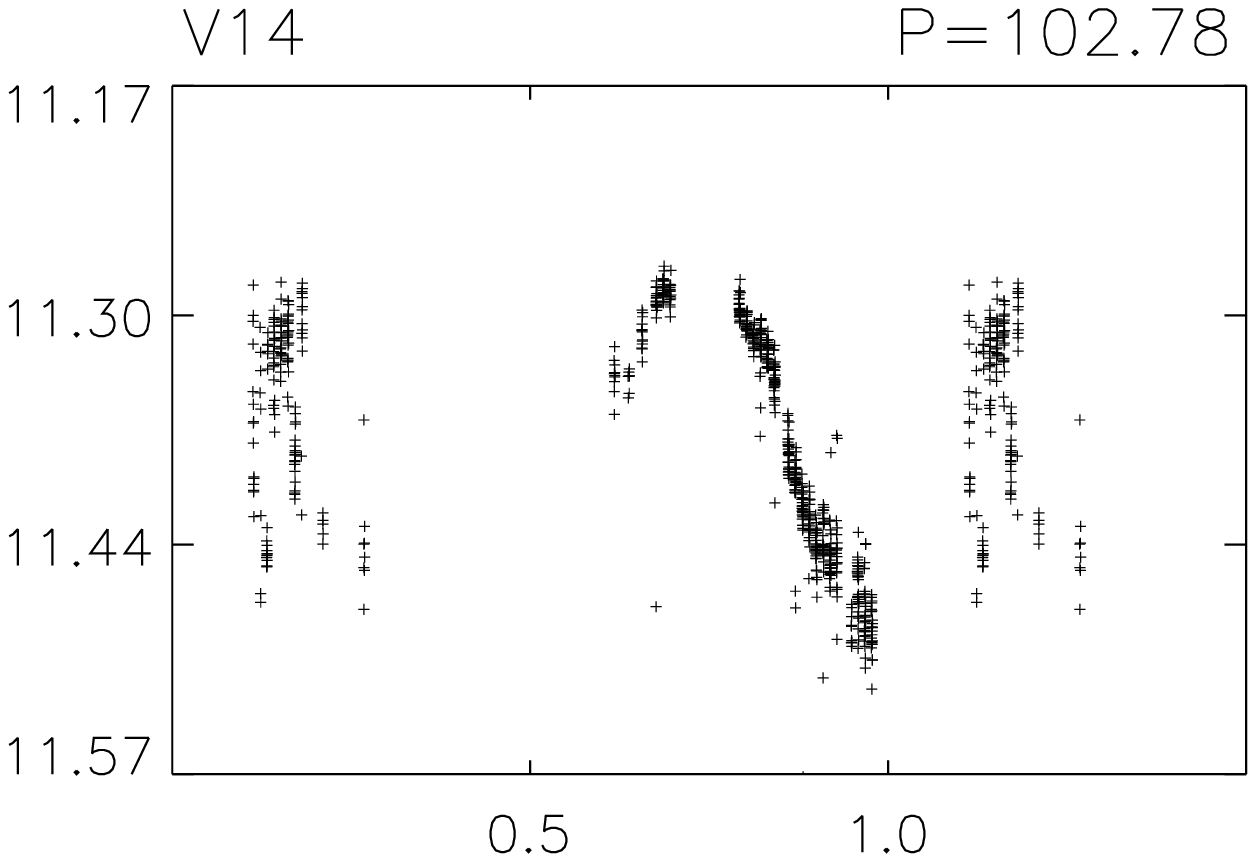}
   \hspace{-0.7cm}\includegraphics[width=4.65cm]{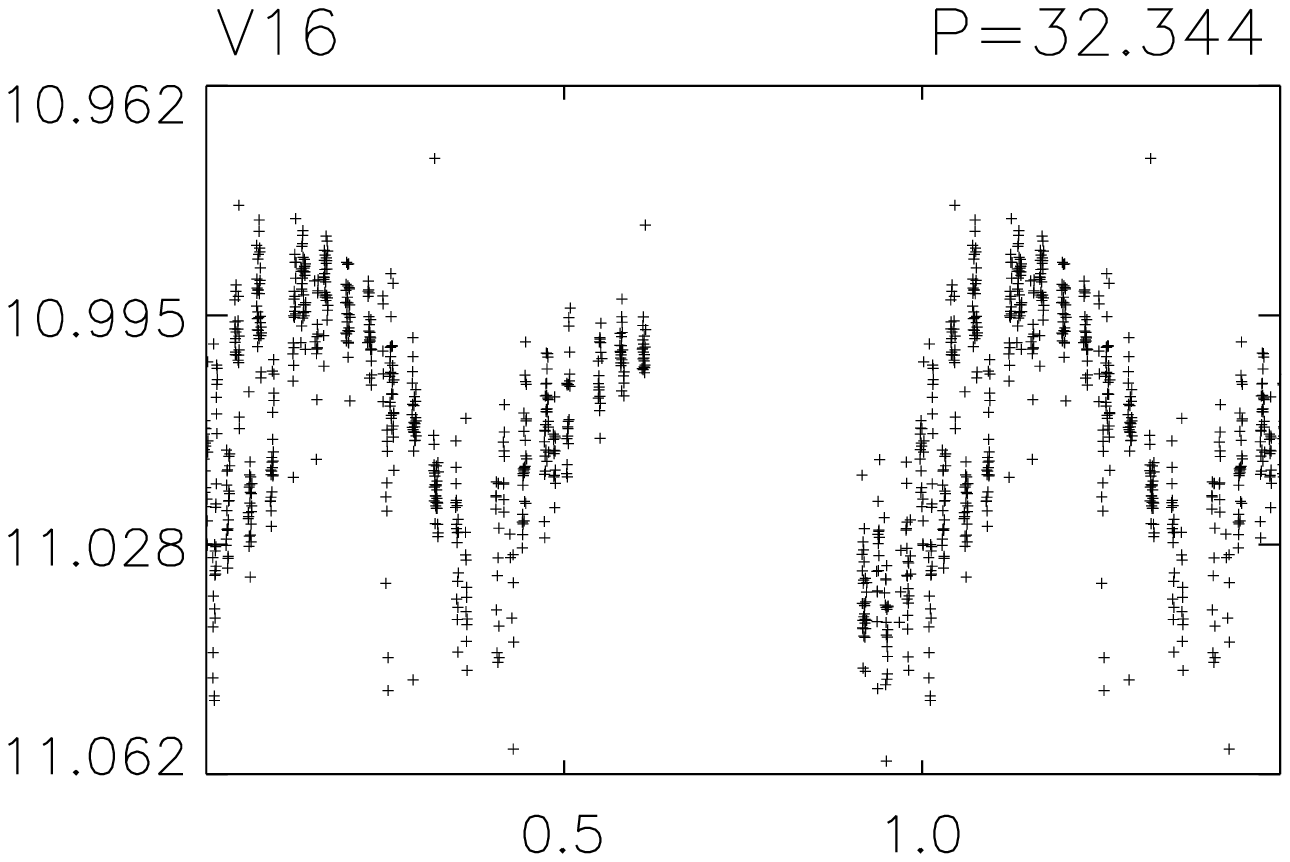}}
\vspace{-0.3cm}
   \hbox{\includegraphics[width=4.65cm]{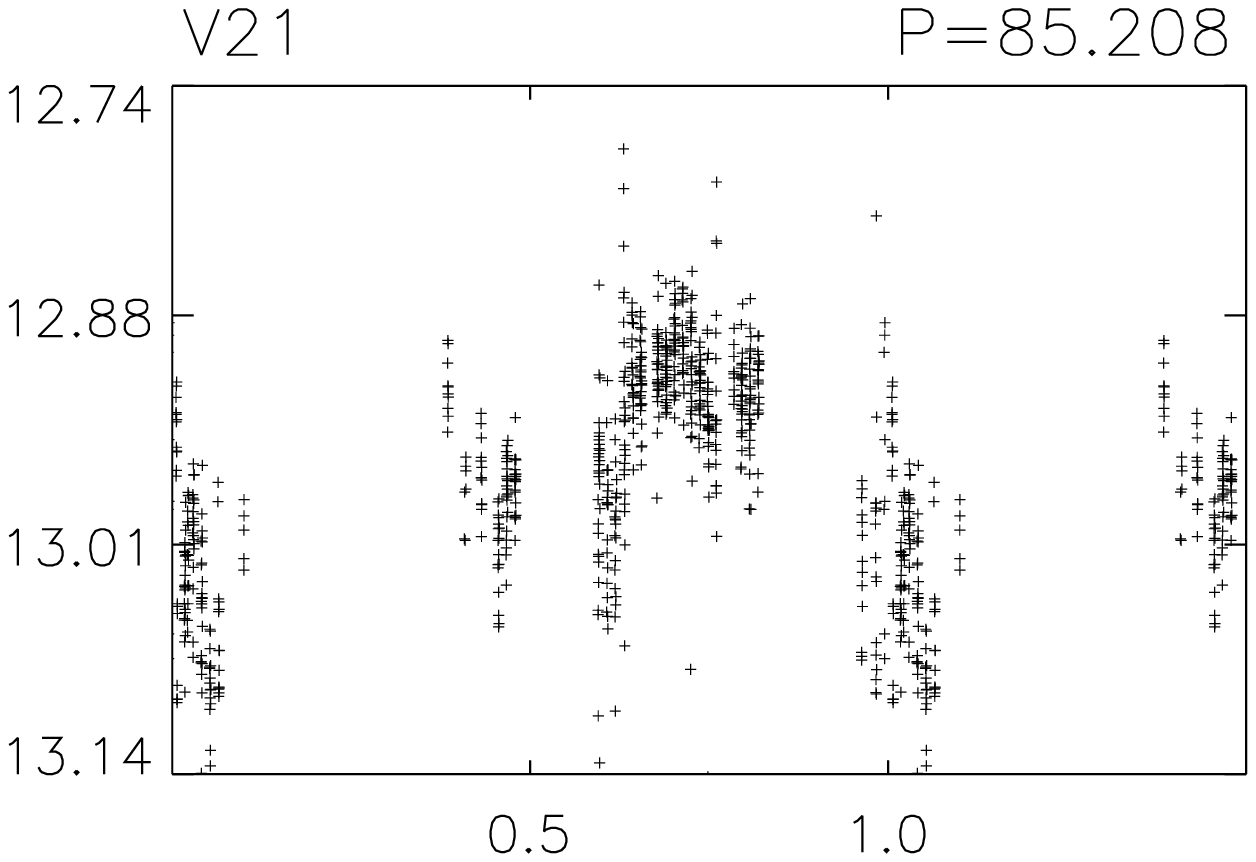}
   \hspace{-0.7cm}\includegraphics[width=4.65cm]{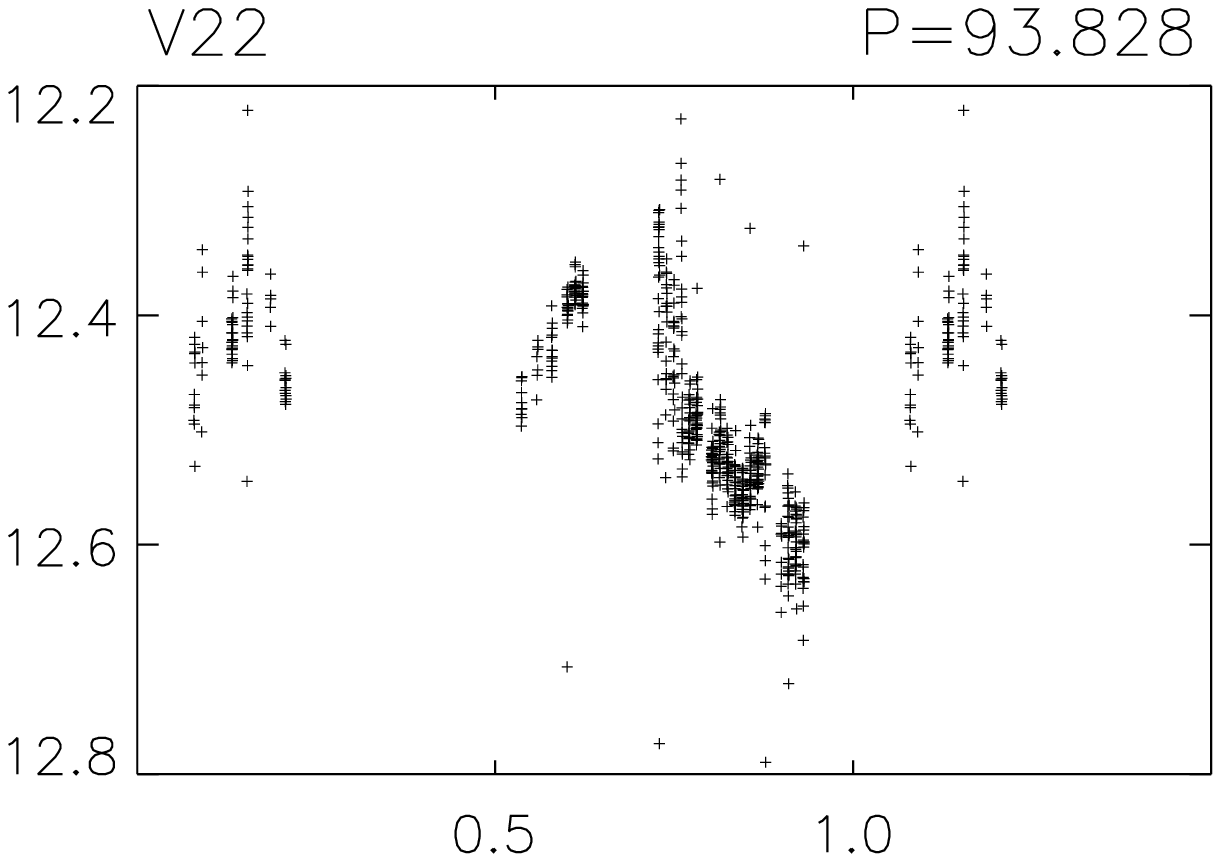}
   \hspace{-0.7cm}\includegraphics[width=4.65cm]{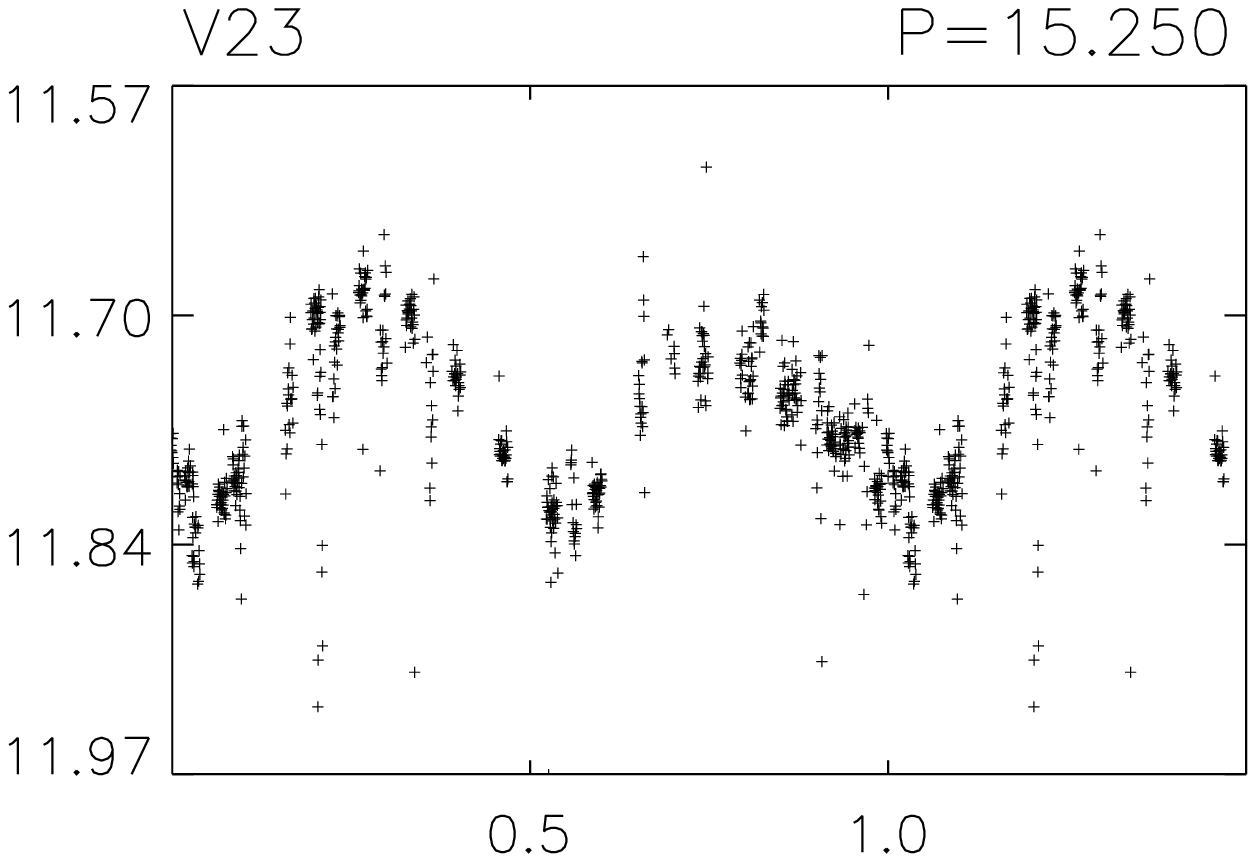}}
\vspace{-0.3cm}
   \hbox{\includegraphics[width=4.65cm]{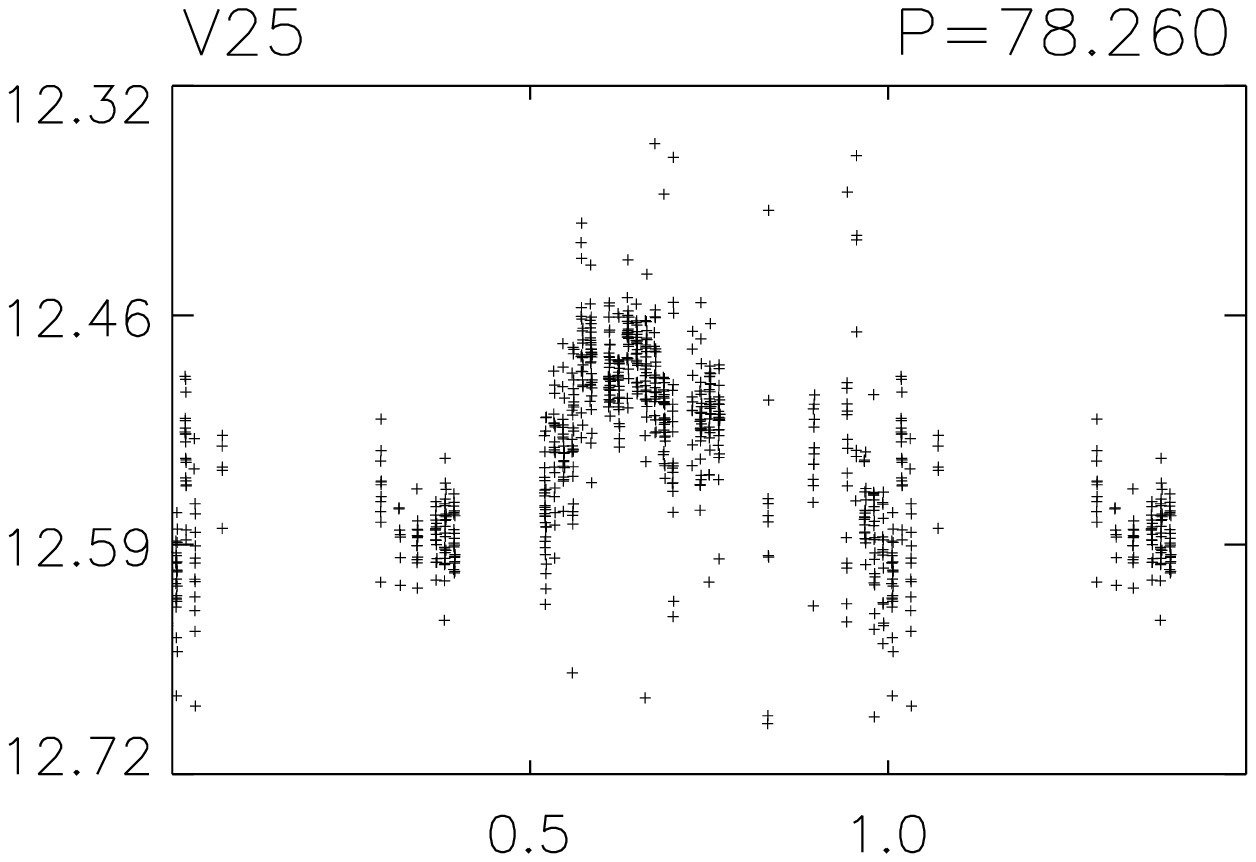}   
   \hspace{-0.7cm}\includegraphics[width=4.65cm]{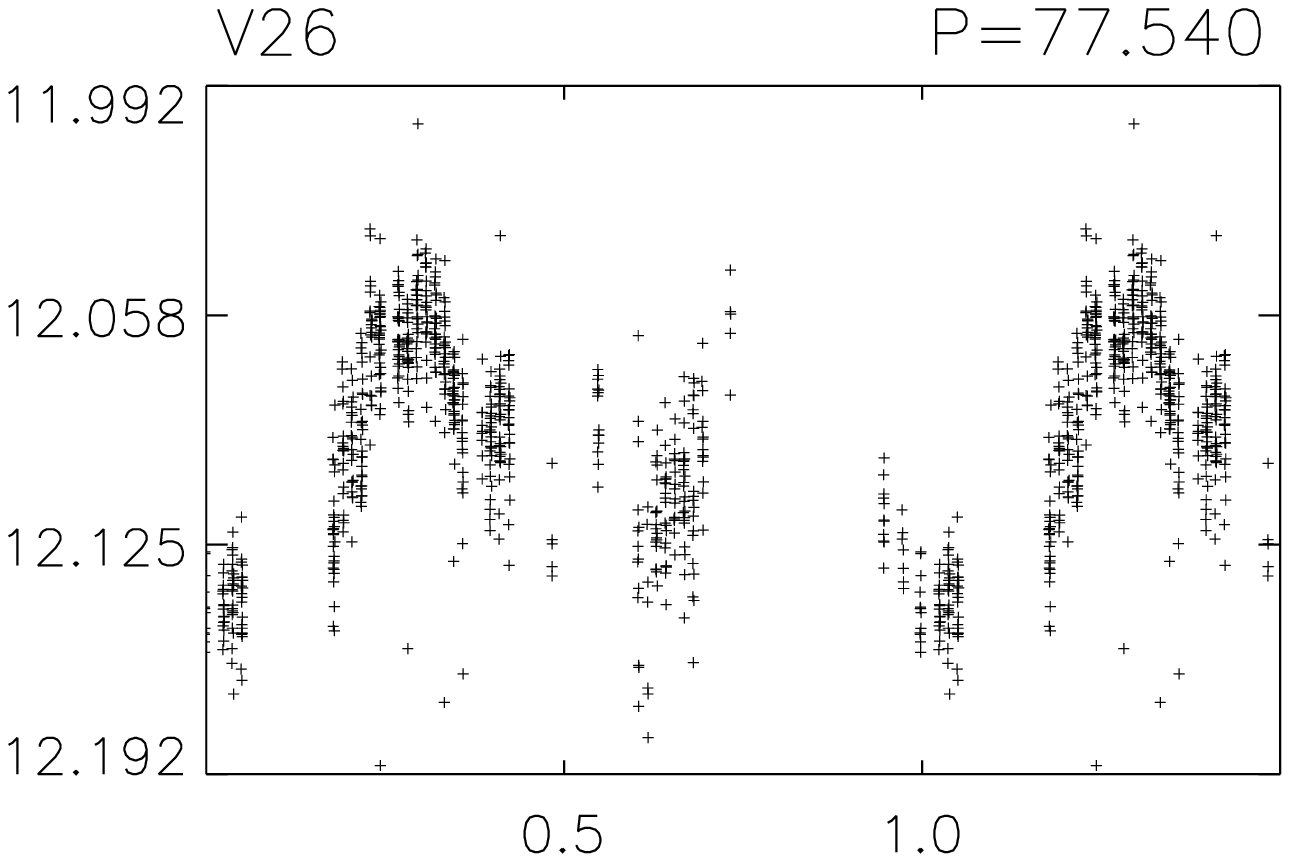}
   \hspace{-0.7cm}\includegraphics[width=4.65cm]{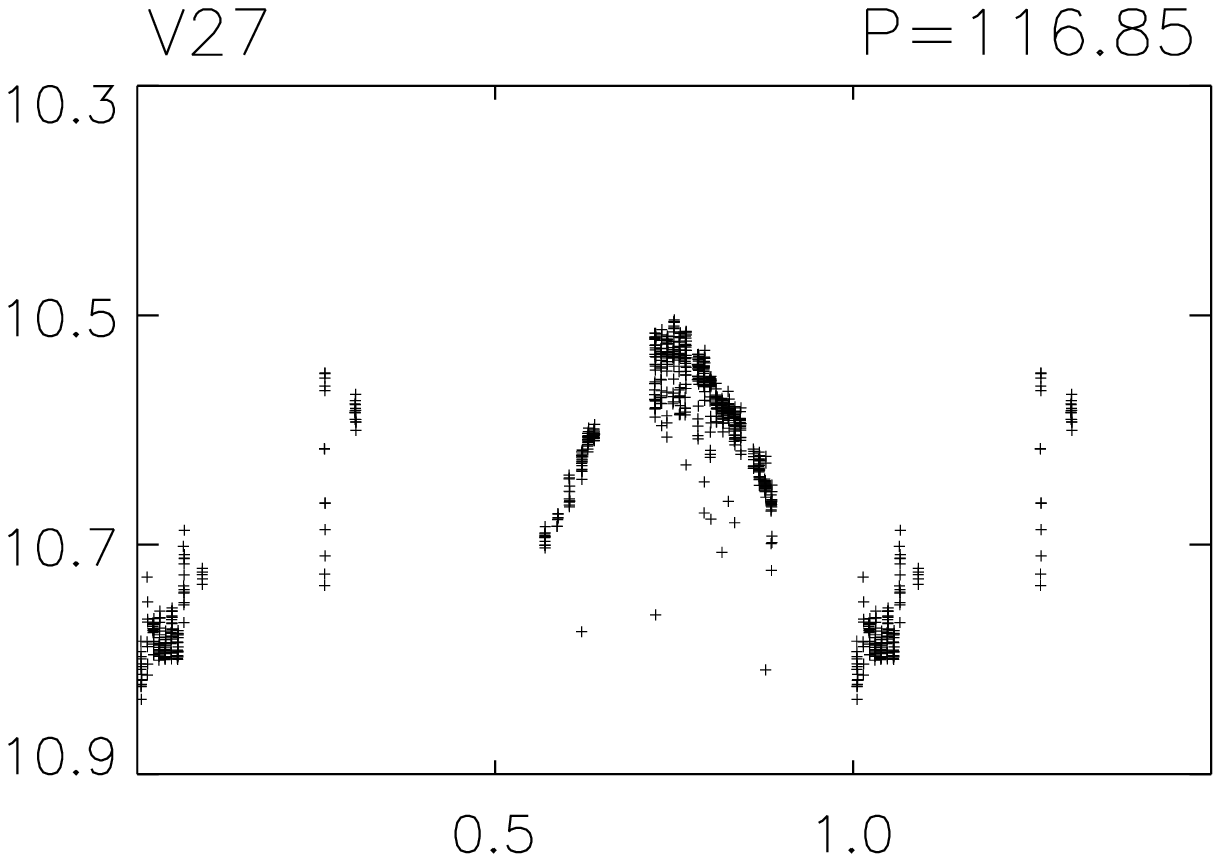}}
\vspace{-0.3cm}
   \hbox{\includegraphics[width=4.65cm]{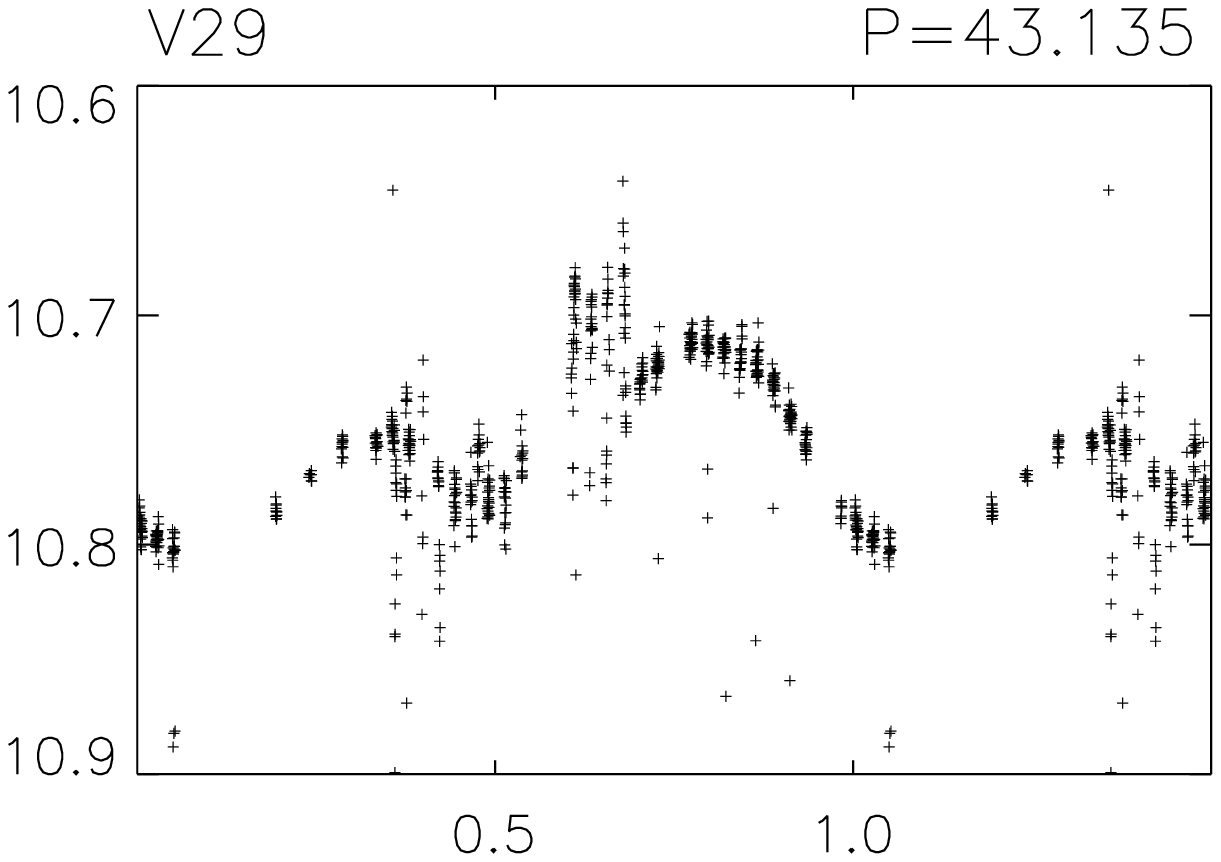}
   \hspace{-0.7cm}\includegraphics[width=4.65cm]{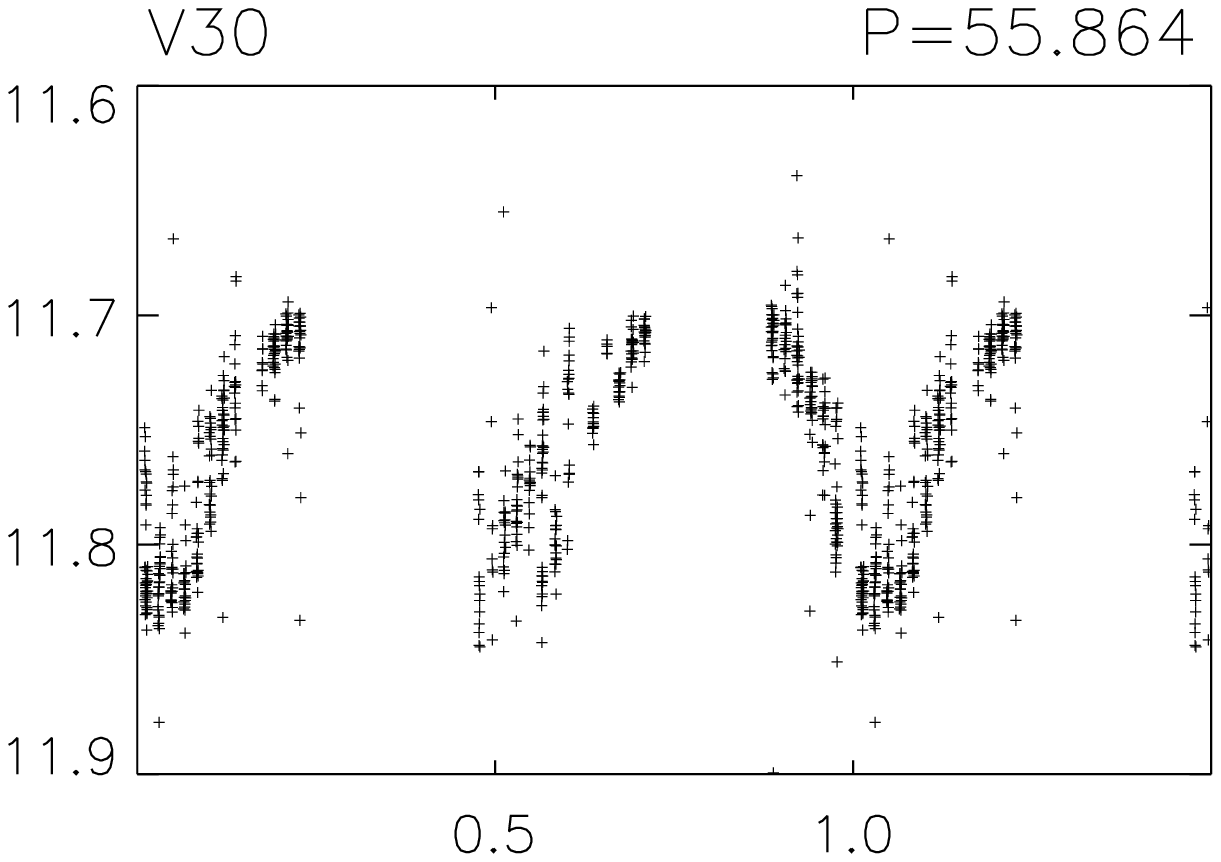}
   \hspace{-0.7cm}\includegraphics[width=4.65cm]{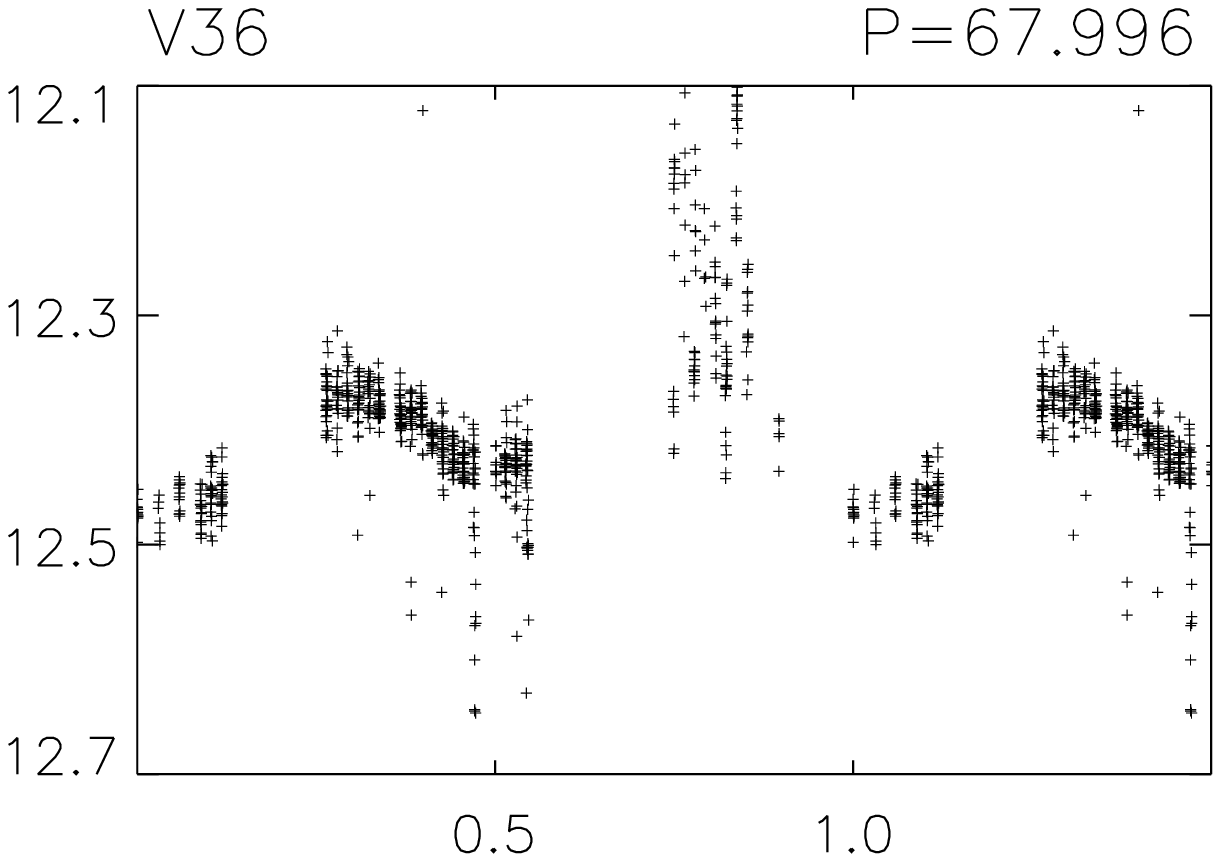}}
\caption{Light curves of 43 periodic pulsating stars in $R$. Magnitude is shown versus phase. The period is given in the upper right of each diagram is days.}
\end{figure*} 
\addtocounter{figure}{-1}
\begin{figure*}[!ht]
\centering
   \hbox{\includegraphics[width=4.65cm]{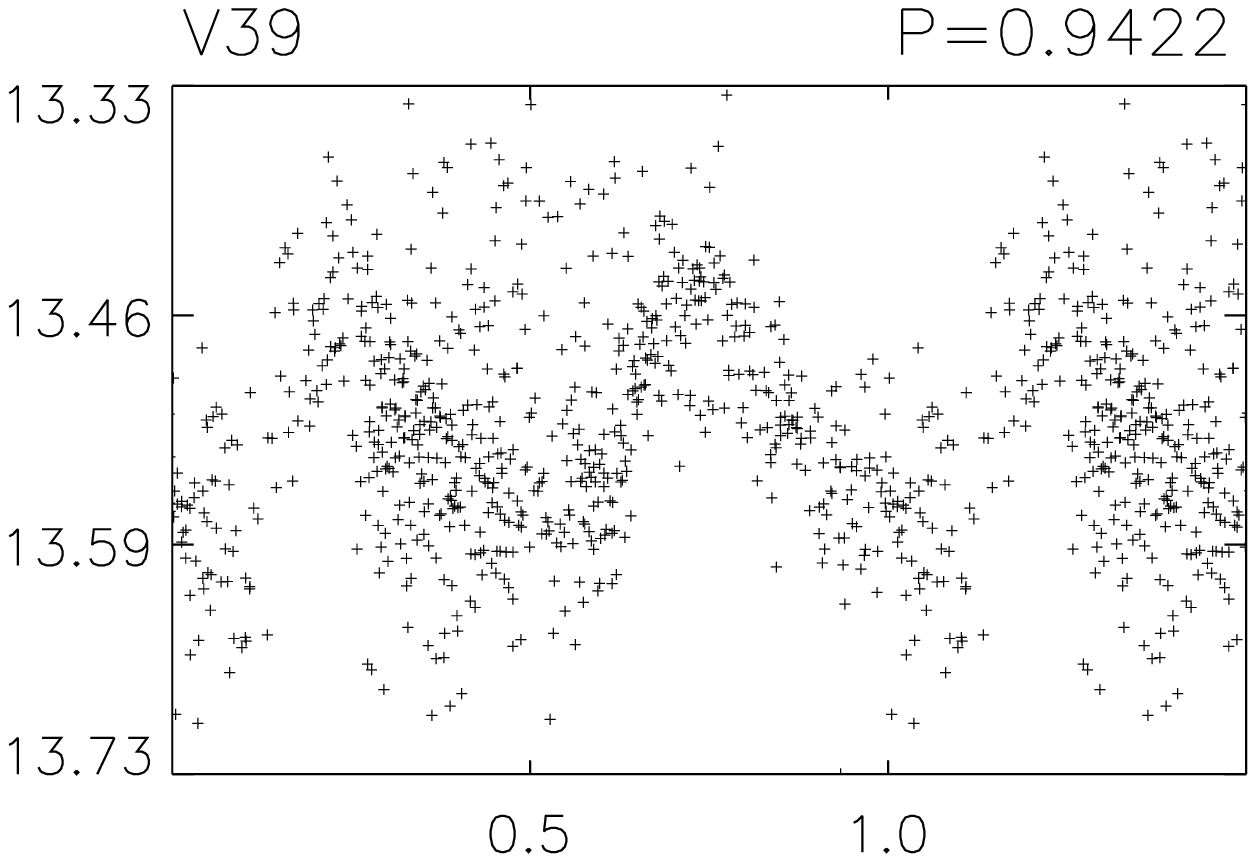}
   \hspace{-0.7cm}\includegraphics[width=4.65cm]{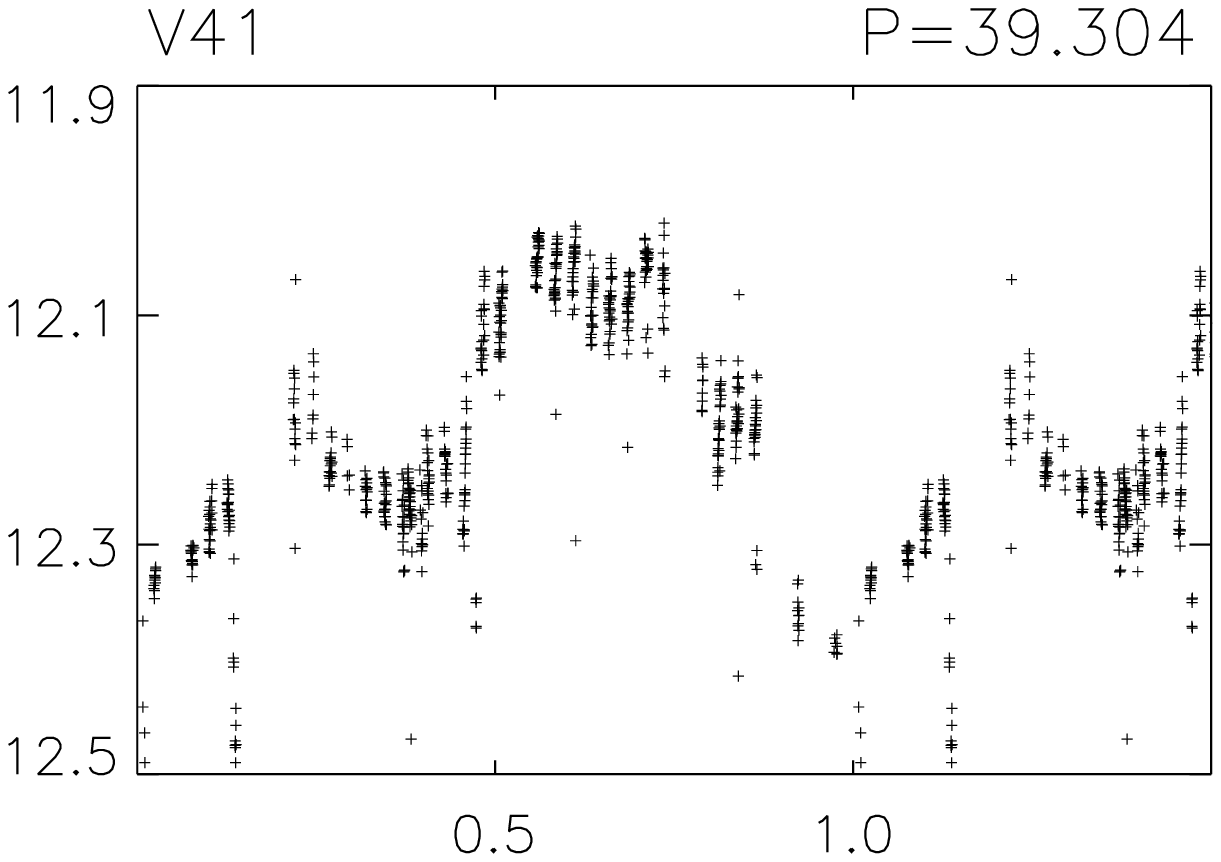}
   \hspace{-0.7cm}\includegraphics[width=4.65cm]{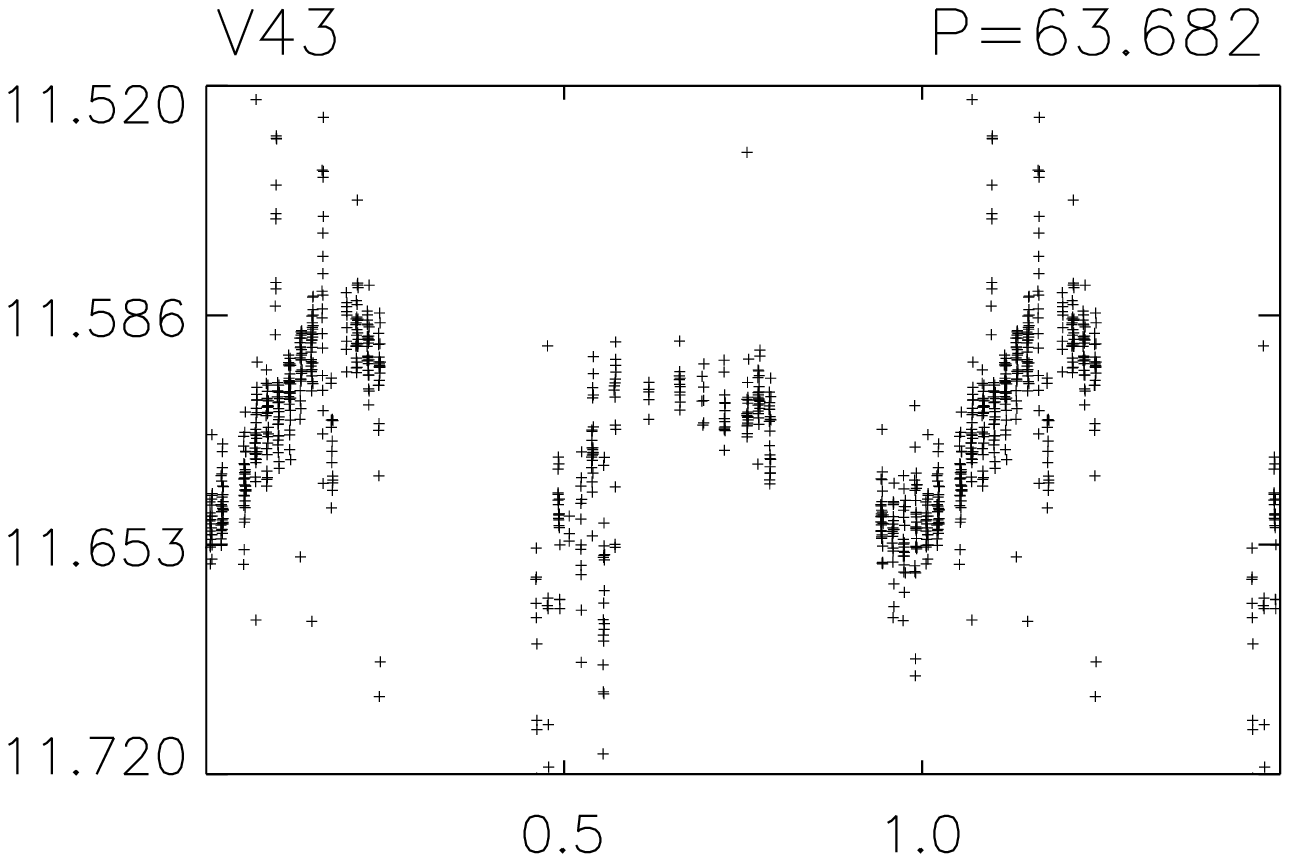}}
\vspace{-0.3cm}
   \hbox{\includegraphics[width=4.65cm]{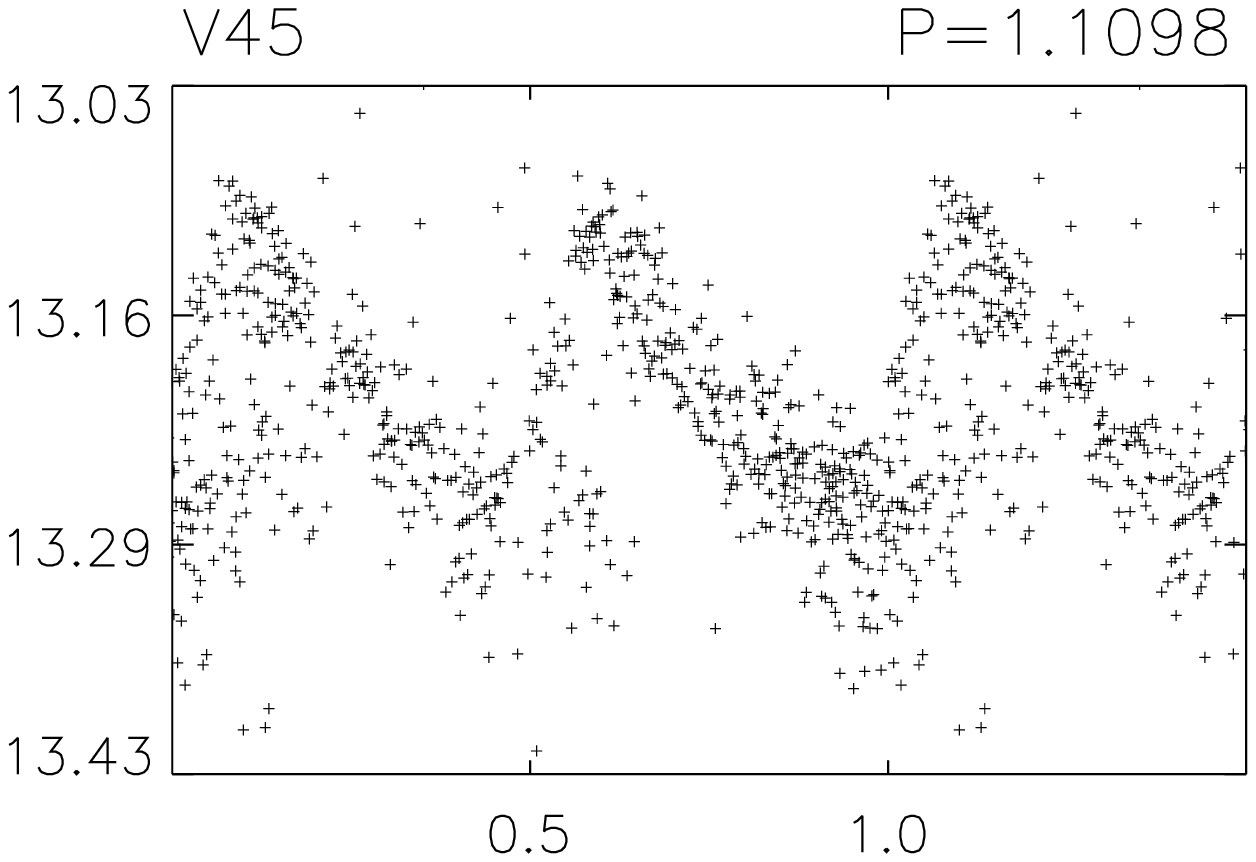}   
   \hspace{-0.7cm}\includegraphics[width=4.65cm]{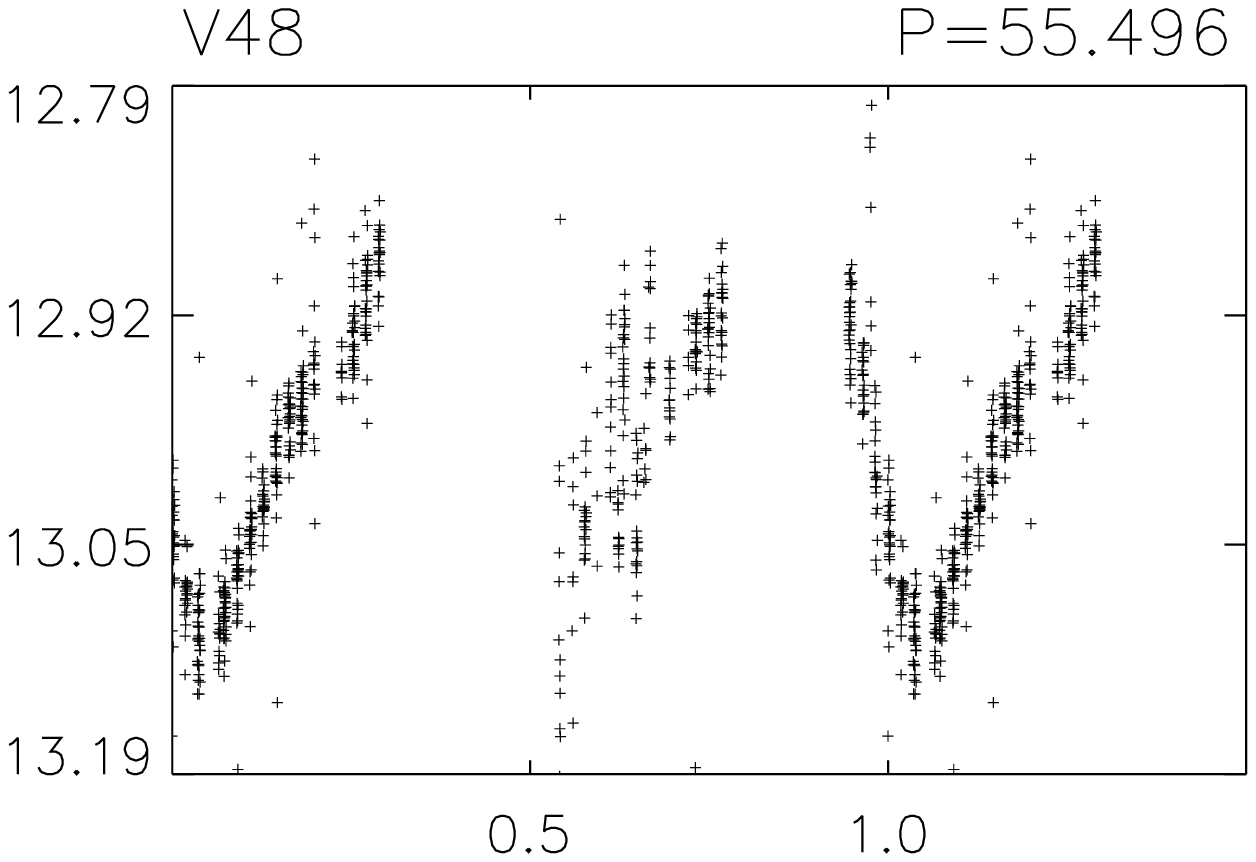}
   \hspace{-0.7cm}\includegraphics[width=4.65cm]{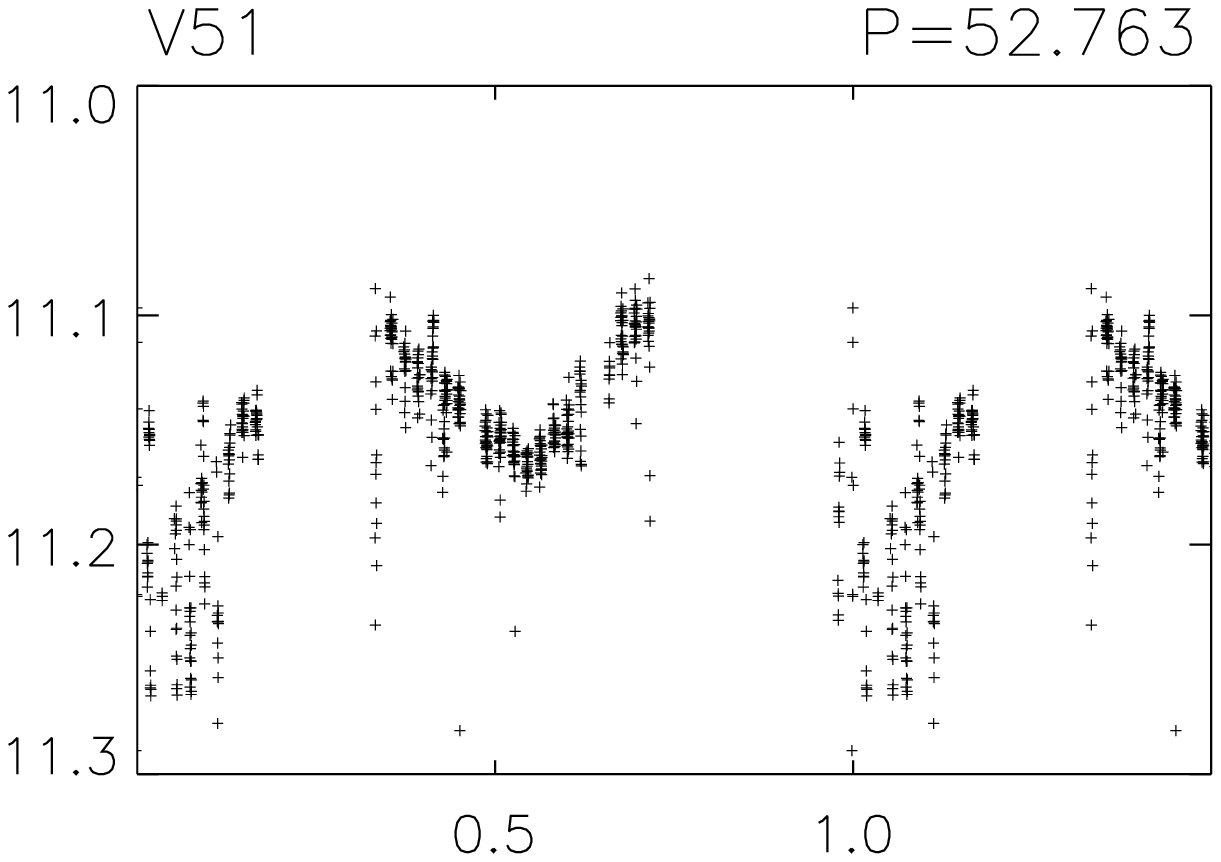}}
\vspace{-0.3cm}
   \hbox{\includegraphics[width=4.65cm]{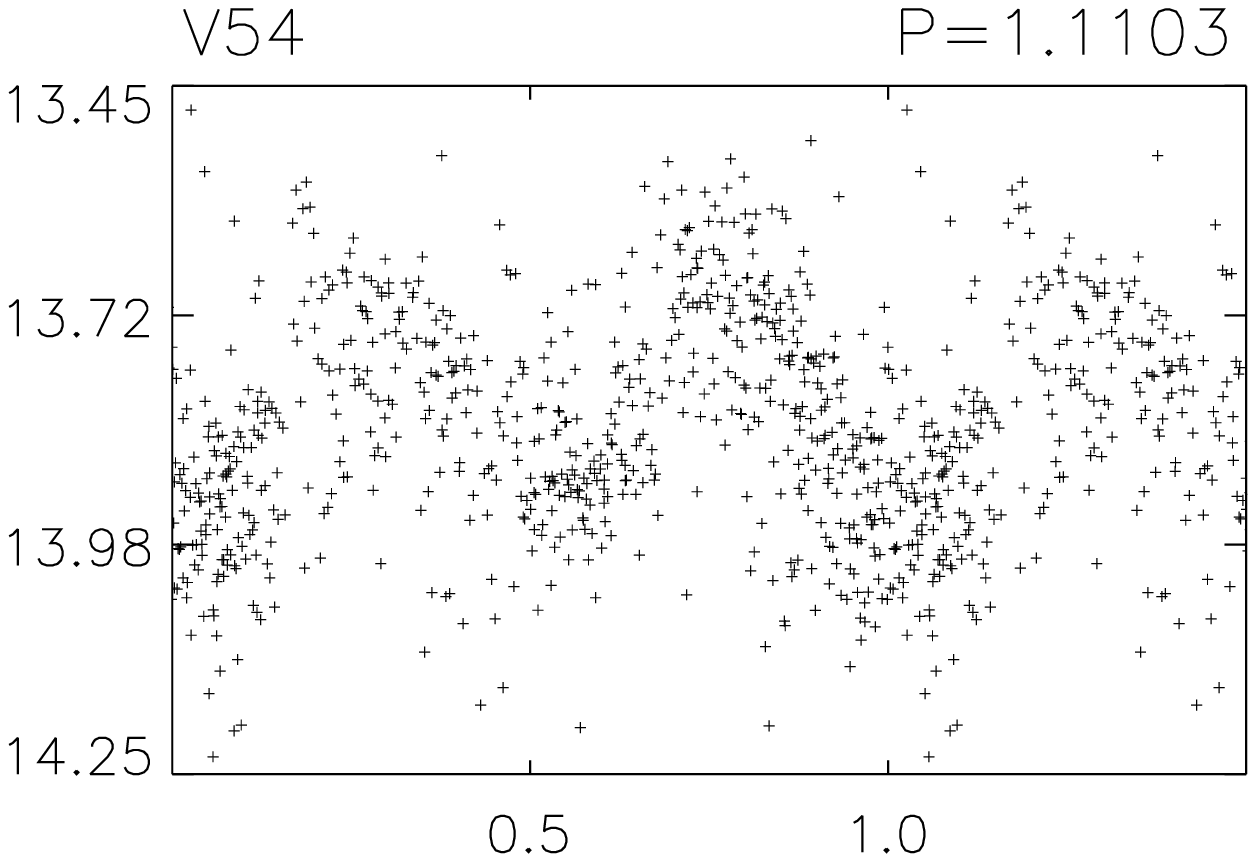}
   \hspace{-0.7cm}\includegraphics[width=4.65cm]{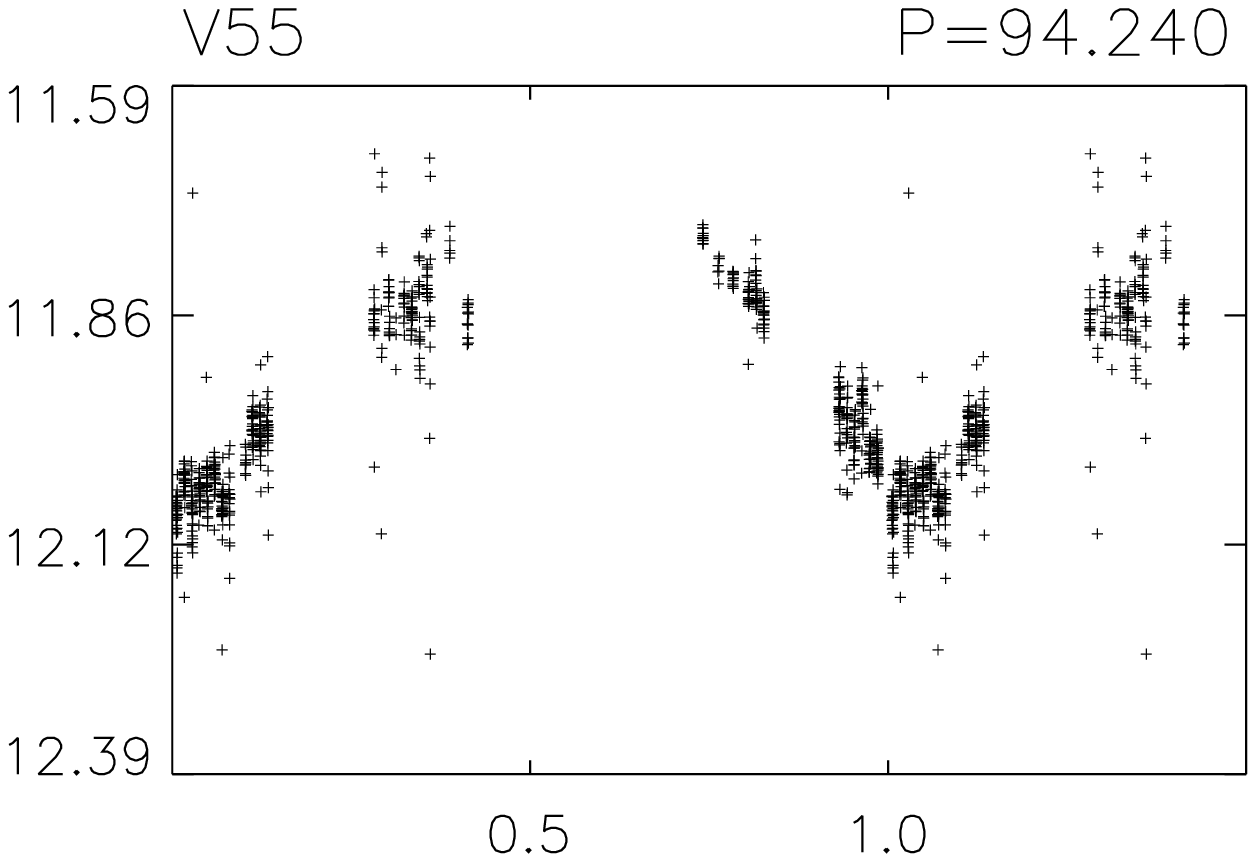}
   \hspace{-0.7cm}\includegraphics[width=4.65cm]{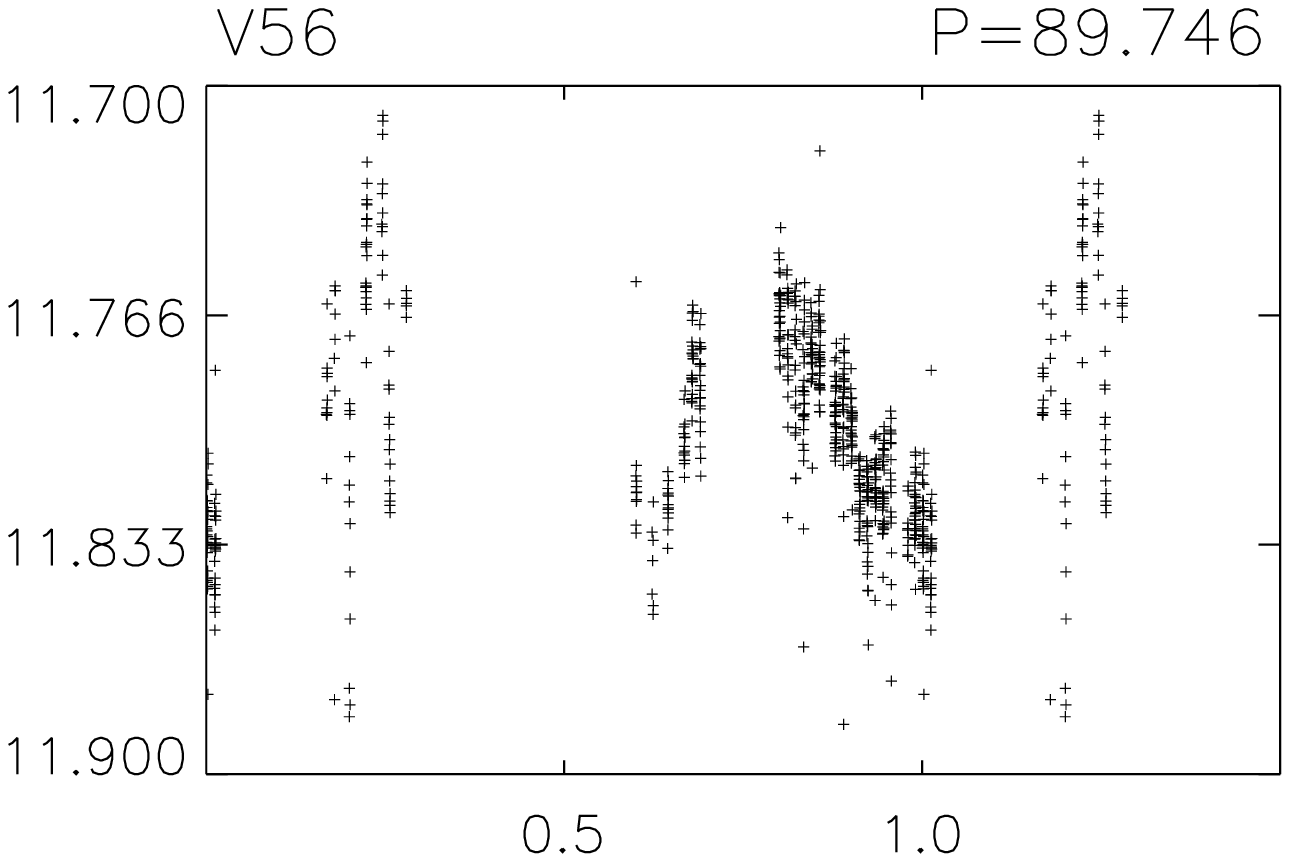}}
\vspace{-0.3cm}
   \hbox{\includegraphics[width=4.65cm]{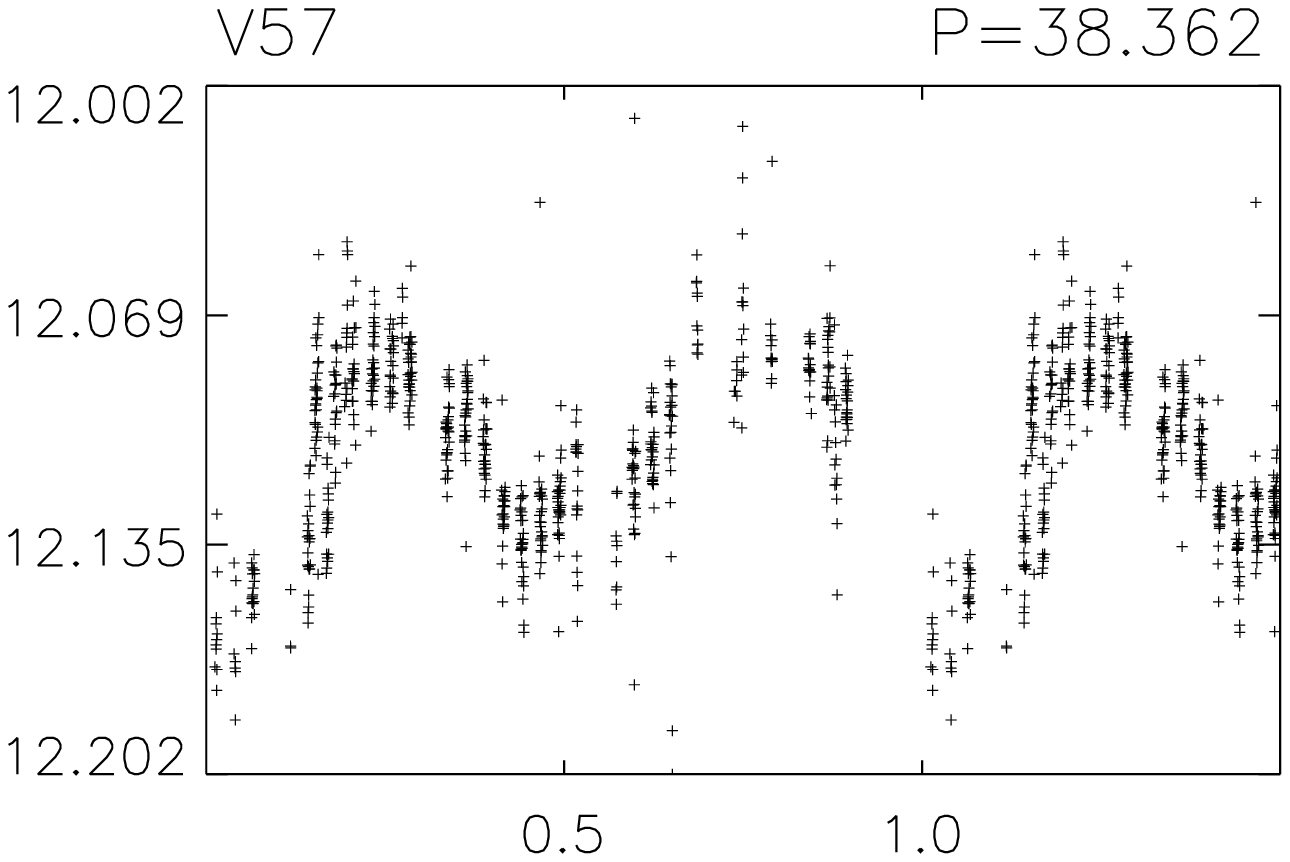}
   \hspace{-0.7cm}\includegraphics[width=4.65cm]{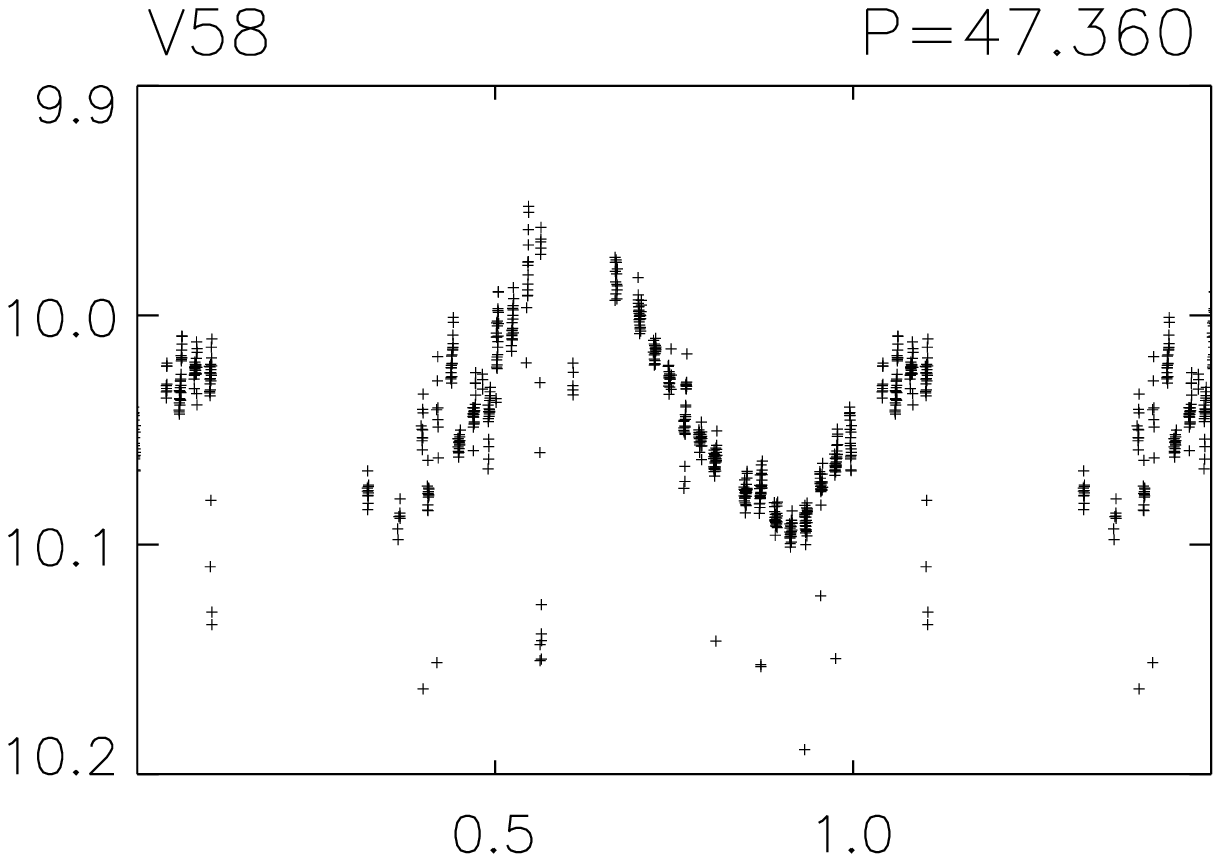}
   \hspace{-0.7cm}\includegraphics[width=4.65cm]{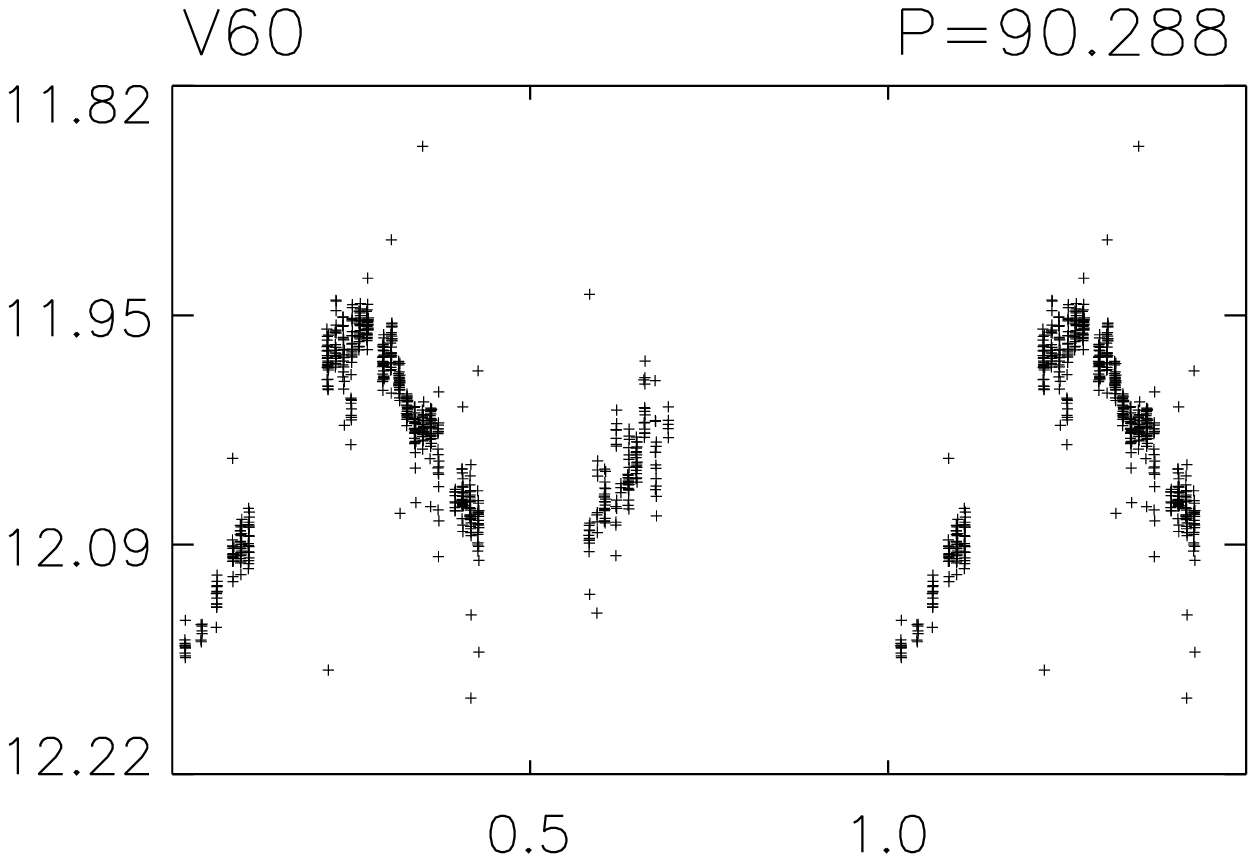}}
\vspace{-0.3cm}
   \hbox{\includegraphics[width=4.65cm]{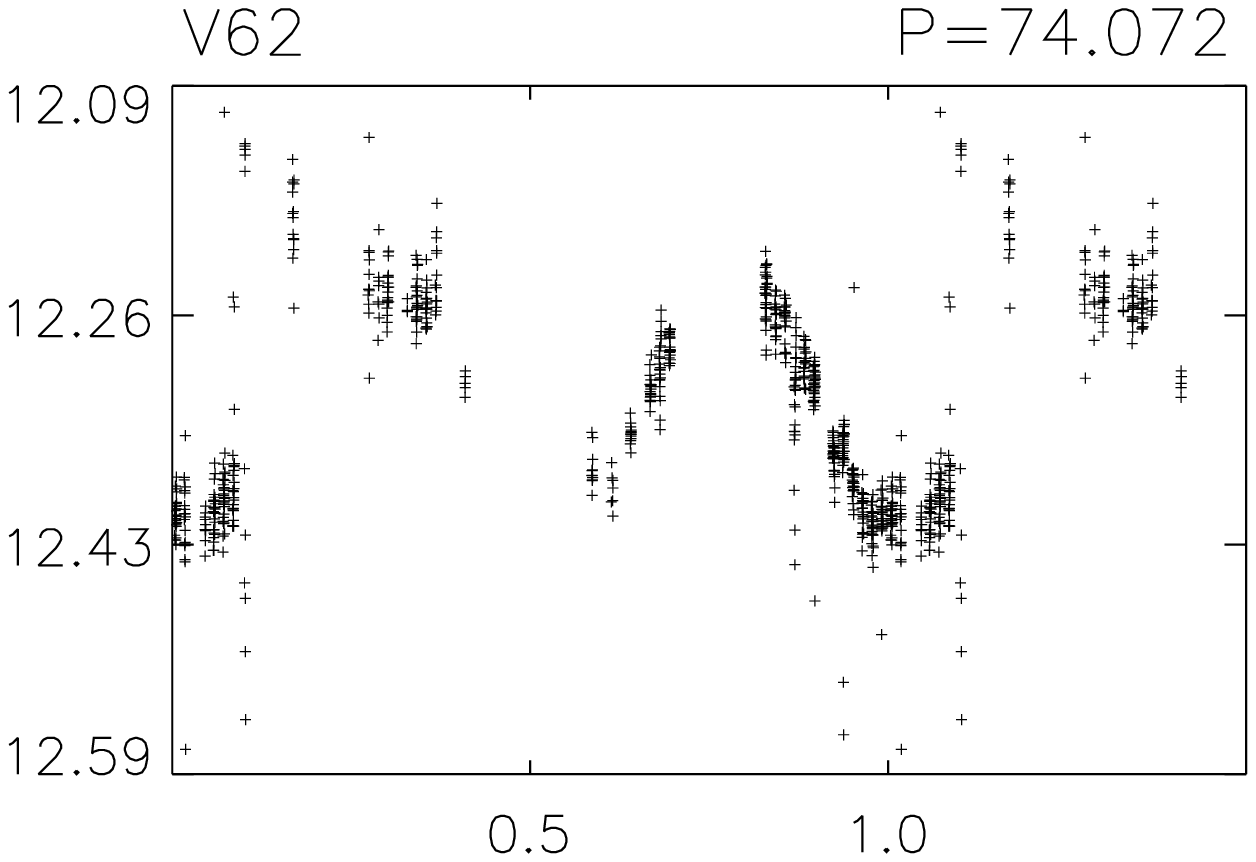}   
   \hspace{-0.7cm}\includegraphics[width=4.65cm]{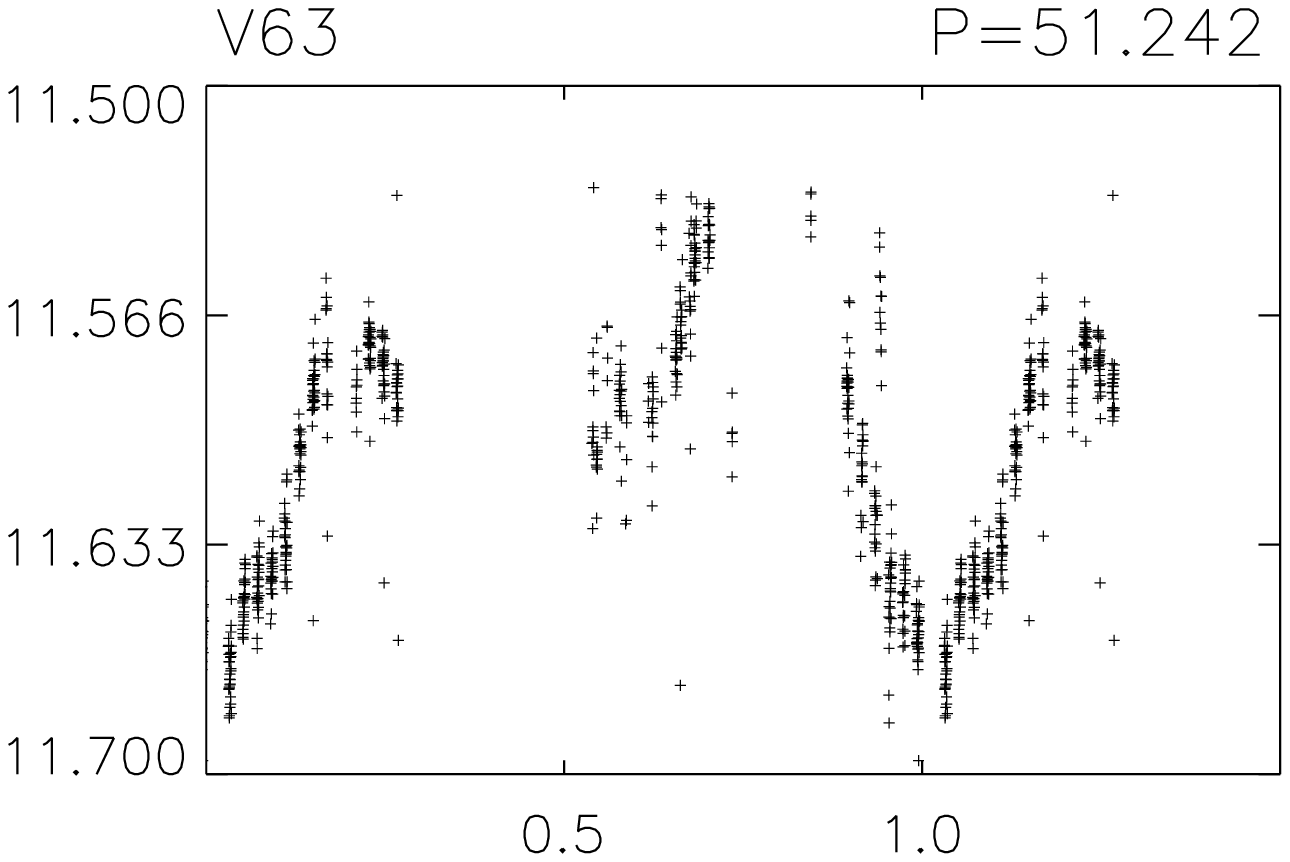}
   \hspace{-0.7cm}\includegraphics[width=4.65cm]{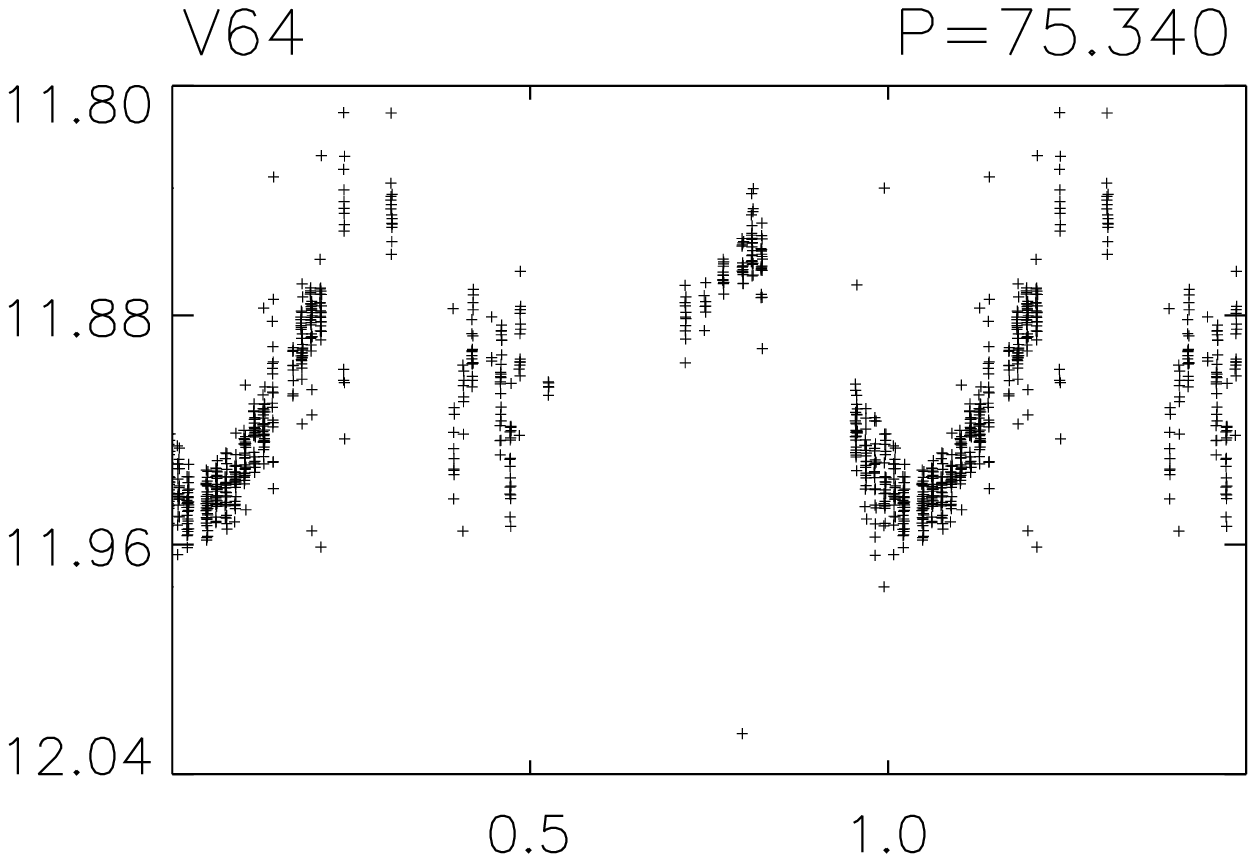}}
\vspace{-0.3cm}
   \hbox{\includegraphics[width=4.65cm]{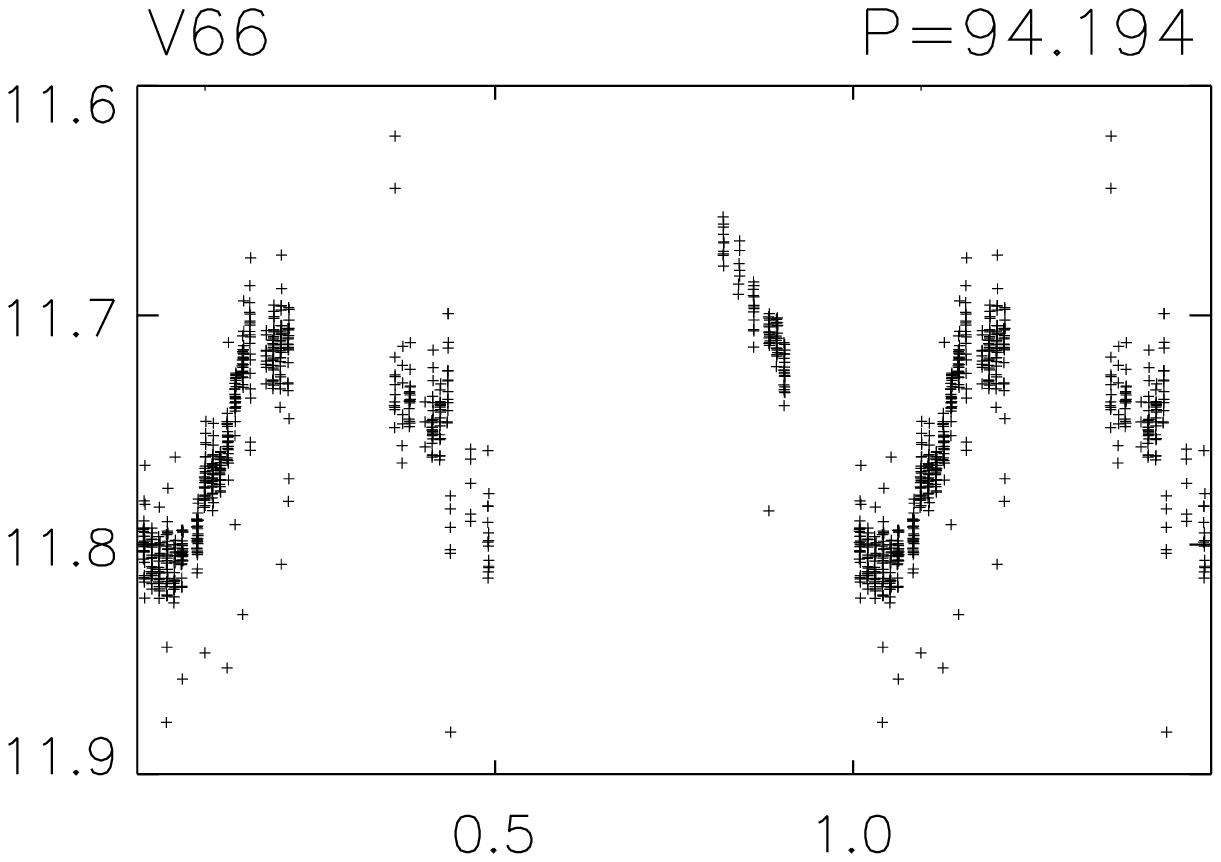}
   \hspace{-0.7cm}\includegraphics[width=4.65cm]{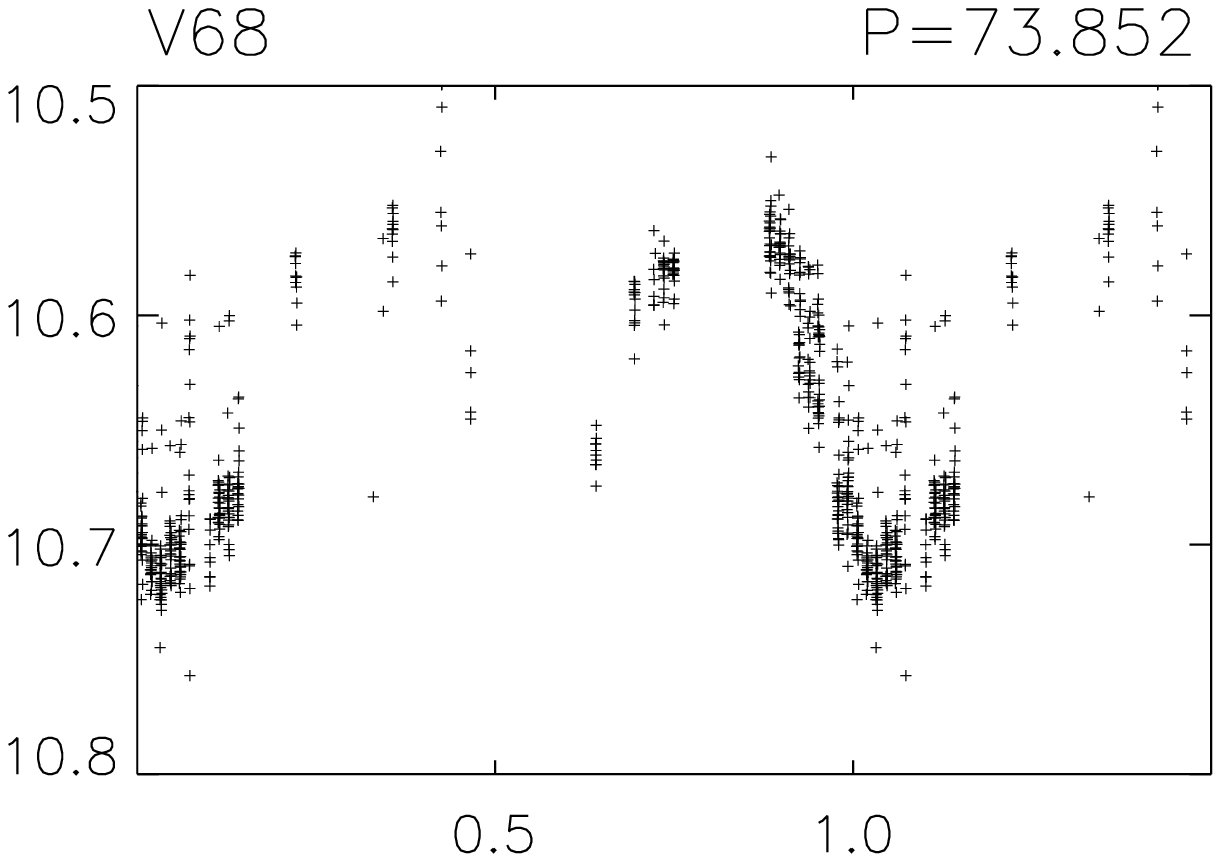}
   \hspace{-0.7cm}\includegraphics[width=4.65cm]{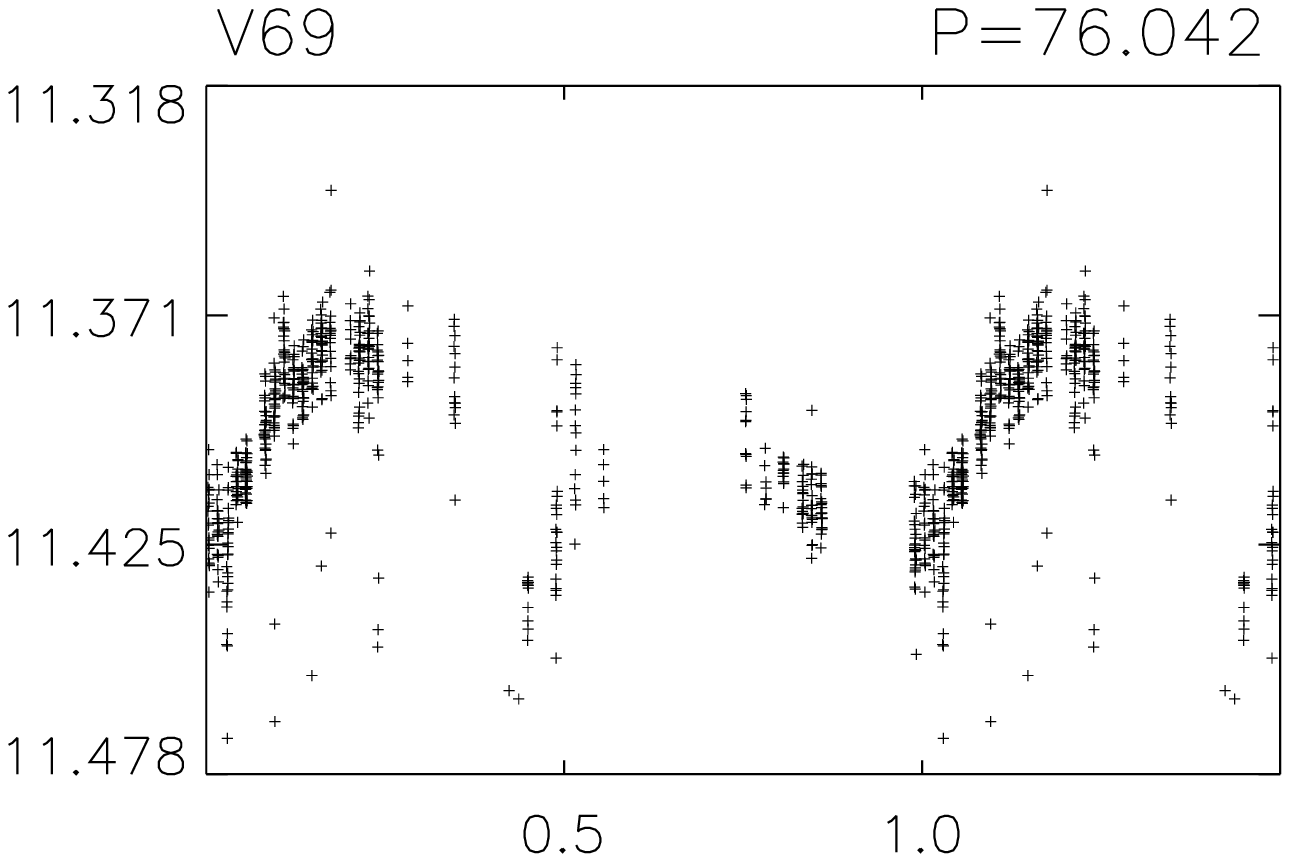}}
\vspace{-0.3cm}
   \hbox{\includegraphics[width=4.65cm]{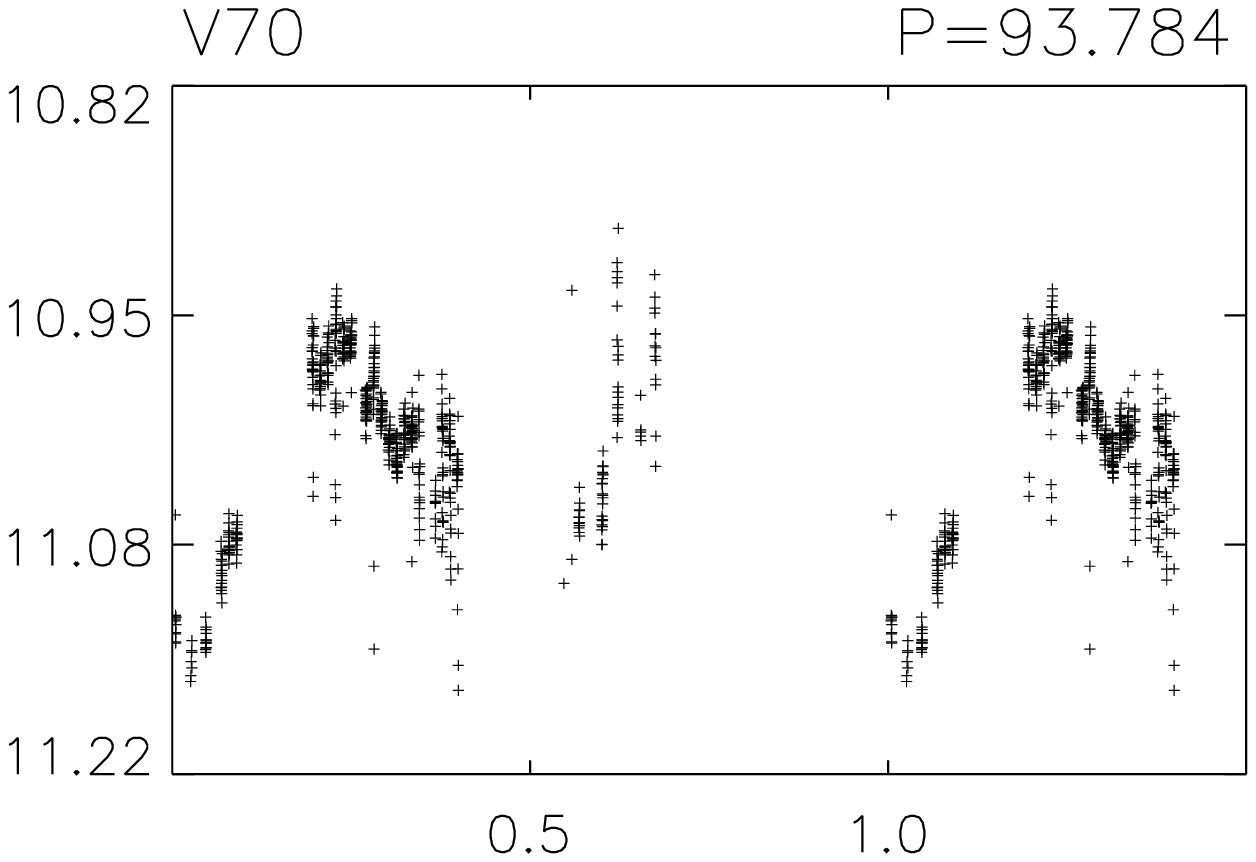}
   \hspace{-0.7cm}\includegraphics[width=4.65cm]{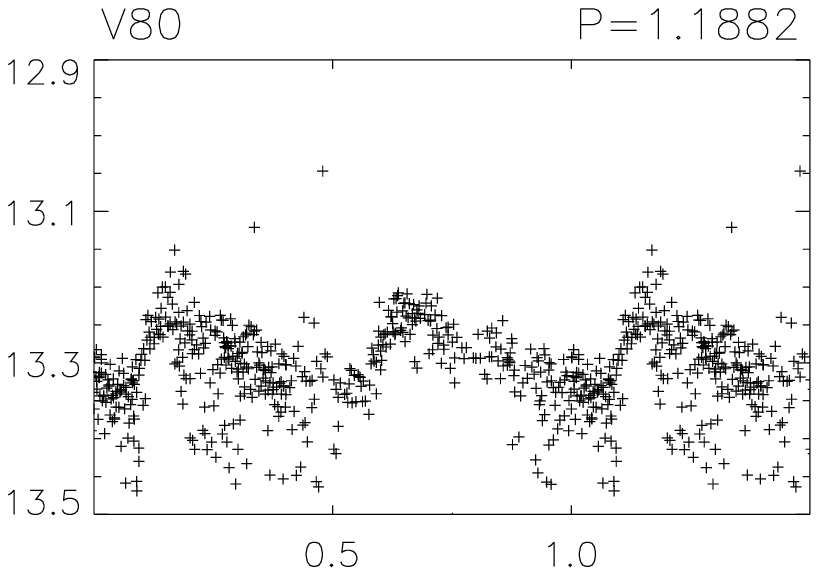}
   \hspace{-0.7cm}\includegraphics[width=4.65cm]{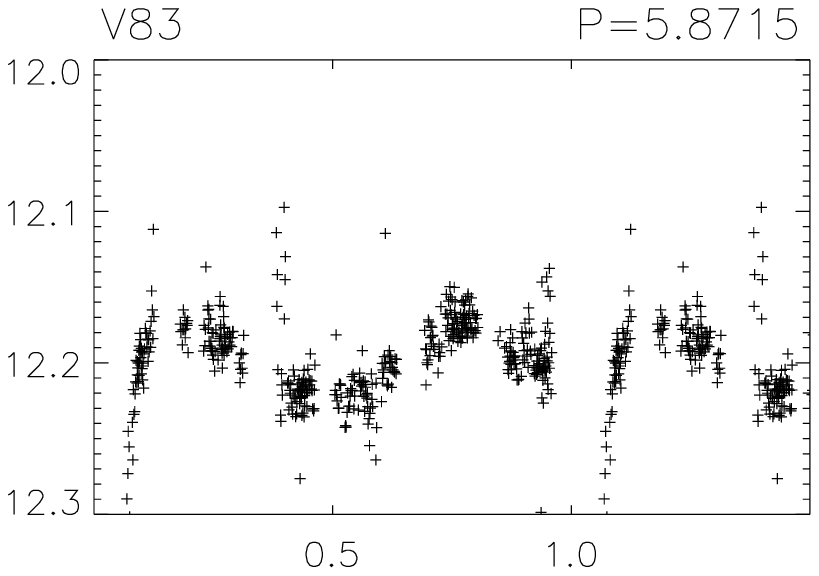}}
   \caption{Continued}
\end{figure*} 
\addtocounter{figure}{-1}
\begin{figure*}[!h]
\centering
\includegraphics[width=4.65cm]{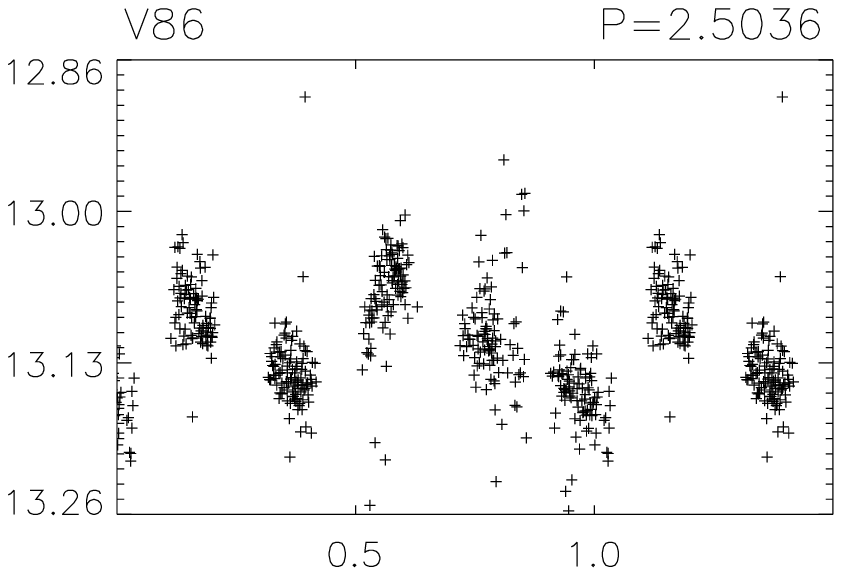}
\caption{Continued}
\label{fig:lcv1}
\end{figure*} 

\begin{figure*}[!ht]
\centering
\hbox{\includegraphics[width=4.65cm]{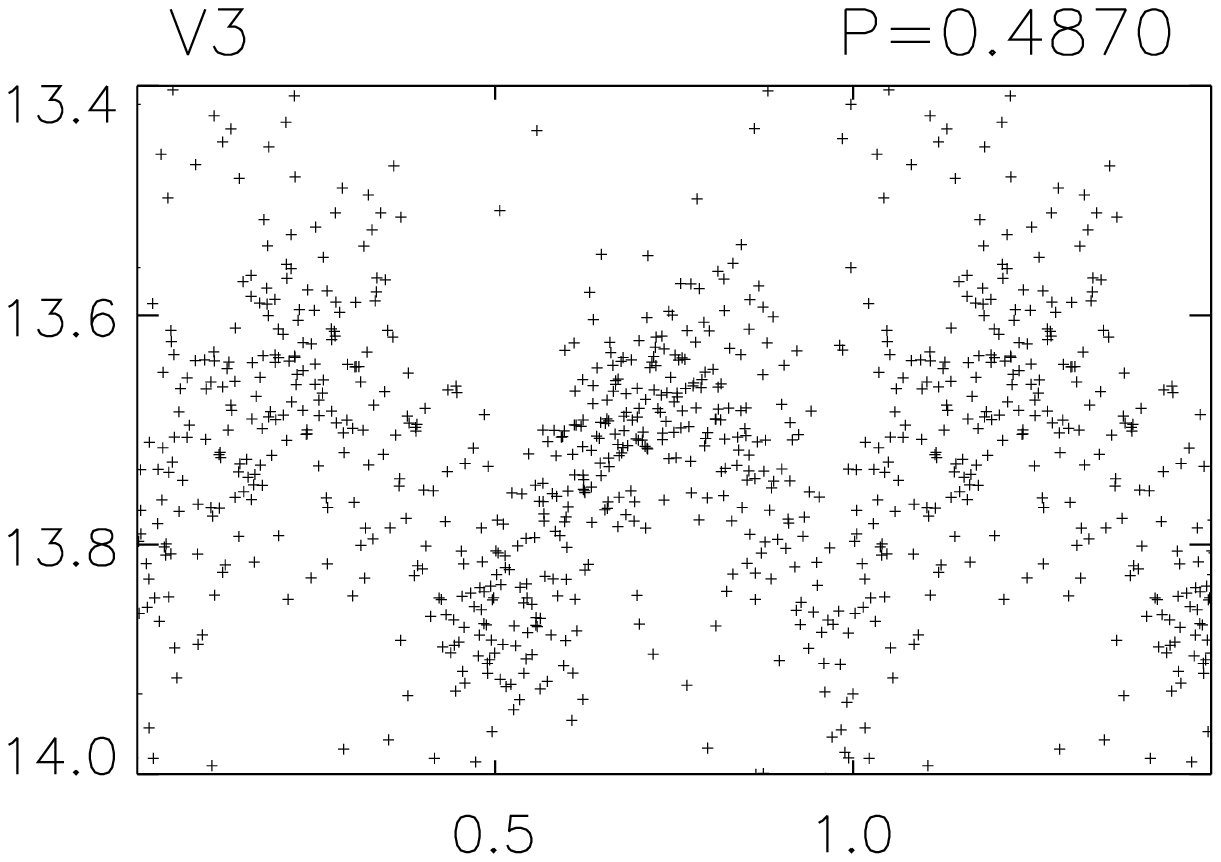}
   \hspace{-0.7cm}\includegraphics[width=4.65cm]{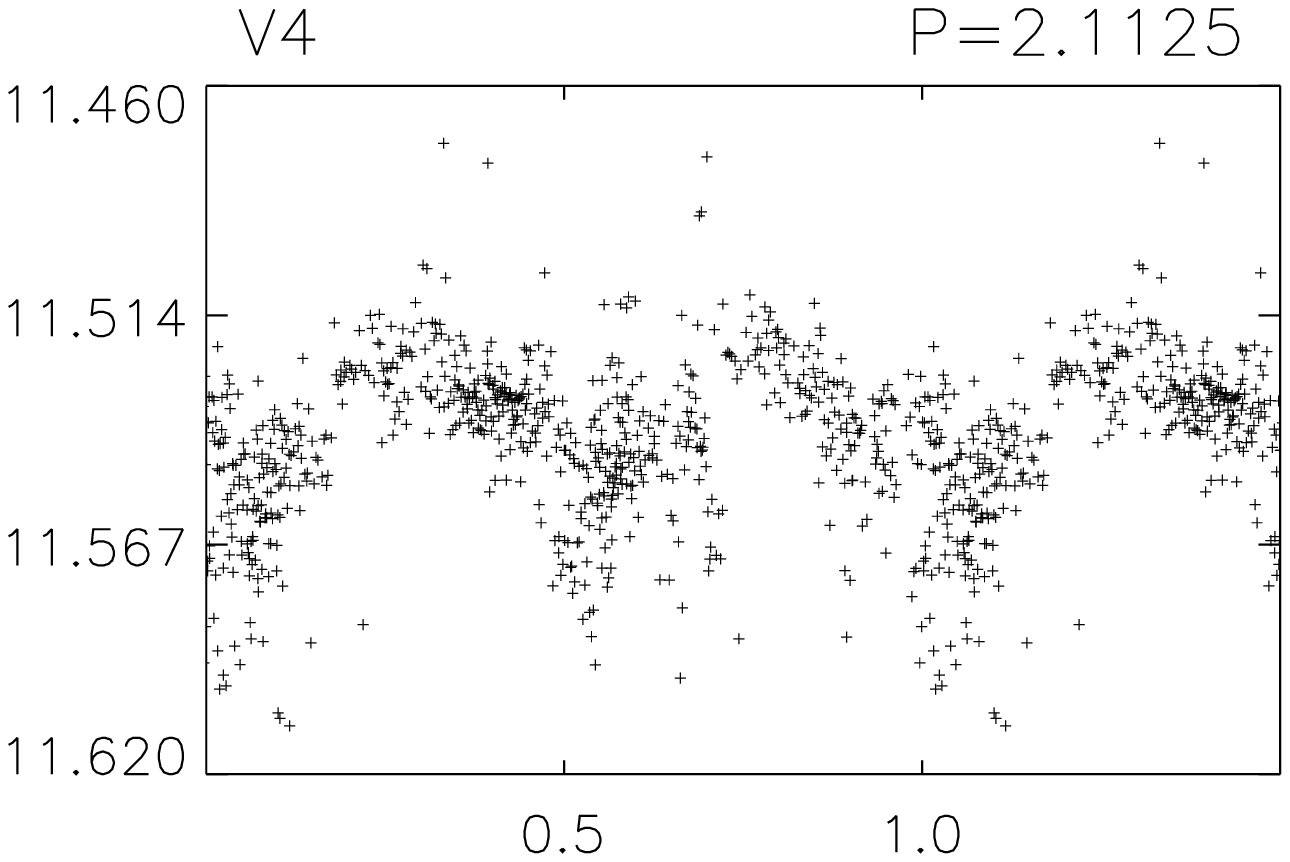}
   \hspace{-0.7cm}\includegraphics[width=4.65cm]{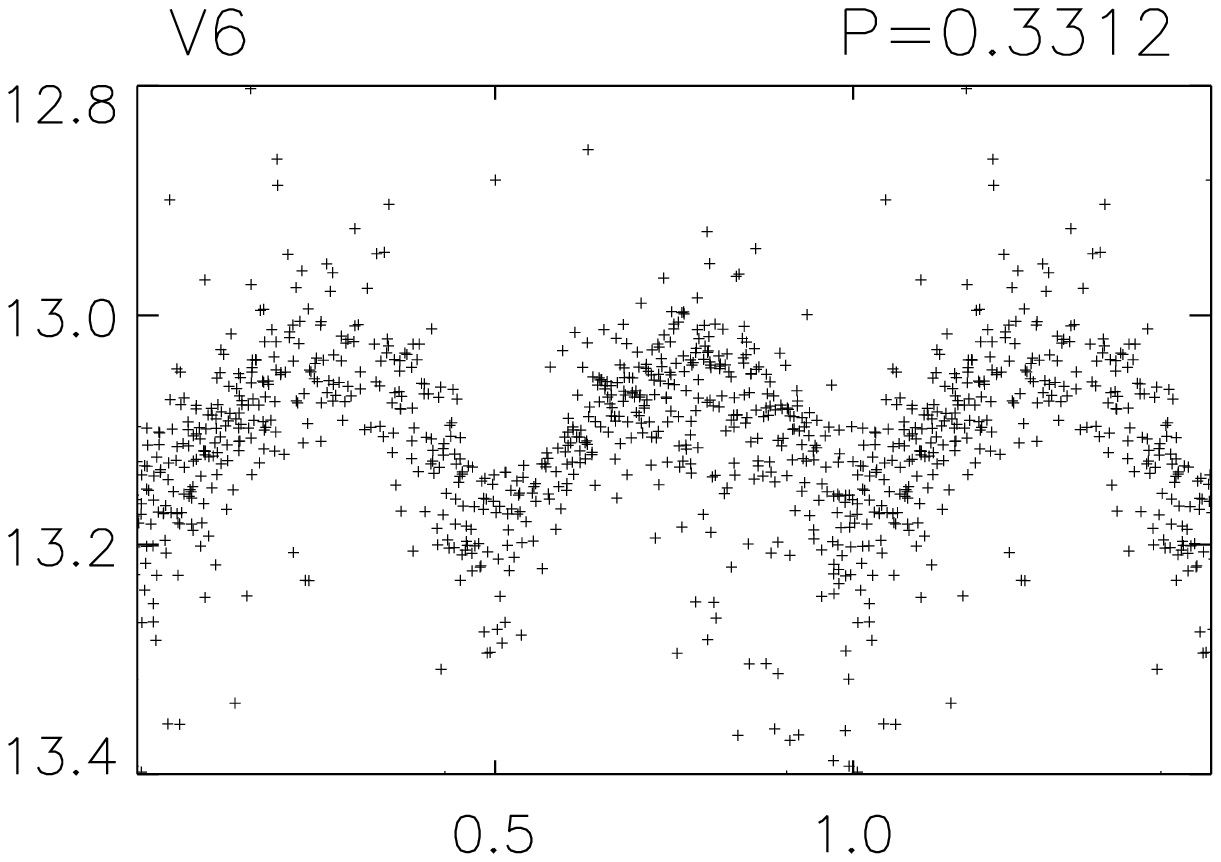}}
\vspace{-0.3cm}
   \hbox{\includegraphics[width=4.65cm]{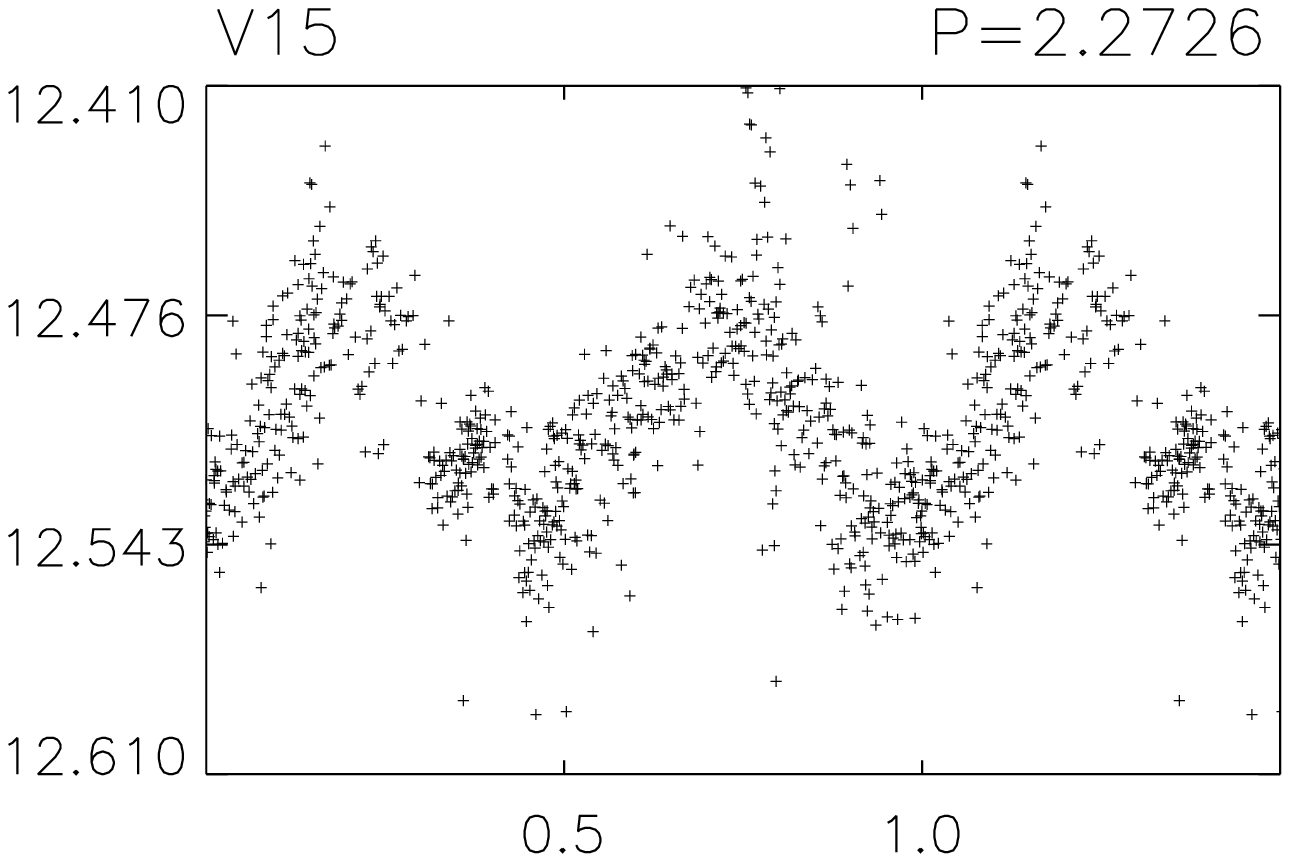}
   \hspace{-0.7cm}\includegraphics[width=4.65cm]{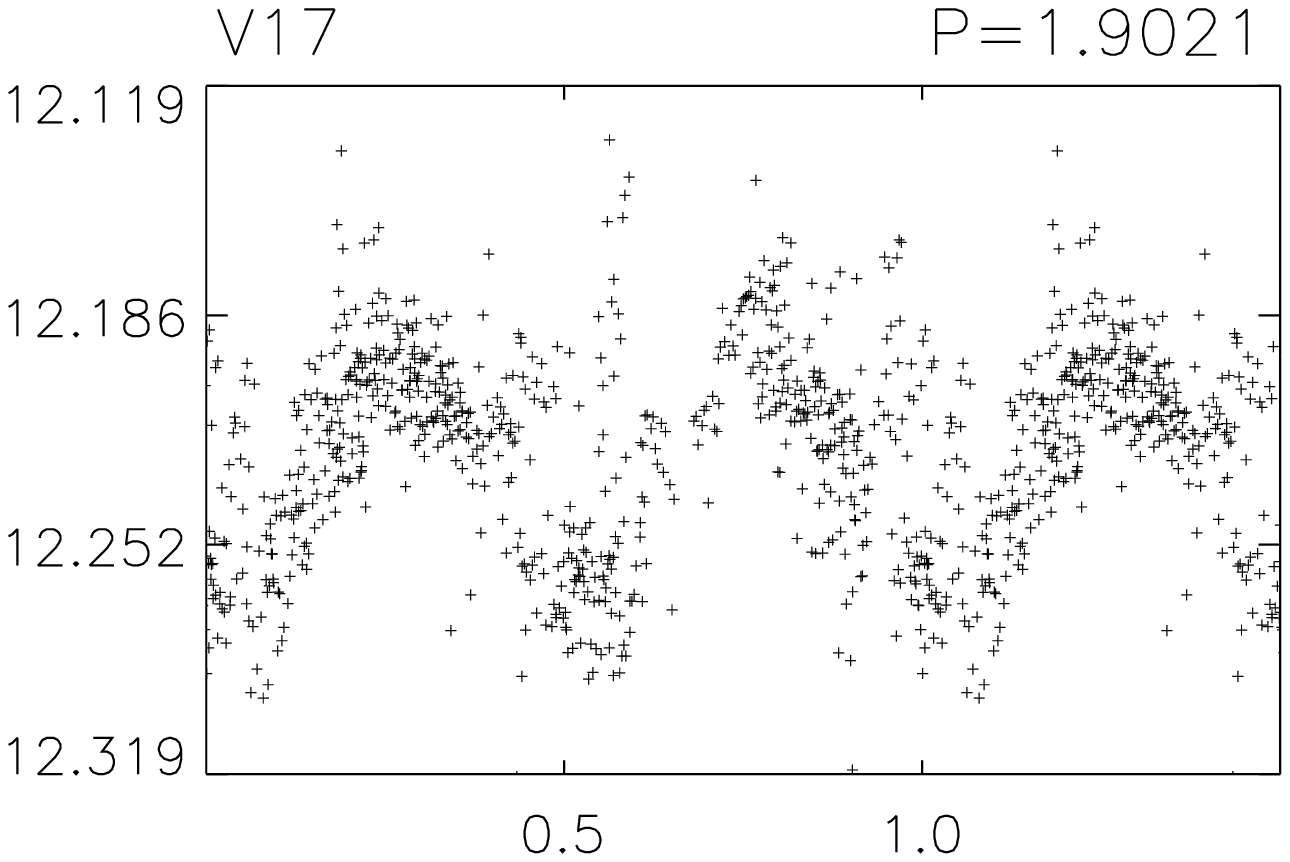}
   \hspace{-0.7cm}\includegraphics[width=4.65cm]{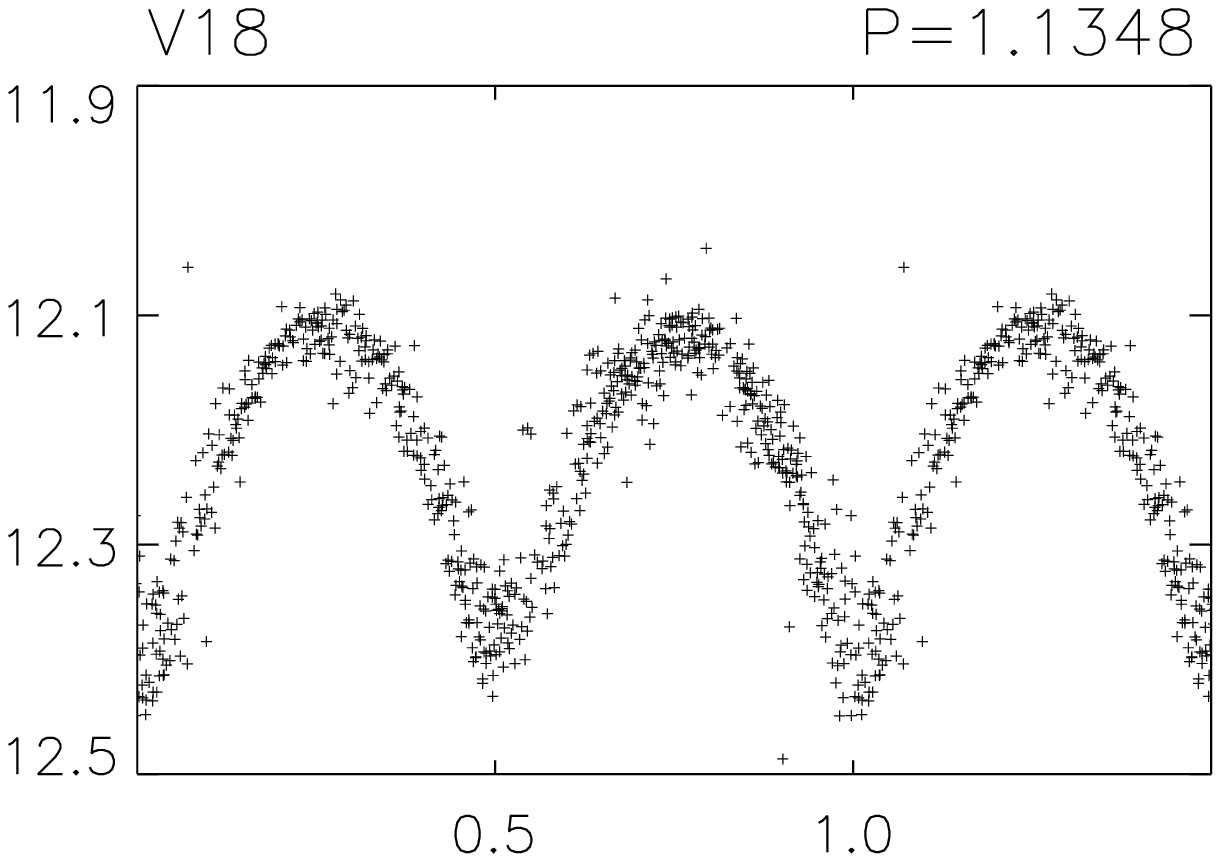}}
\vspace{-0.3cm}
   \hbox{\includegraphics[width=4.65cm]{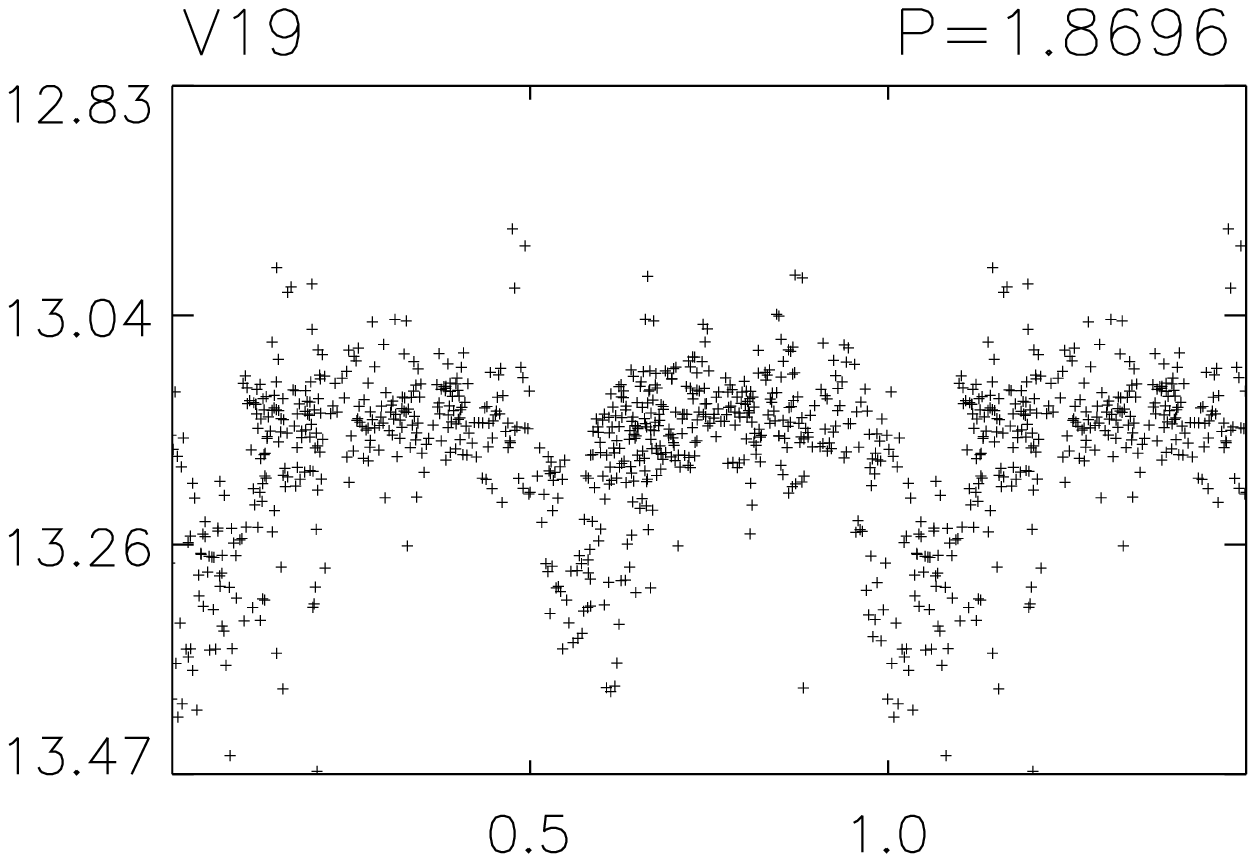}   
   \hspace{-0.7cm}\includegraphics[width=4.65cm]{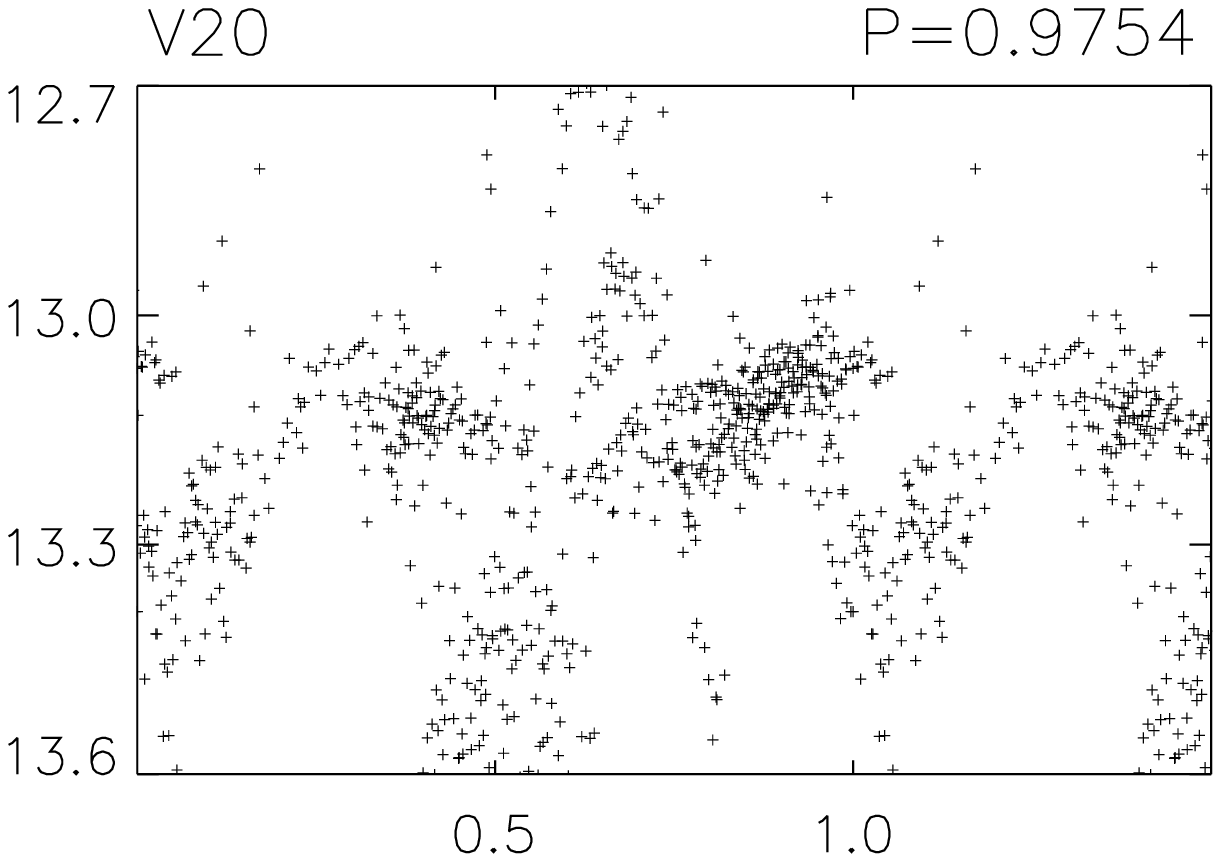}
   \hspace{-0.7cm}\includegraphics[width=4.65cm]{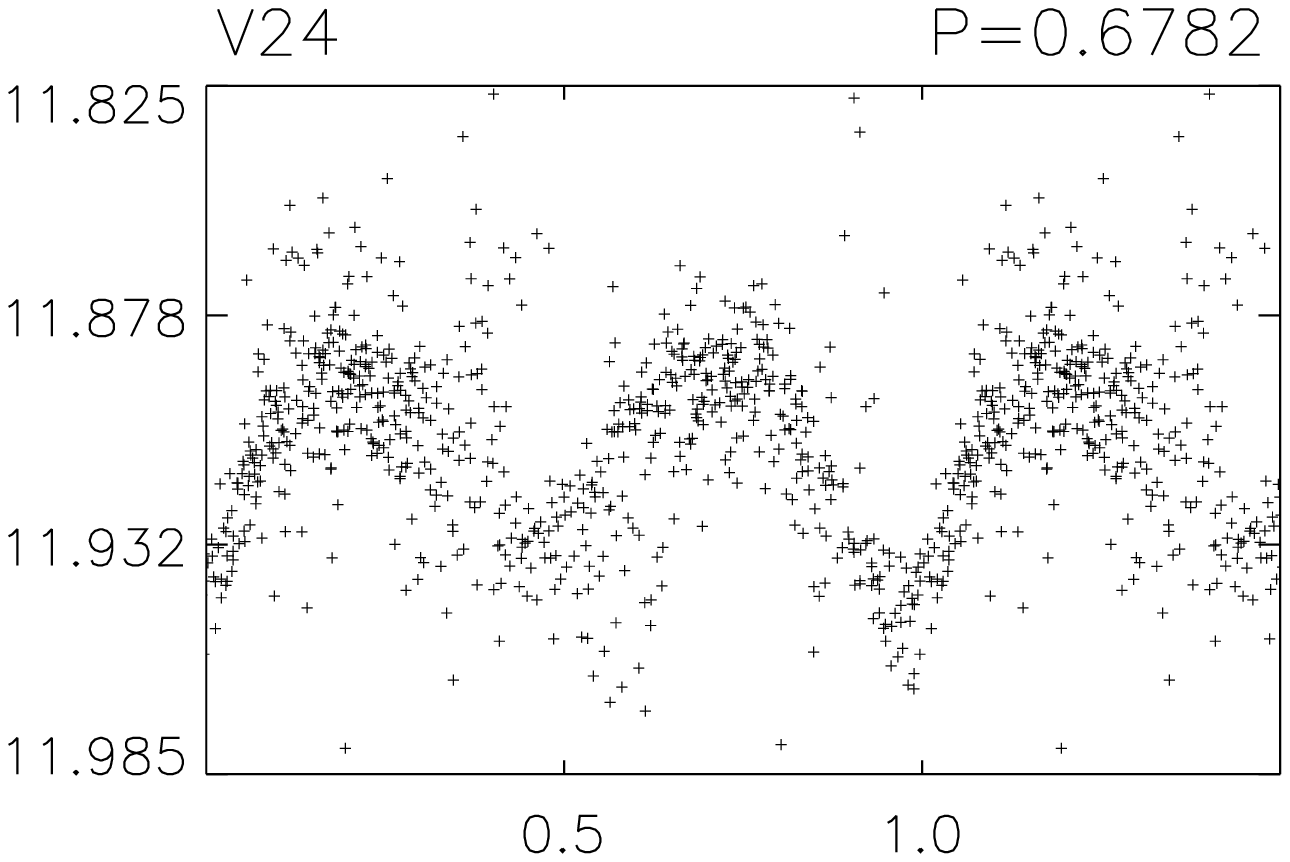}}
\vspace{-0.3cm}
\hbox{\includegraphics[width=4.65cm]{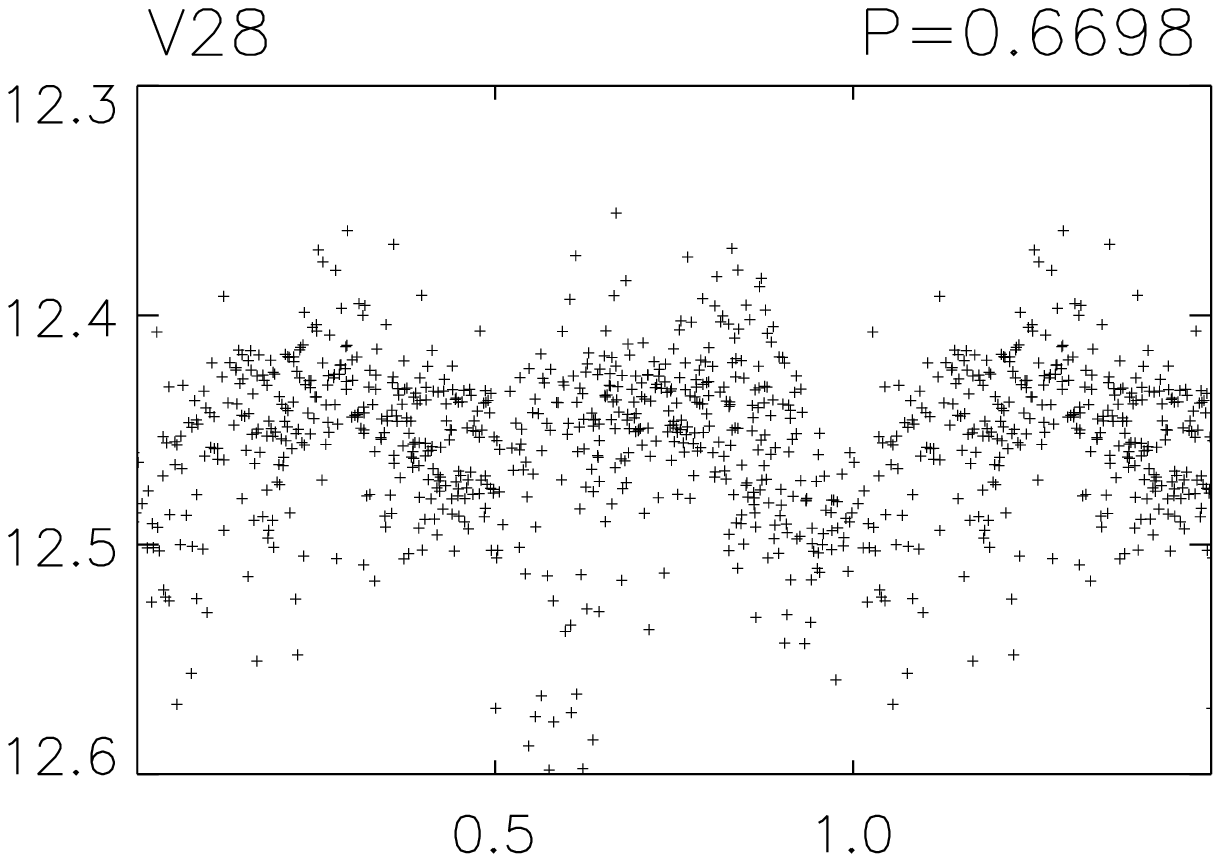}
   \hspace{-0.7cm}\includegraphics[width=4.65cm]{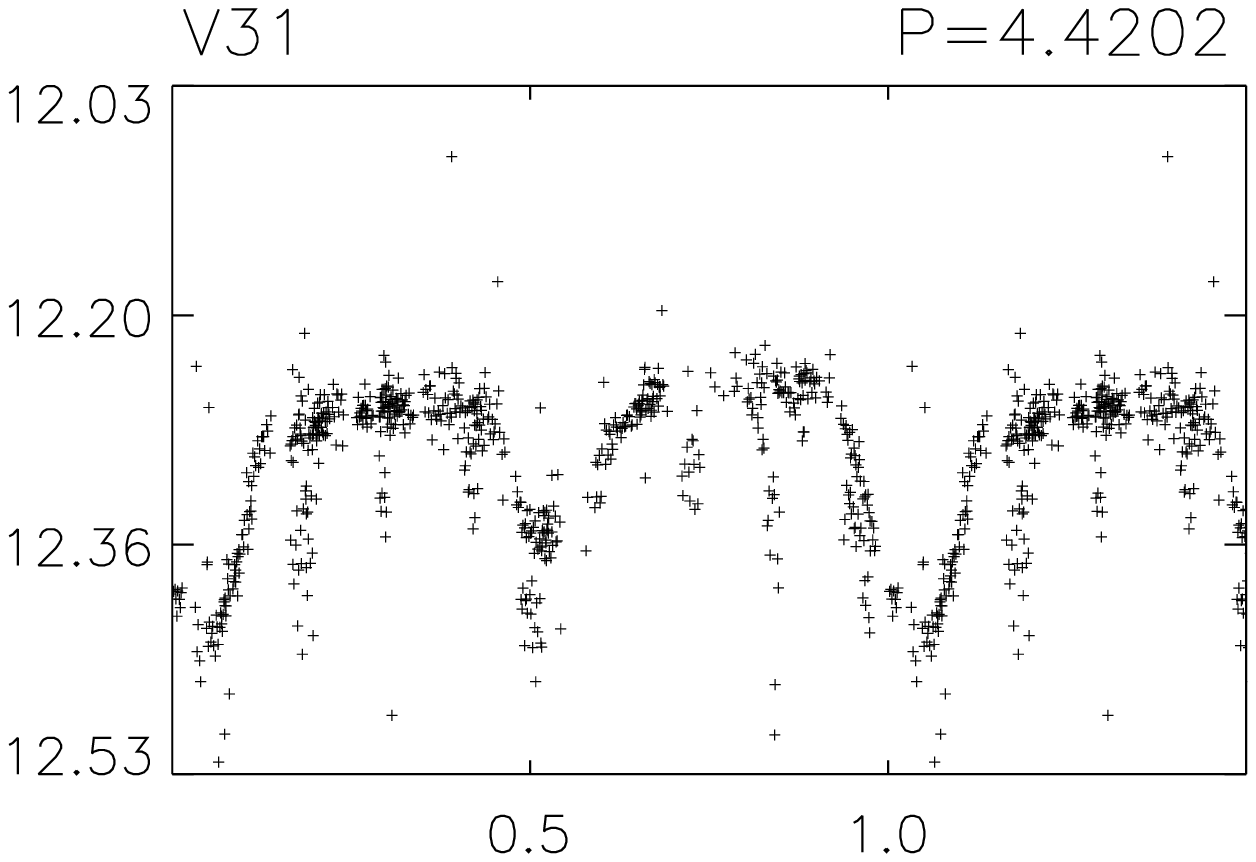}
   \hspace{-0.7cm}\includegraphics[width=4.65cm]{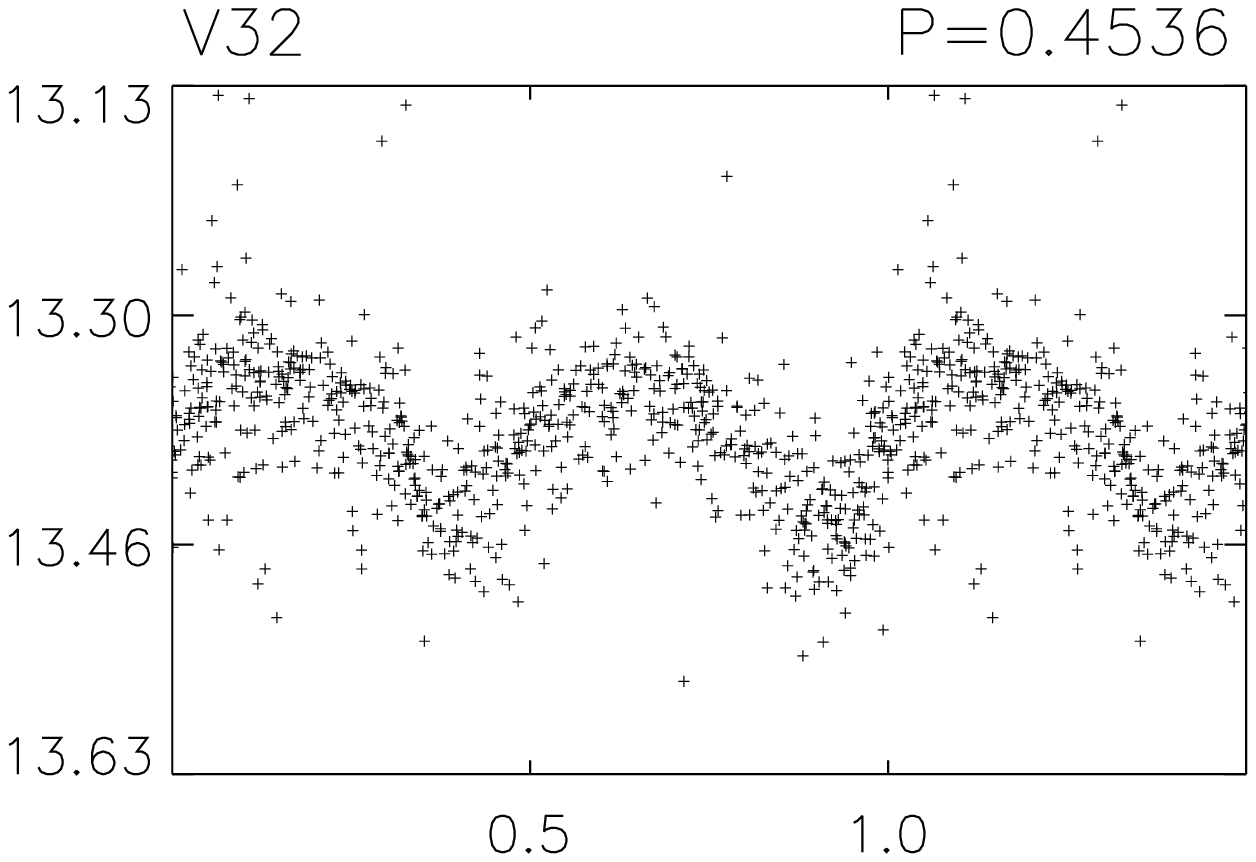}}
\vspace{-0.3cm}
   \hbox{\includegraphics[width=4.65cm]{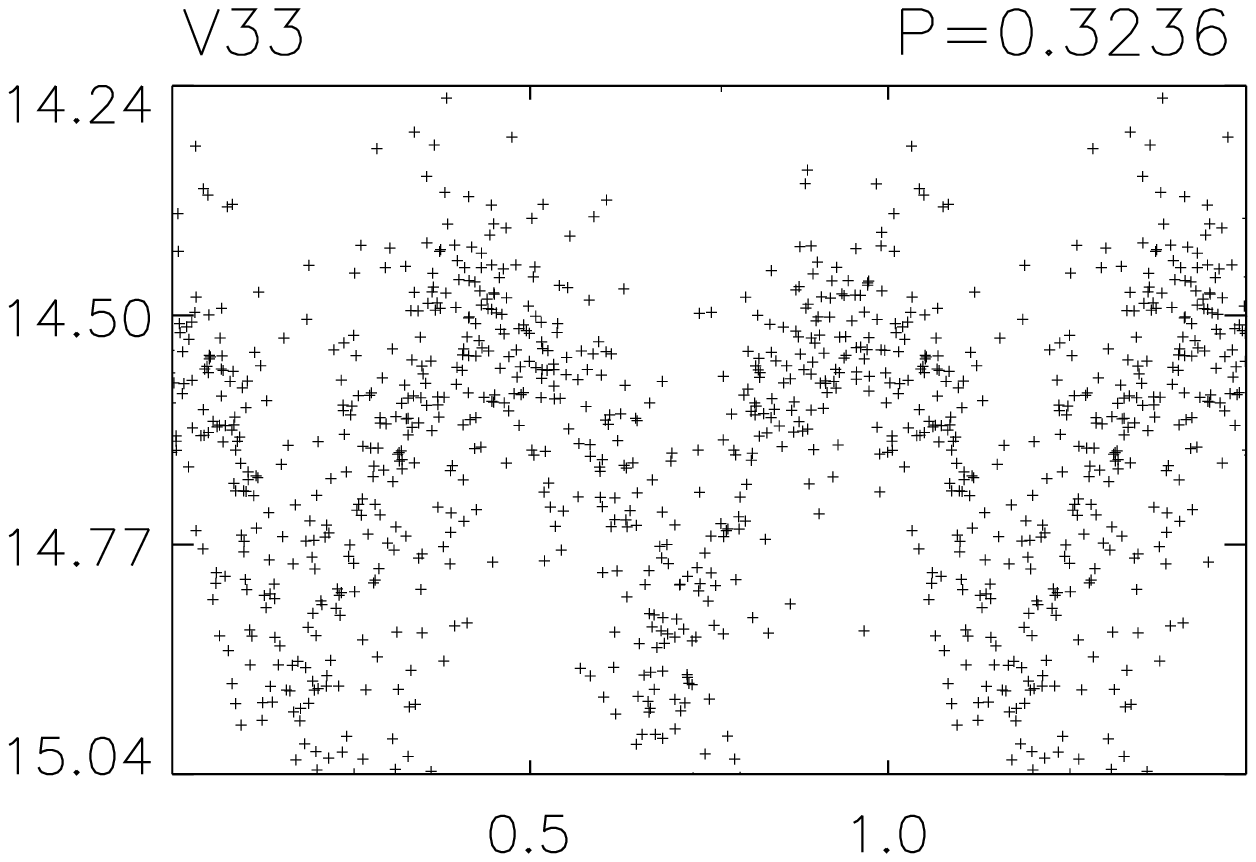}
   \hspace{-0.7cm}\includegraphics[width=4.65cm]{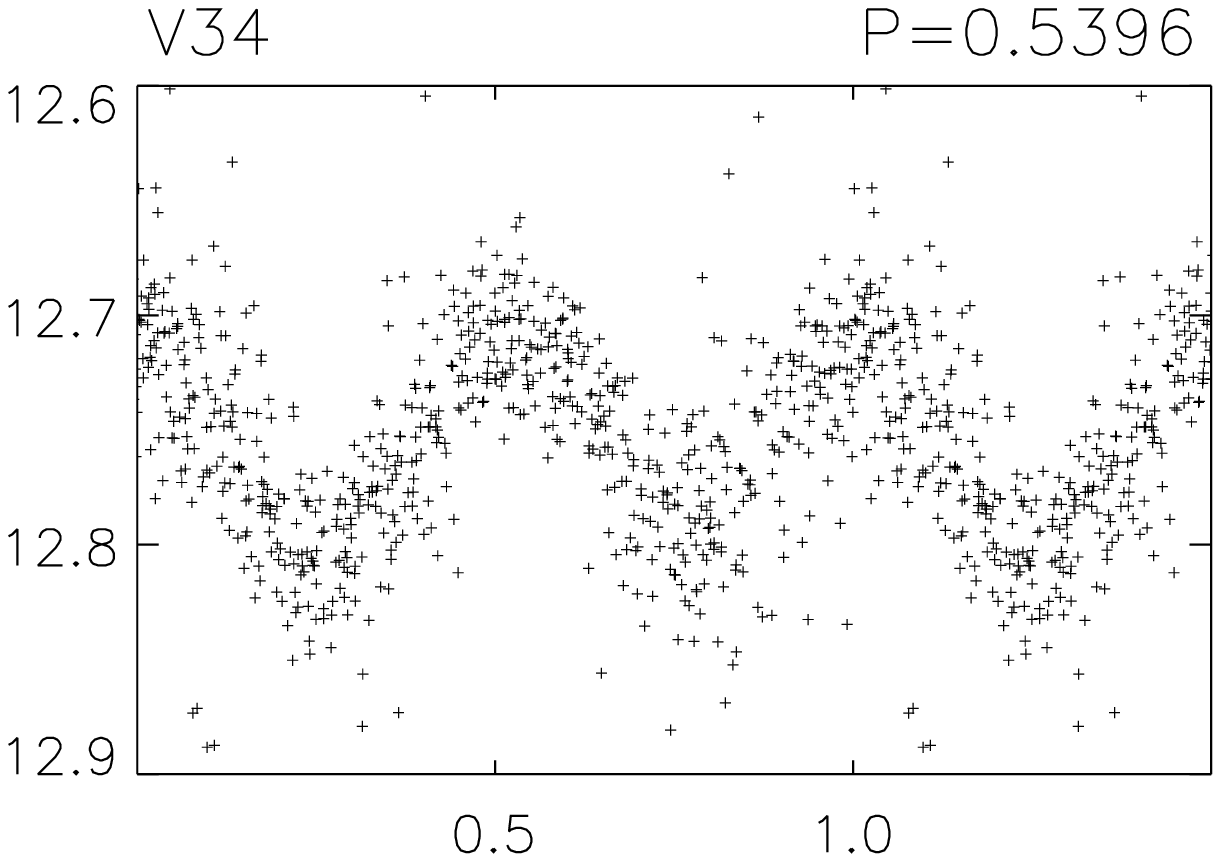}
   \hspace{-0.7cm}\includegraphics[width=4.65cm]{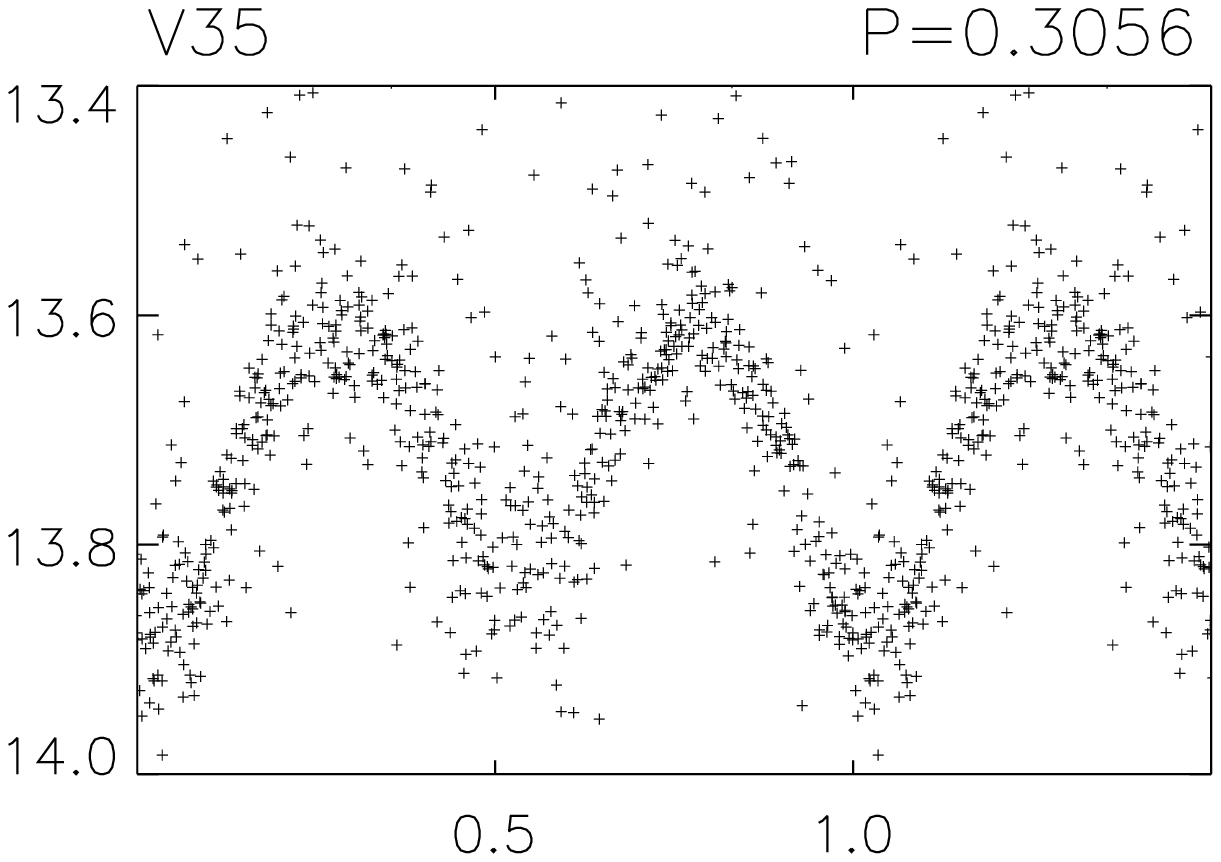}}
\vspace{-0.3cm}
   \hbox{\includegraphics[width=4.65cm]{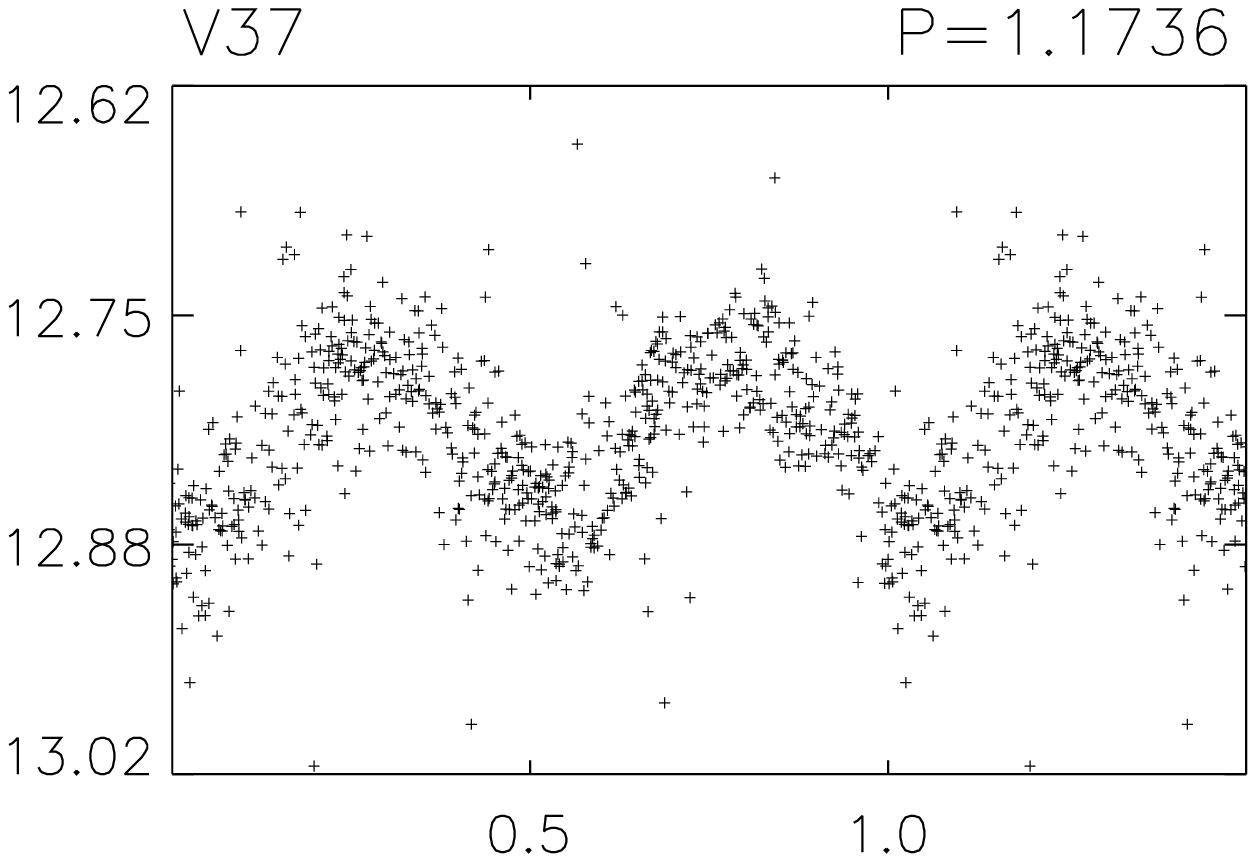}   
   \hspace{-0.7cm}\includegraphics[width=4.65cm]{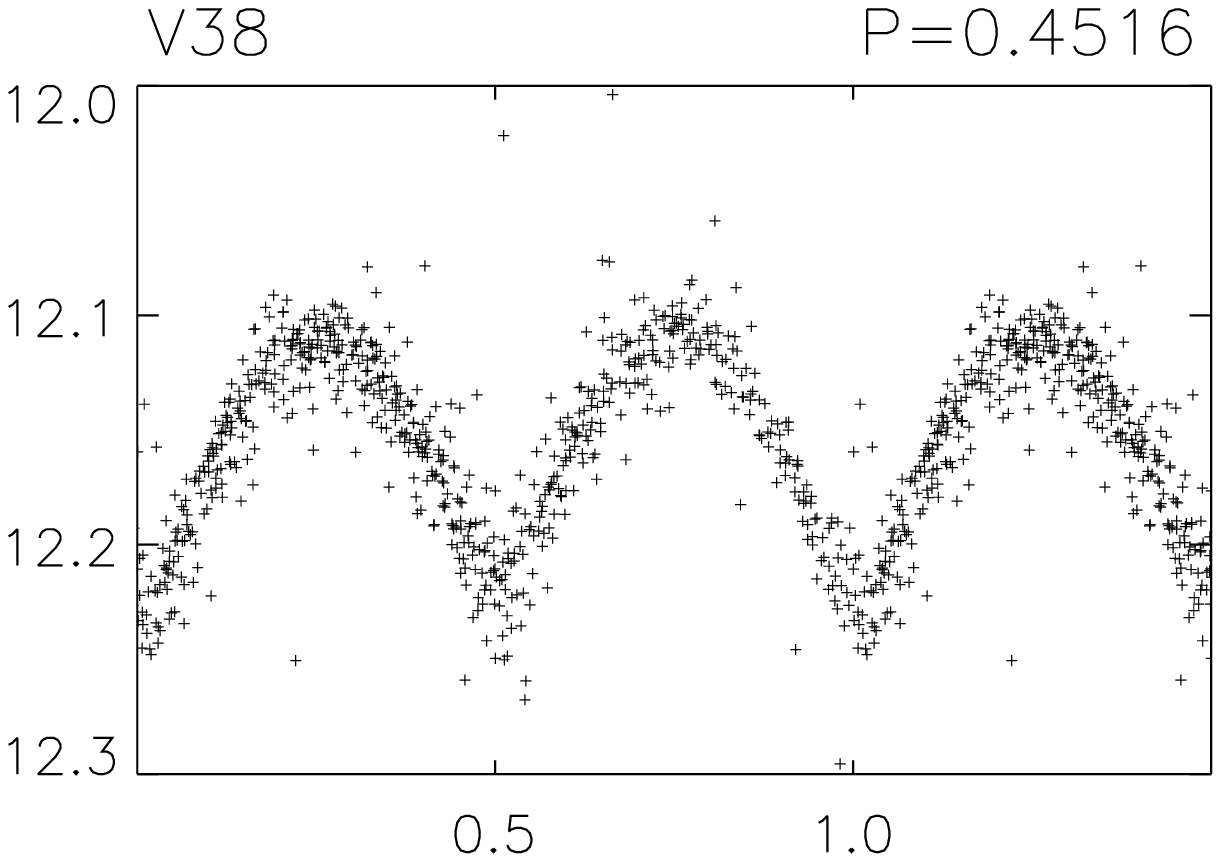}
   \hspace{-0.7cm}\includegraphics[width=4.65cm]{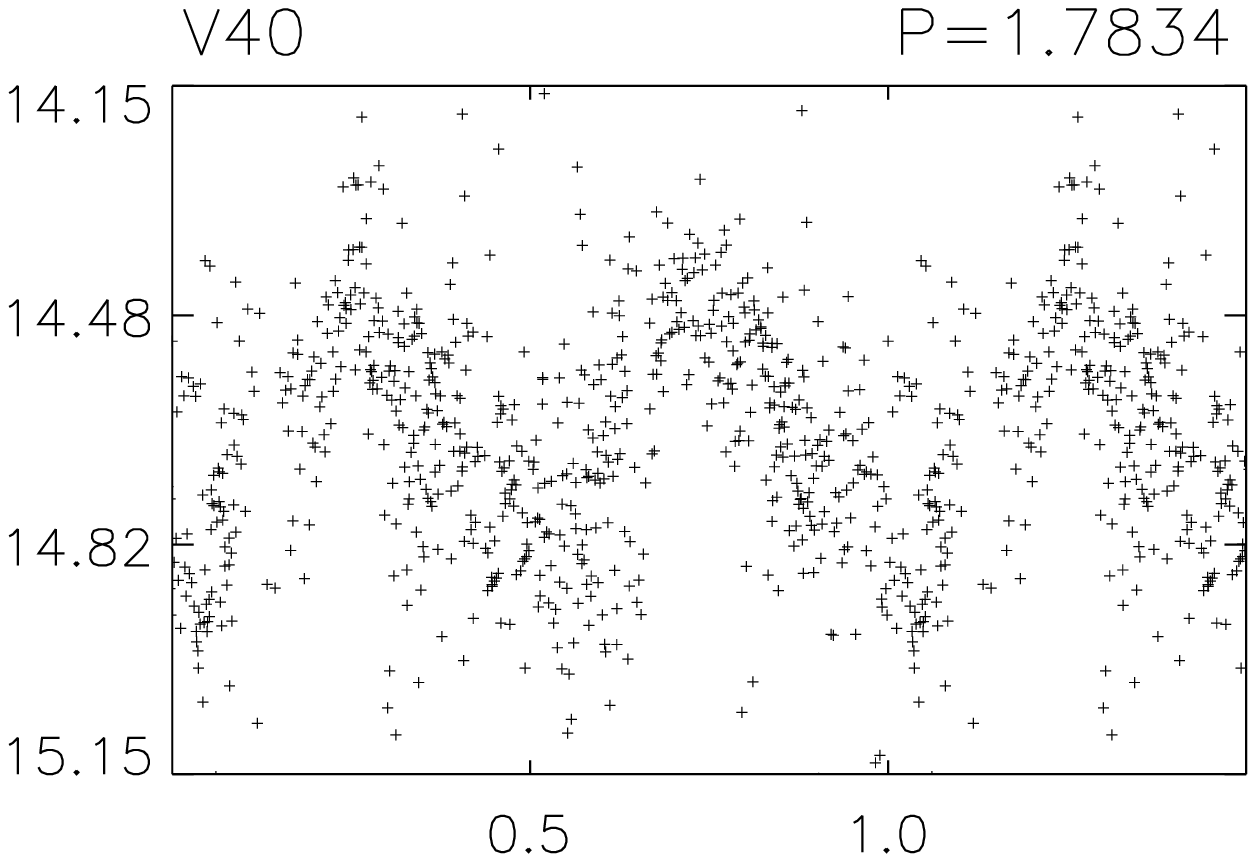}}
\vspace{-0.3cm}
   \hbox{\includegraphics[width=4.65cm]{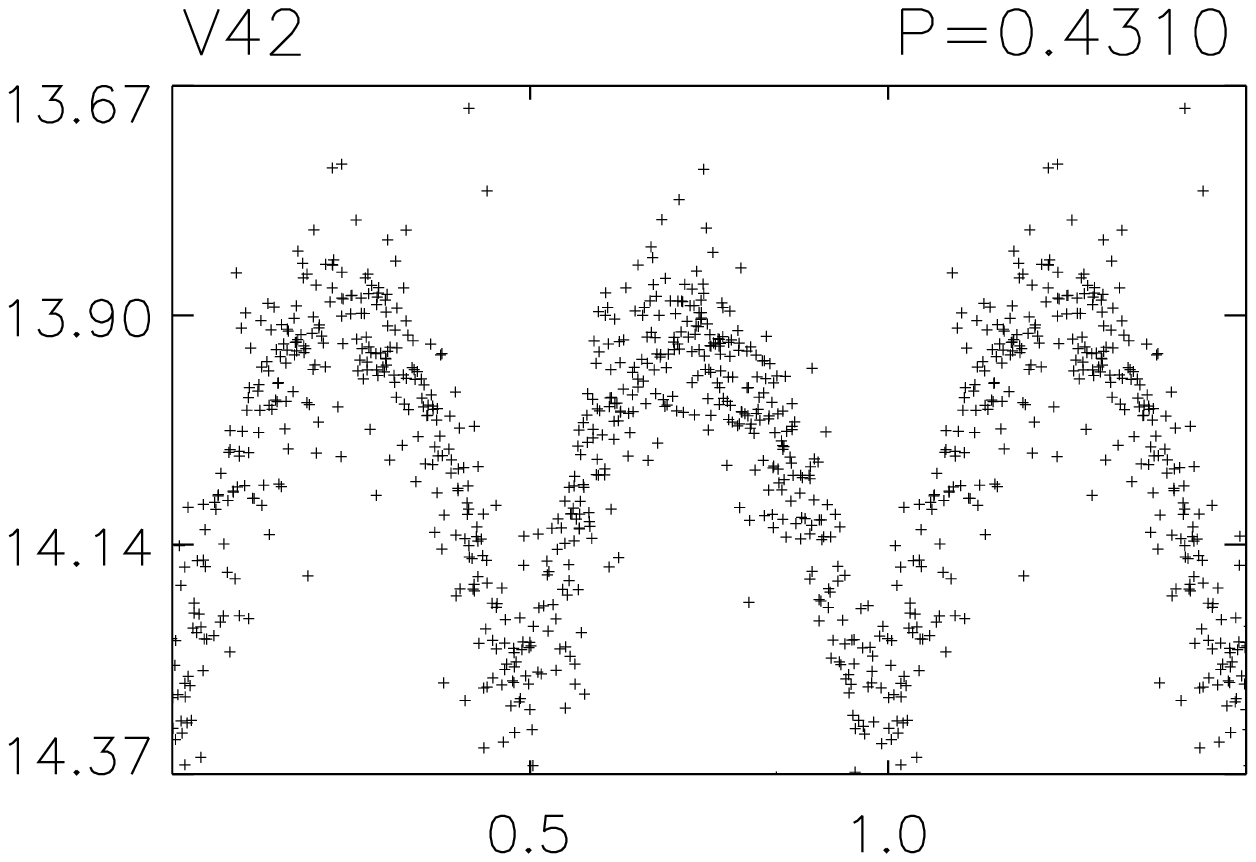}
   \hspace{-0.7cm}\includegraphics[width=4.65cm]{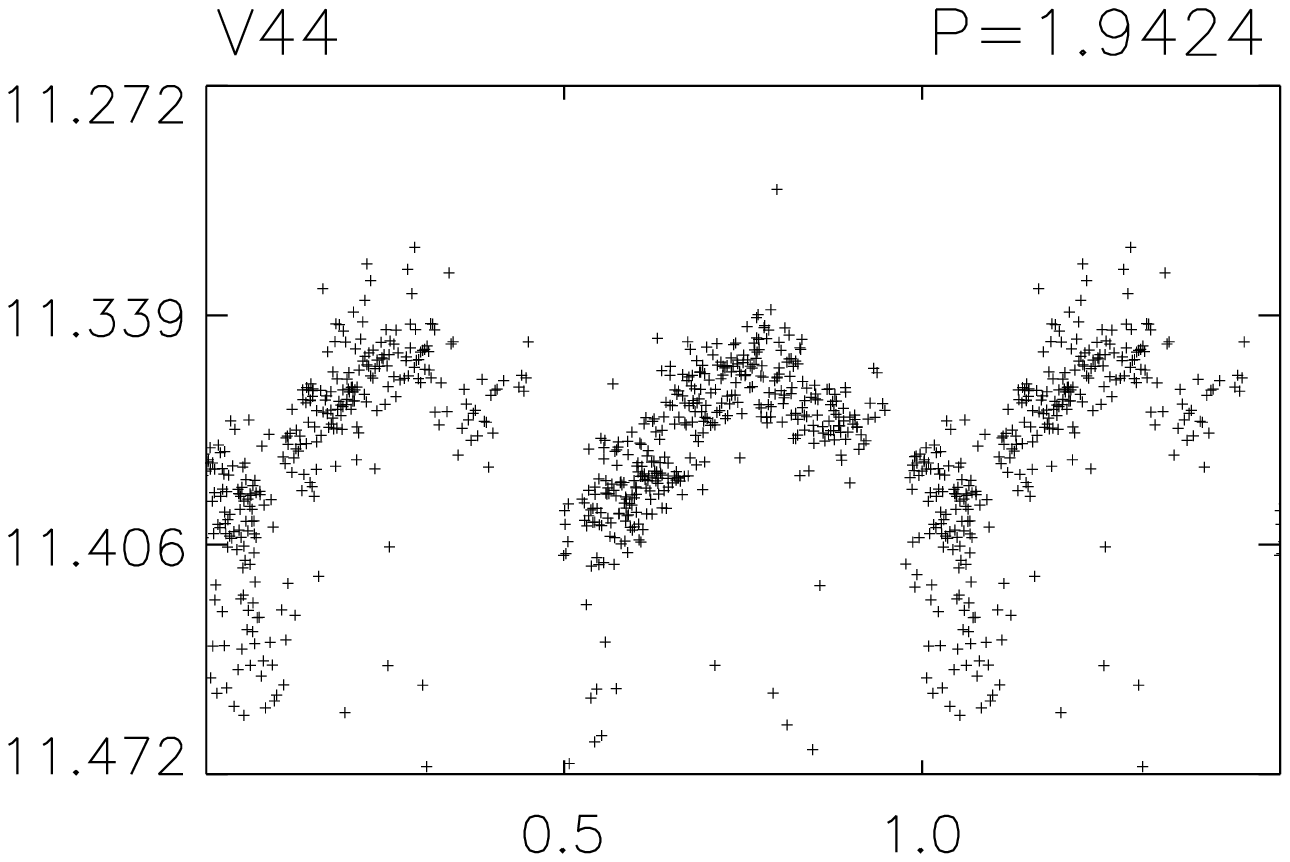}
   \hspace{-0.7cm}\includegraphics[width=4.65cm]{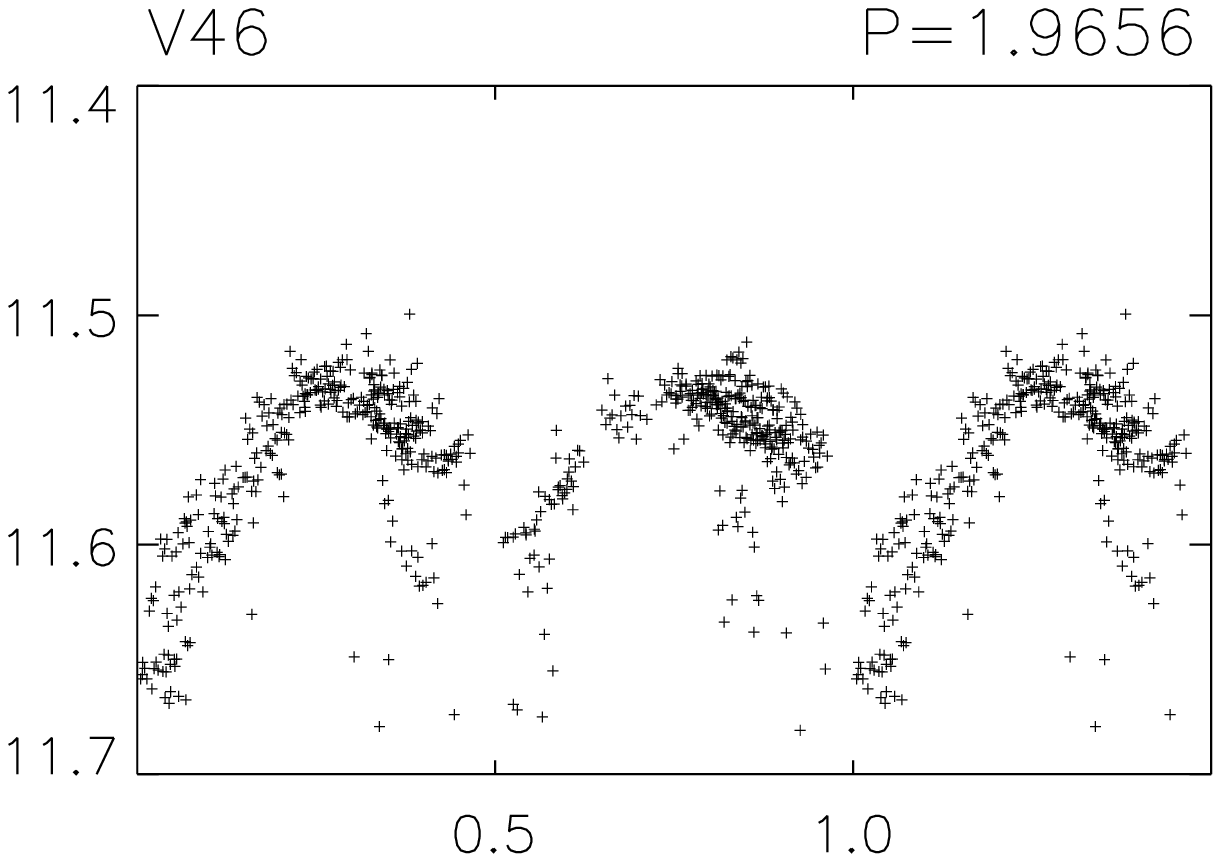}}
\caption{Light curves of 43 eclipsing binaries in $R$. See caption in Fig. \ref{fig:lcv1}.}
\end{figure*} 
\addtocounter{figure}{-1}
\begin{figure*}[!ht]
\centering
   \hbox{\includegraphics[width=4.65cm]{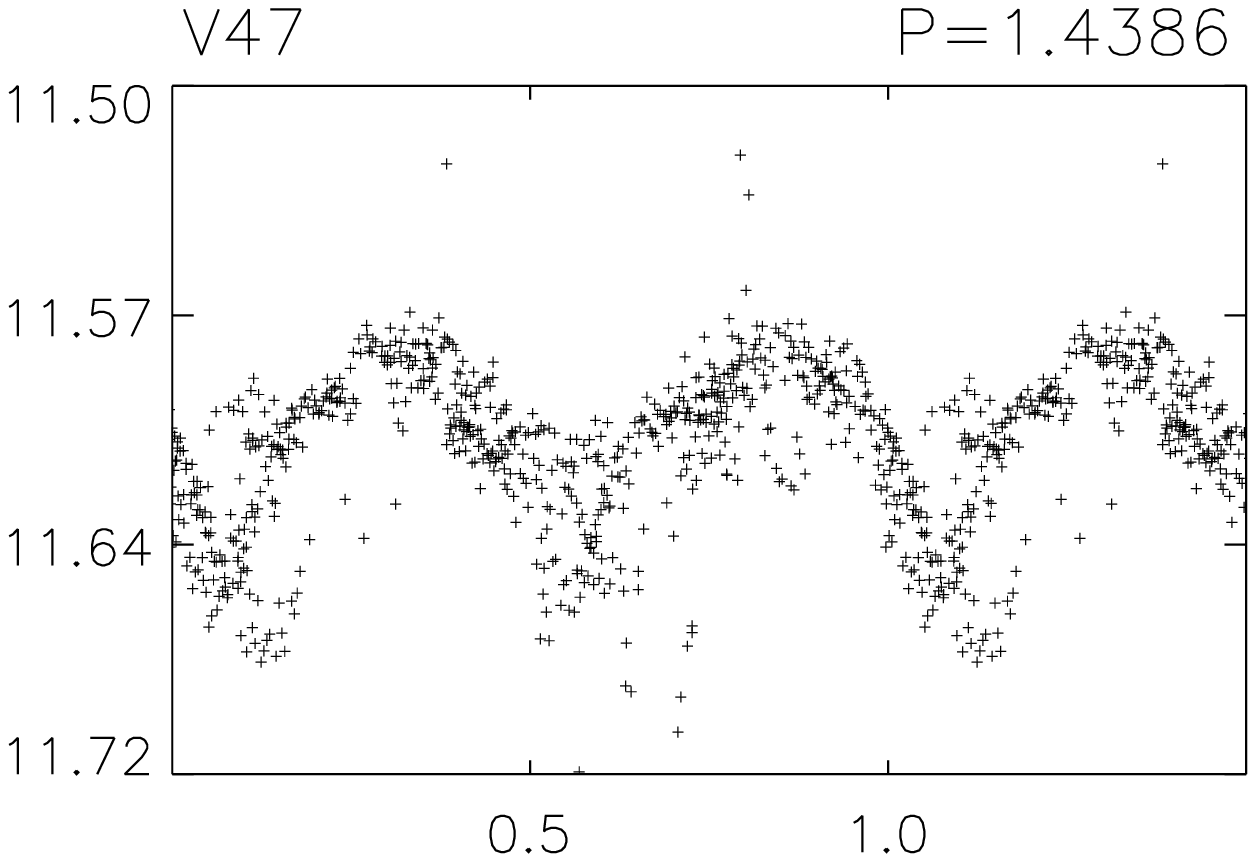}
   \hspace{-0.7cm}\includegraphics[width=4.65cm]{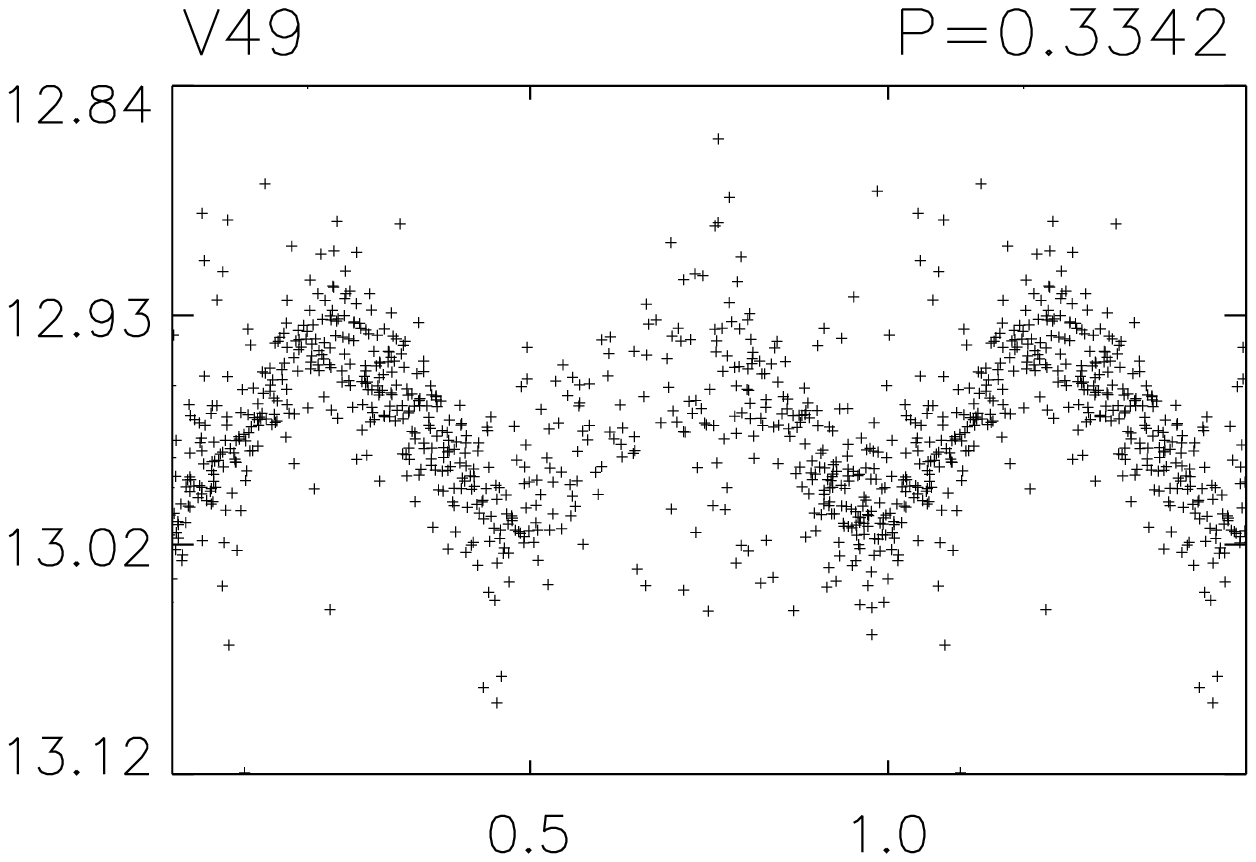}
   \hspace{-0.7cm}\includegraphics[width=4.65cm]{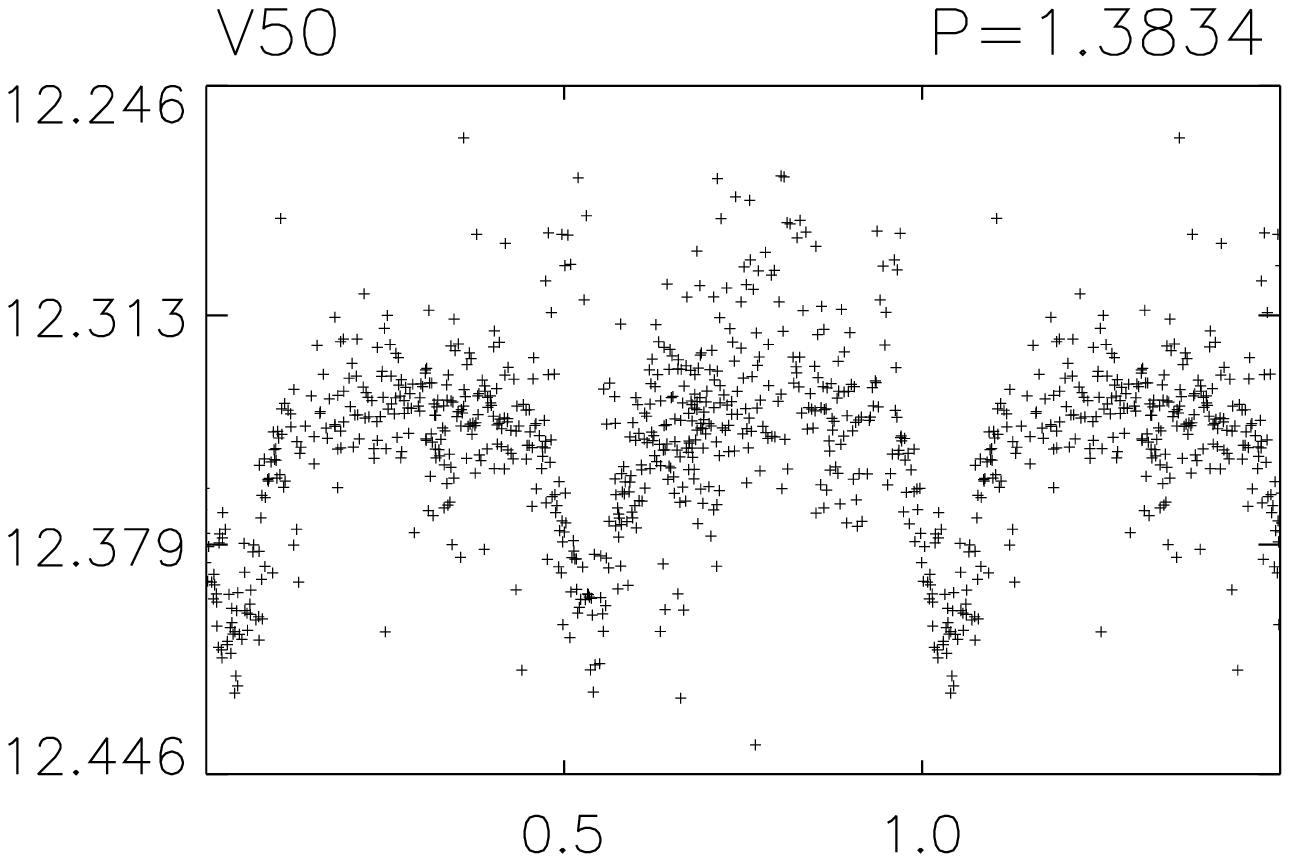}}
\vspace{-0.3cm}
   \hbox{\includegraphics[width=4.65cm]{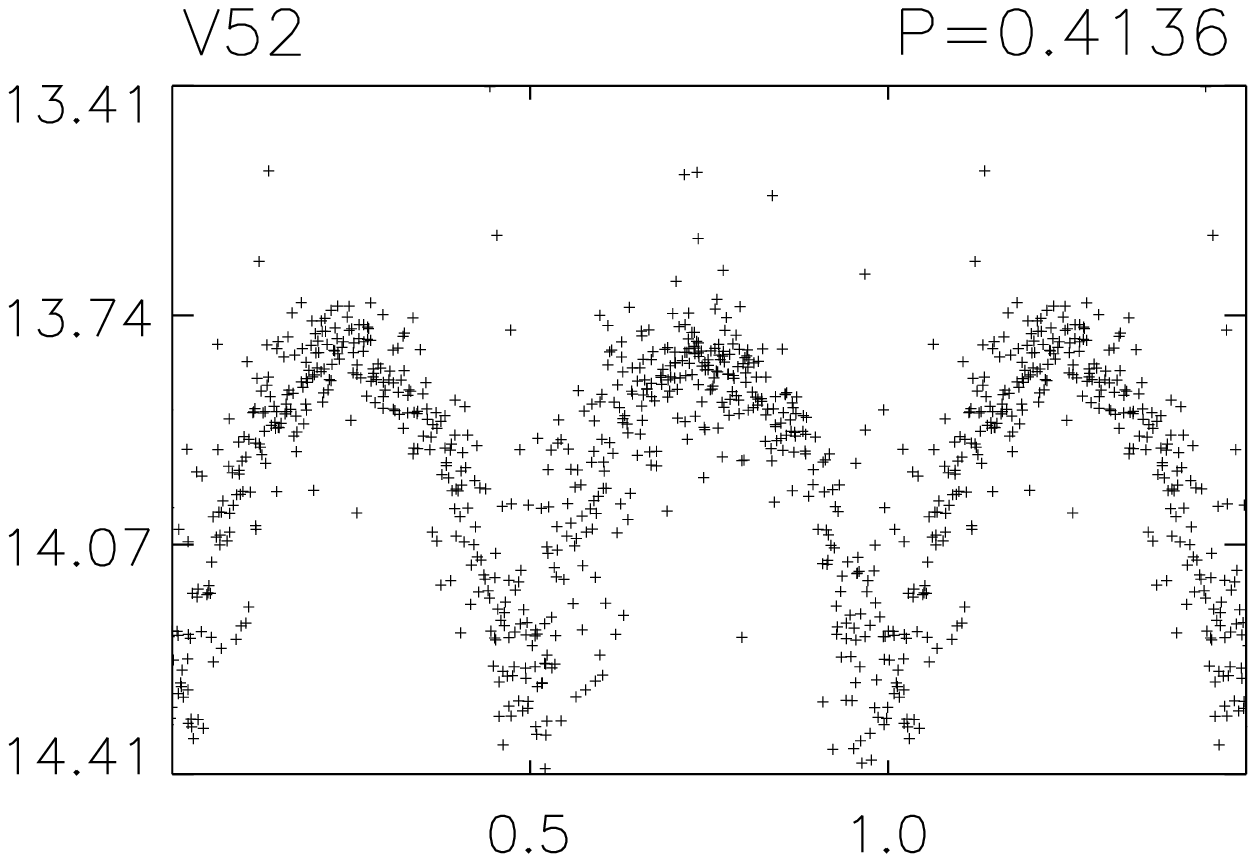}   
   \hspace{-0.7cm}\includegraphics[width=4.65cm]{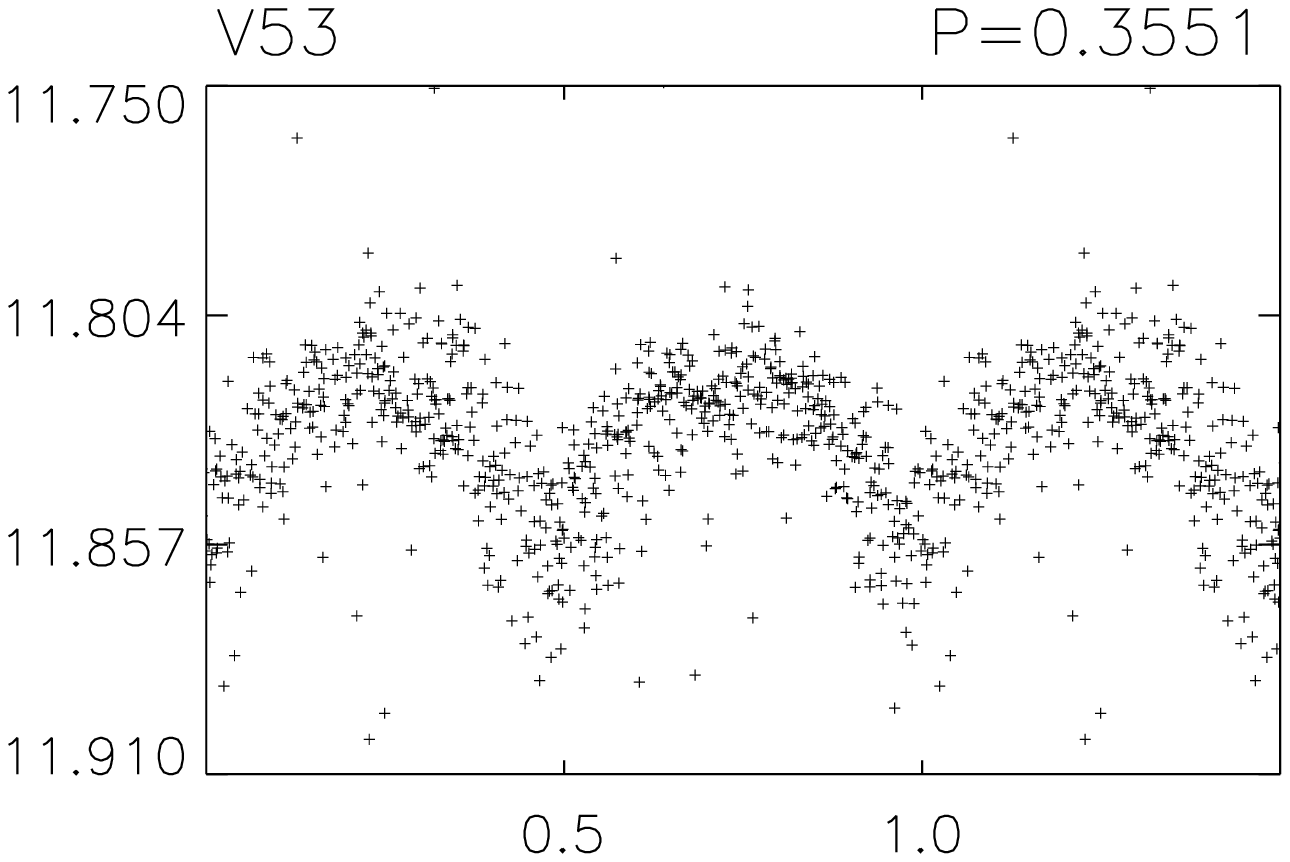}
   \hspace{-0.7cm}\includegraphics[width=4.65cm]{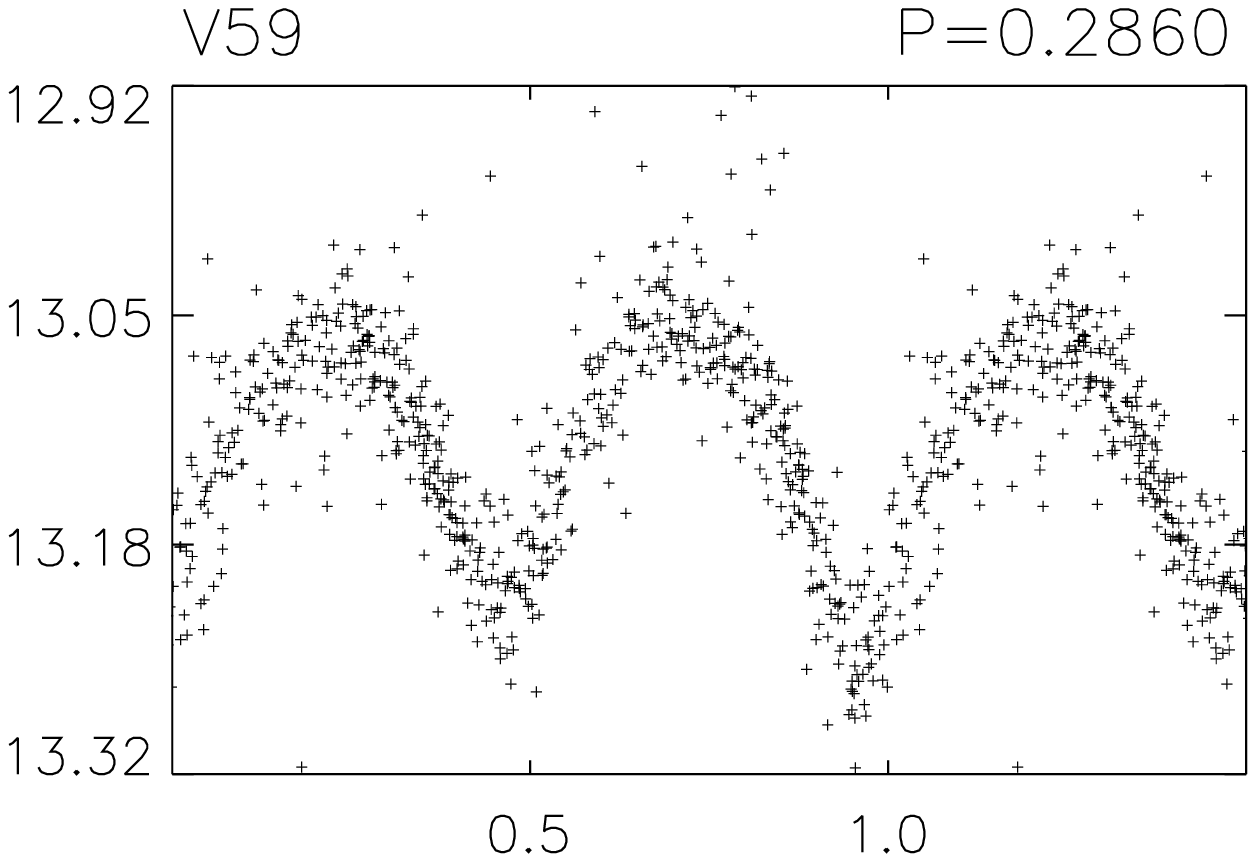}}
\vspace{-0.3cm}
   \hbox{\includegraphics[width=4.65cm]{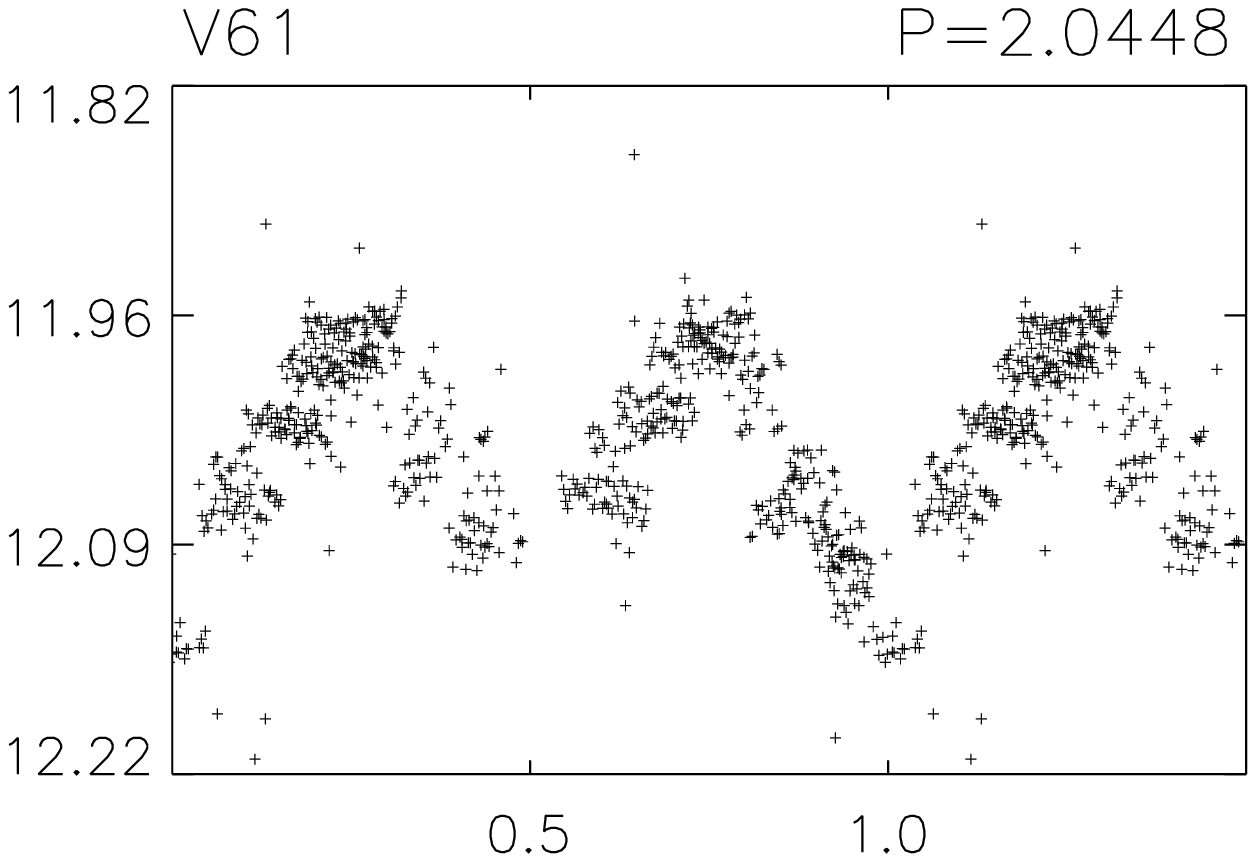}
   \hspace{-0.7cm}\includegraphics[width=4.65cm]{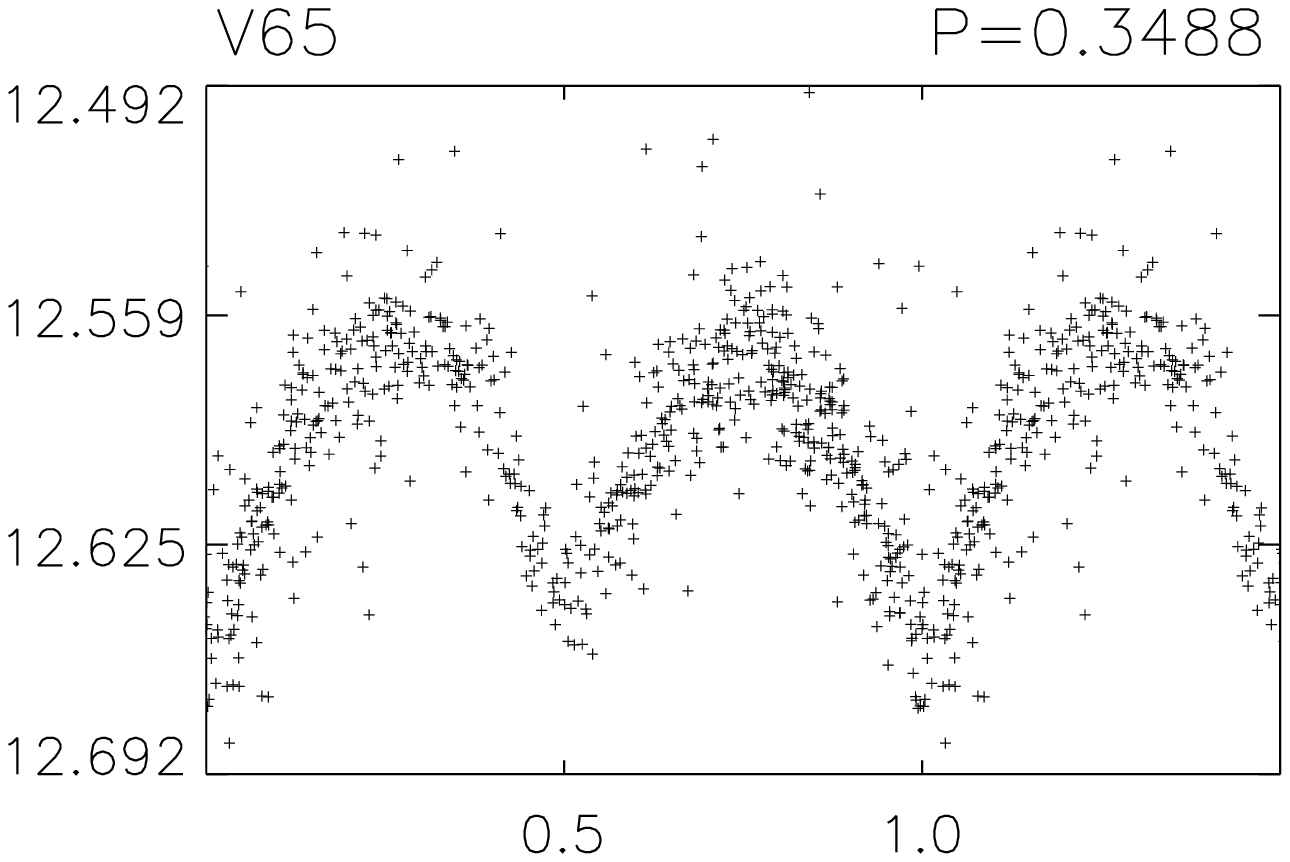}
   \hspace{-0.7cm}\includegraphics[width=4.65cm]{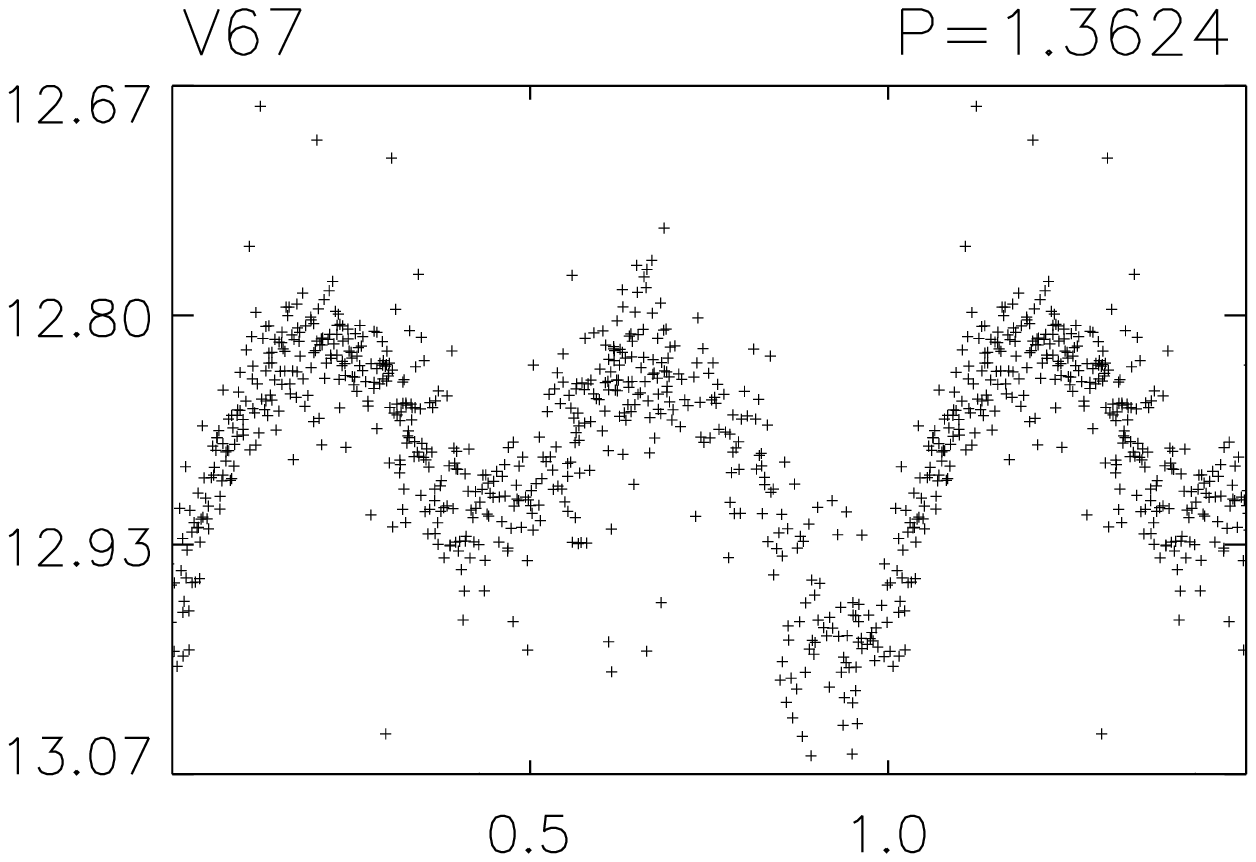}}
\vspace{-0.3cm}
   \hbox{\includegraphics[width=4.65cm]{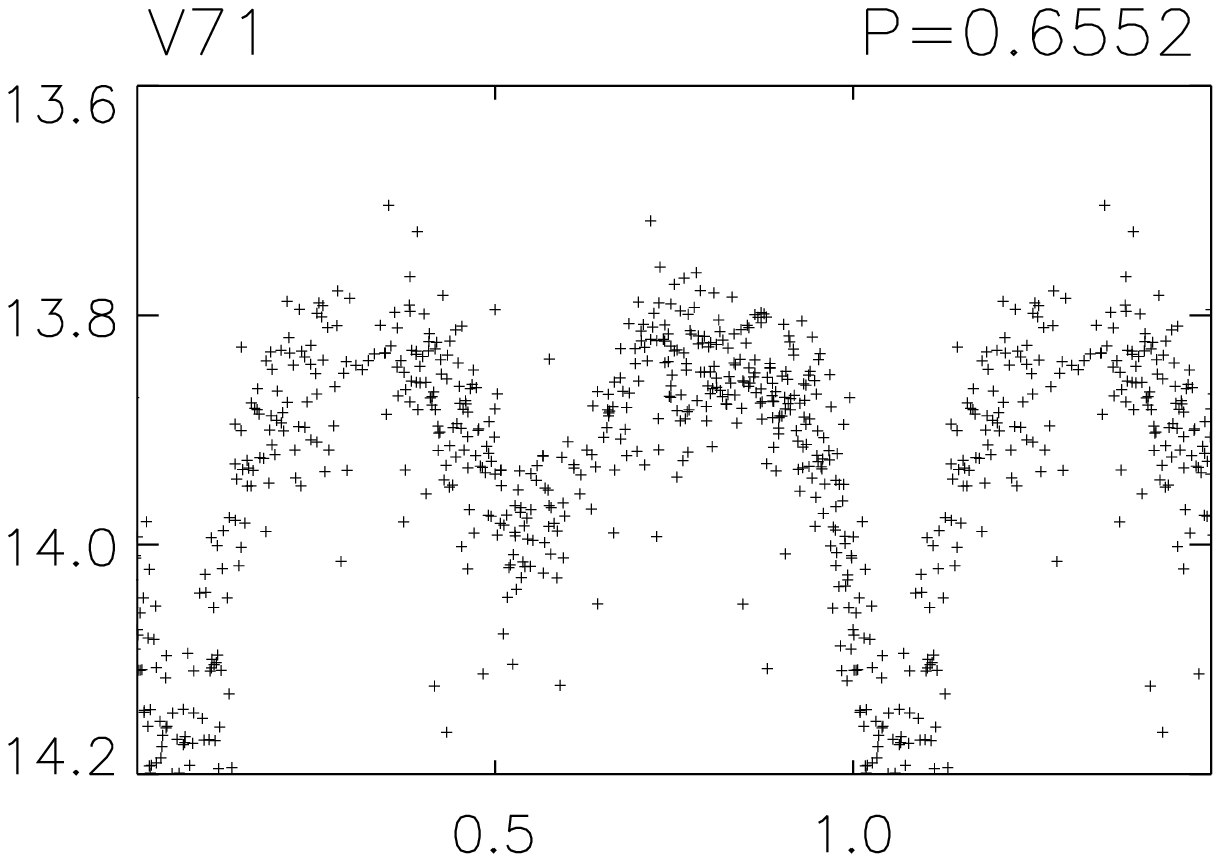}
   \hspace{-0.7cm}\includegraphics[width=4.65cm]{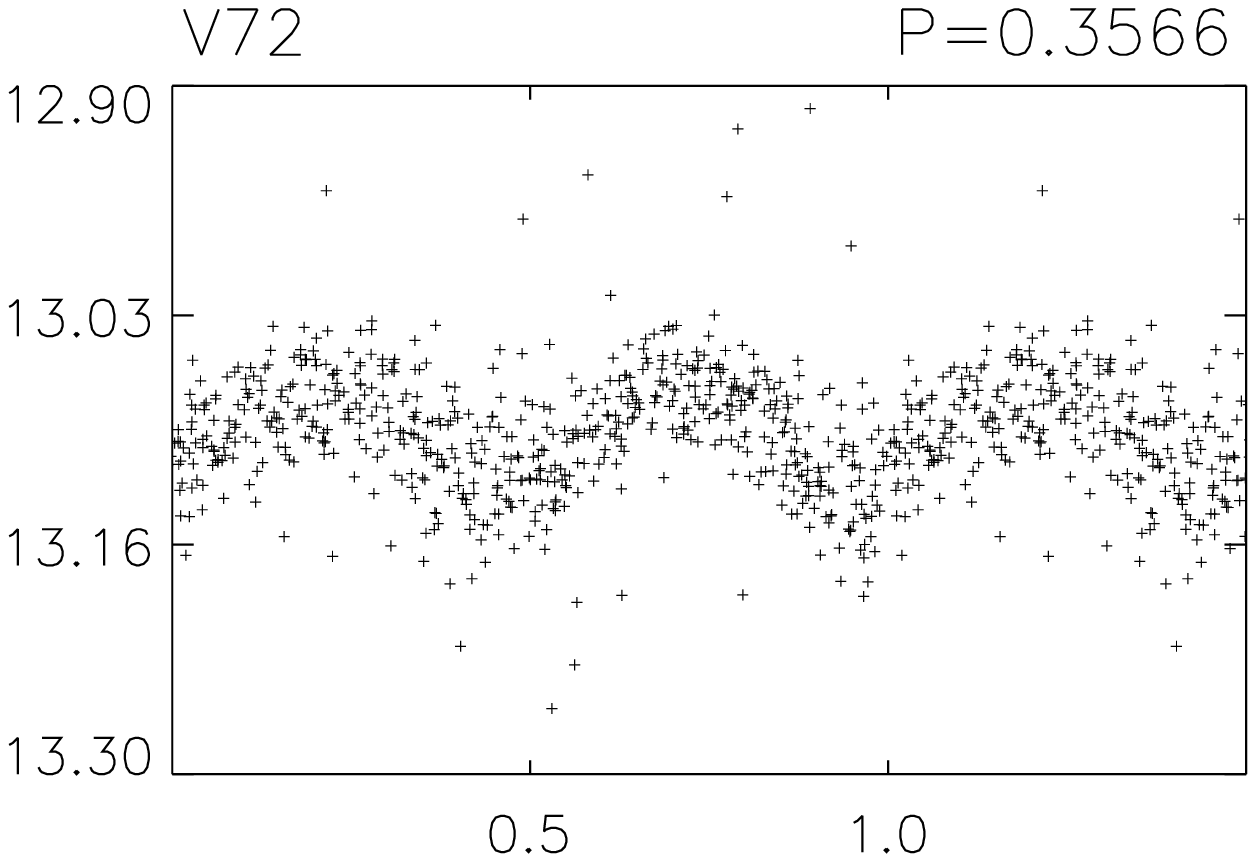}
   \hspace{-0.7cm}\includegraphics[width=4.65cm]{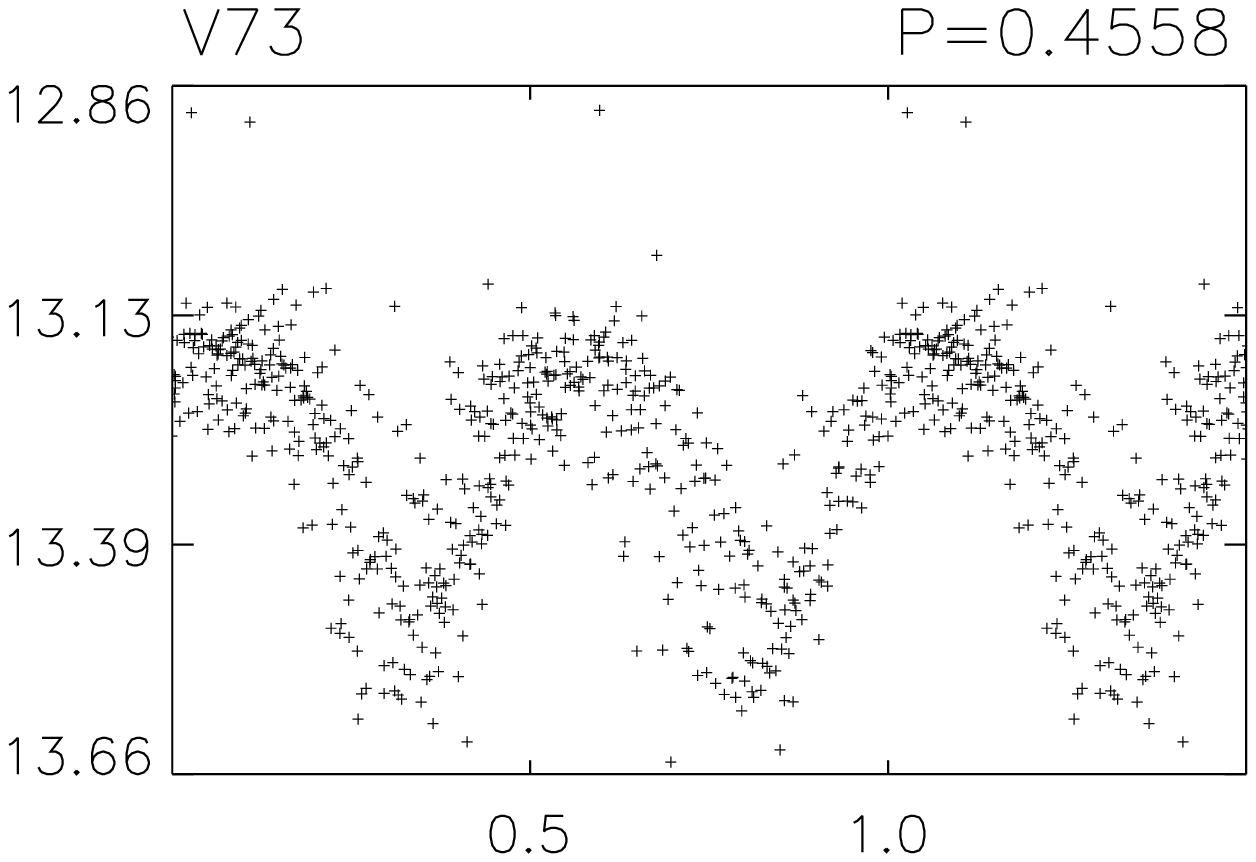}}
\vspace{-0.3cm}
   \hbox{\includegraphics[width=4.65cm]{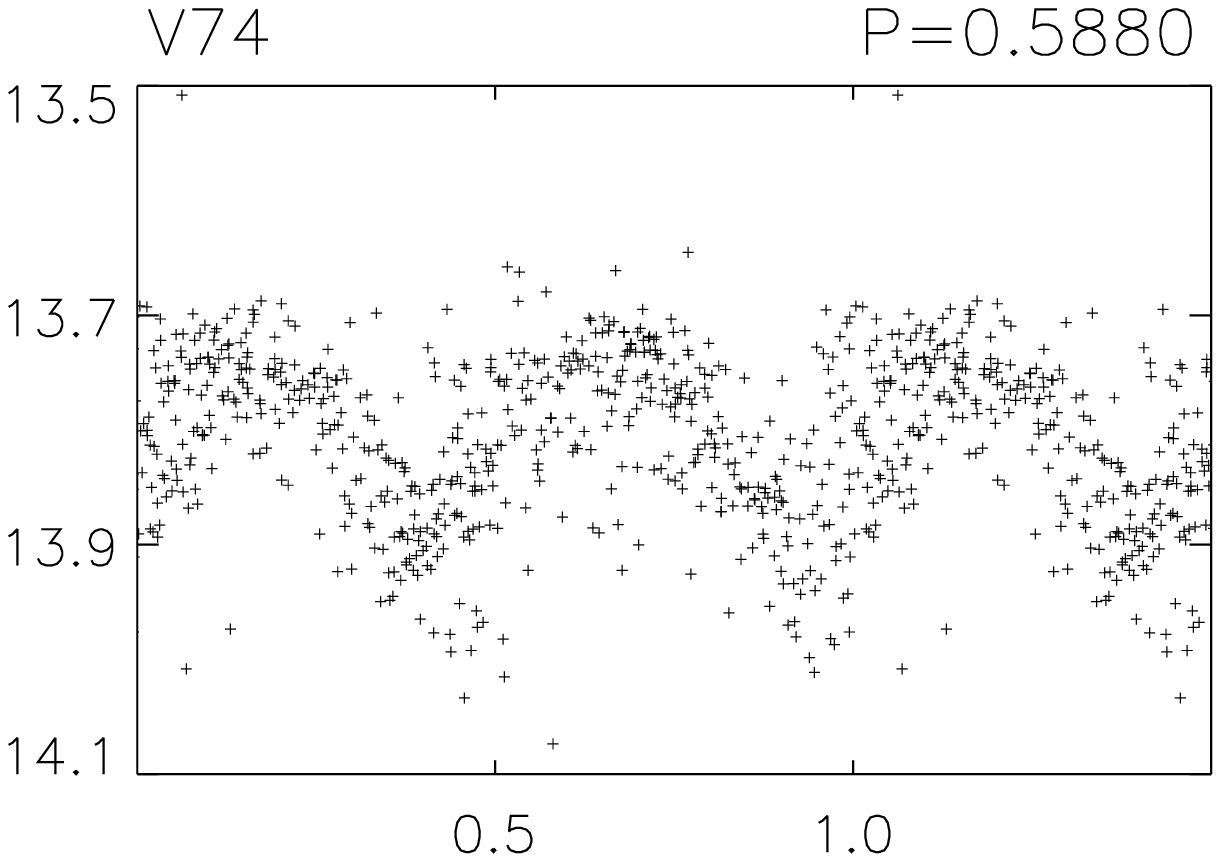}   
   \hspace{-0.7cm}\includegraphics[width=4.65cm]{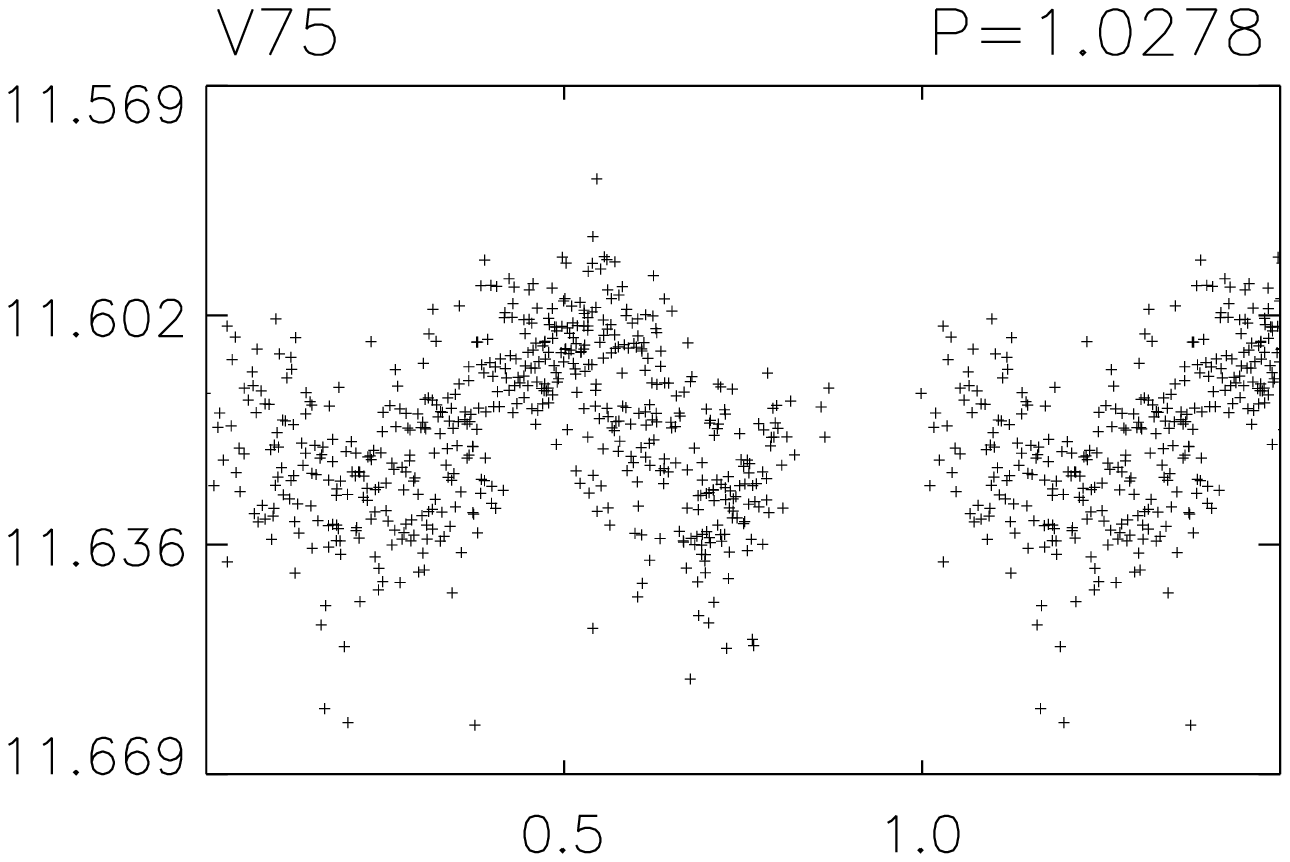}
   \hspace{-0.7cm}\includegraphics[width=4.65cm]{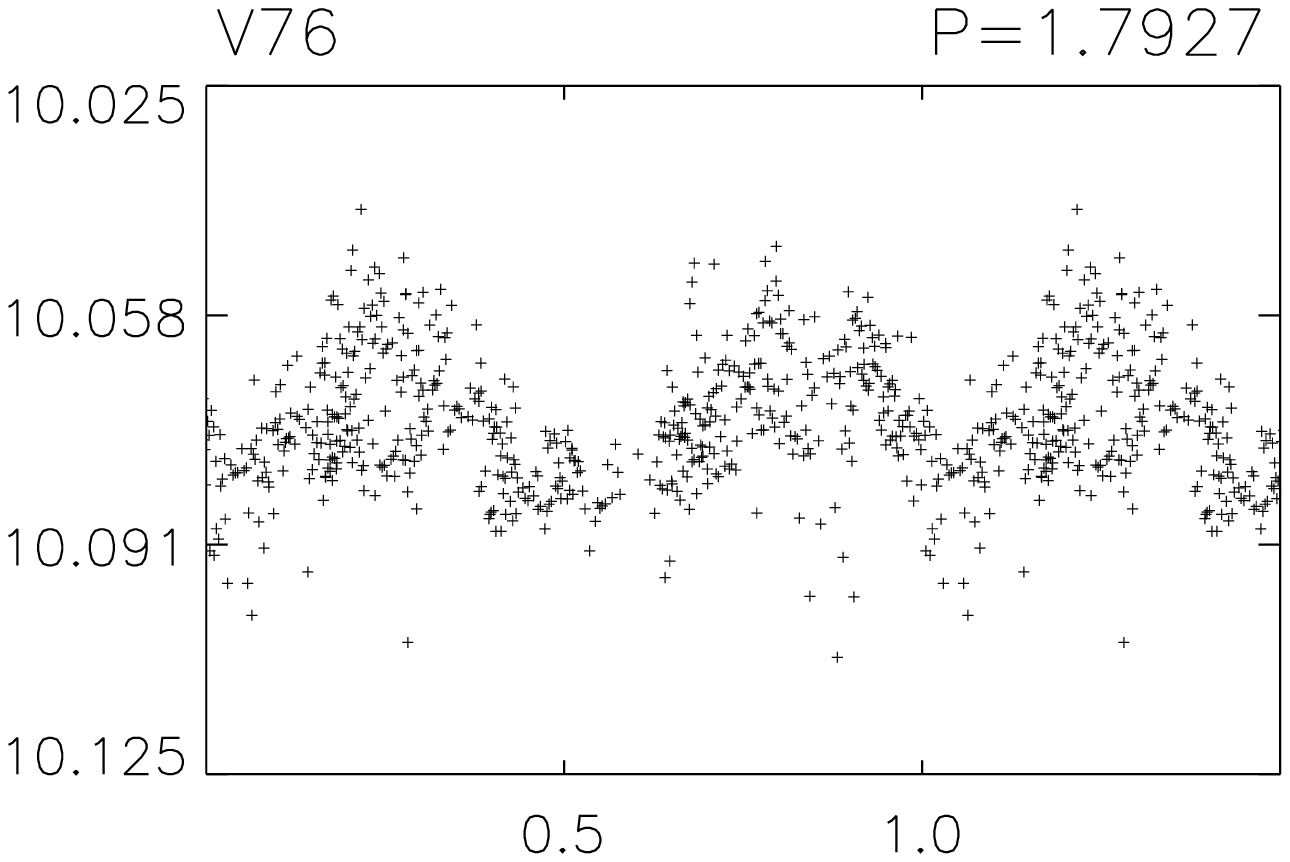}}
\vspace{-0.3cm}
   \hbox{\includegraphics[width=4.65cm]{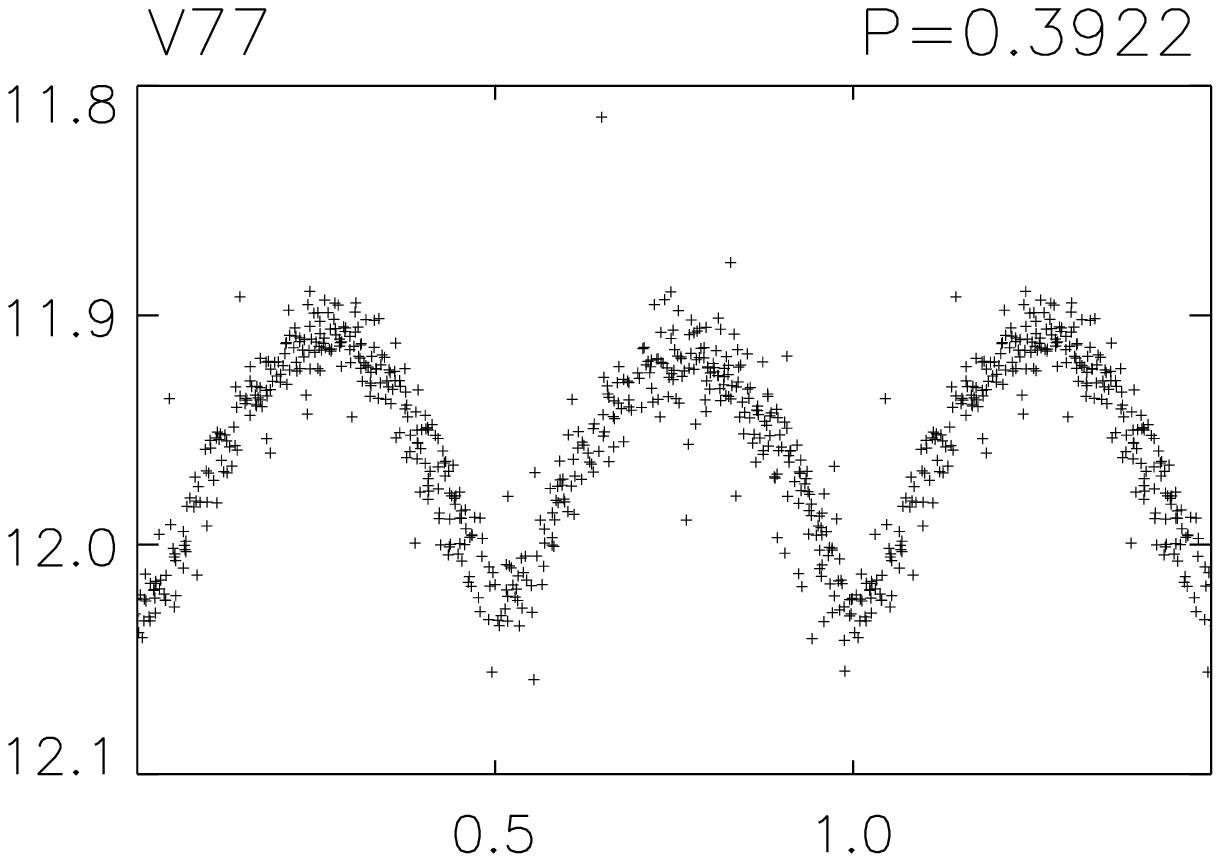}
   \hspace{-0.7cm}\includegraphics[width=4.65cm]{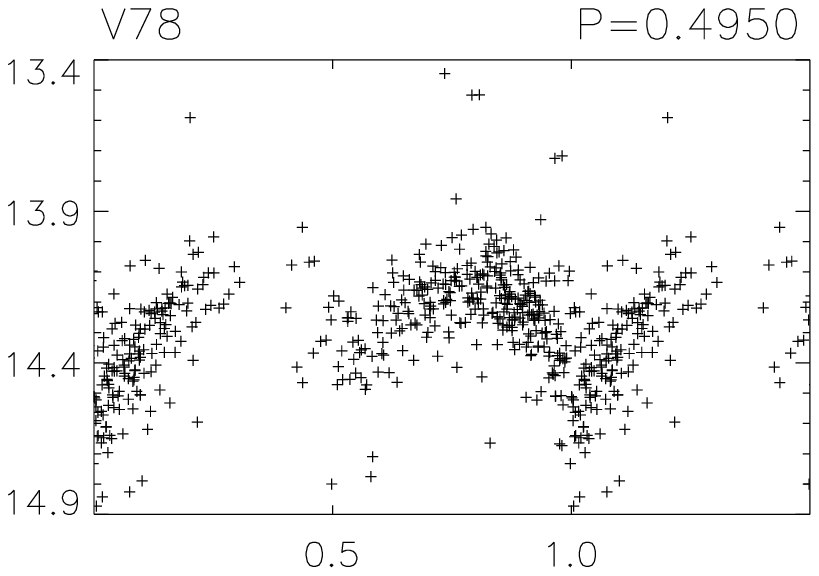}
   \hspace{-0.7cm}\includegraphics[width=4.65cm]{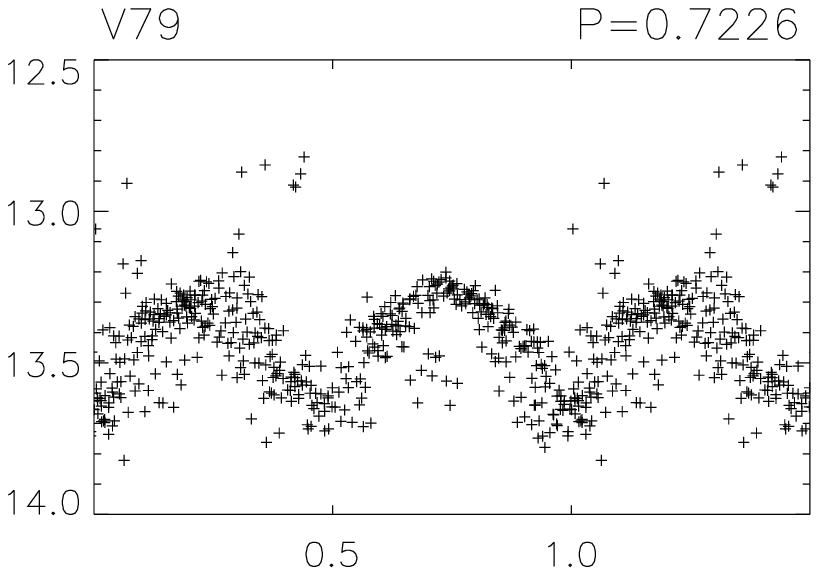}}
\vspace{-0.3cm}
   \hbox{\includegraphics[width=4.65cm]{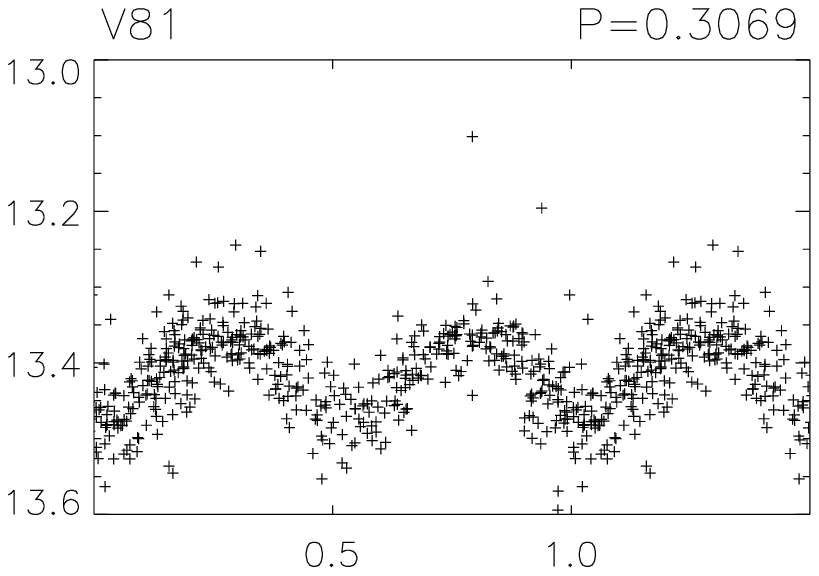}
   \hspace{-0.7cm}\includegraphics[width=4.65cm]{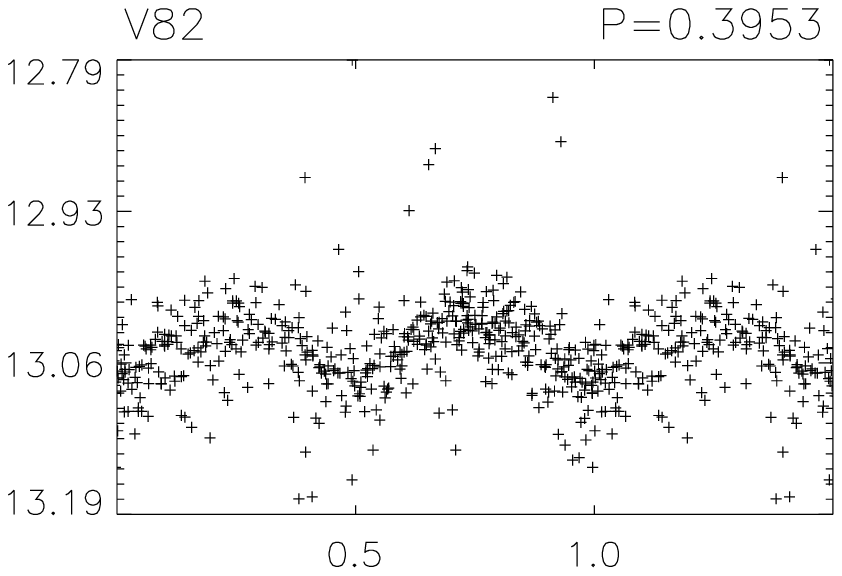}
   \hspace{-0.7cm}\includegraphics[width=4.65cm]{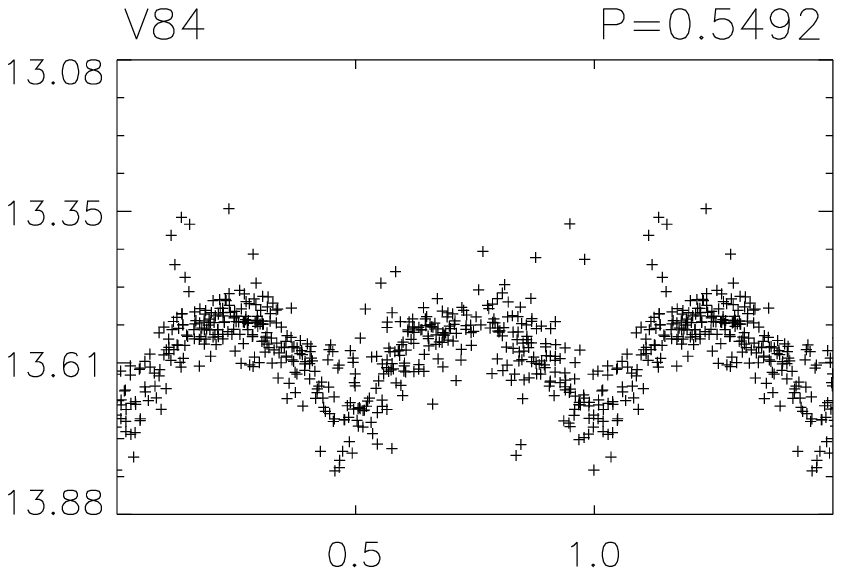}}
   \caption{Continued}
\end{figure*} 
\addtocounter{figure}{-1}
\begin{figure*}[!h]
\centering
\includegraphics[width=4.65cm]{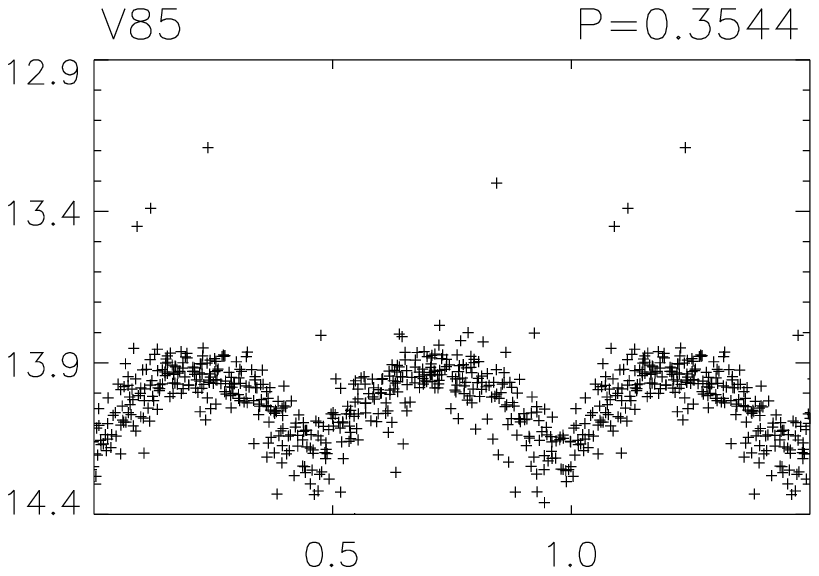}
\caption{Continued}
\label{fig:lcv2}
\end{figure*} 

\begin{figure}
\centering
   \hbox{\includegraphics[width=4.65cm]{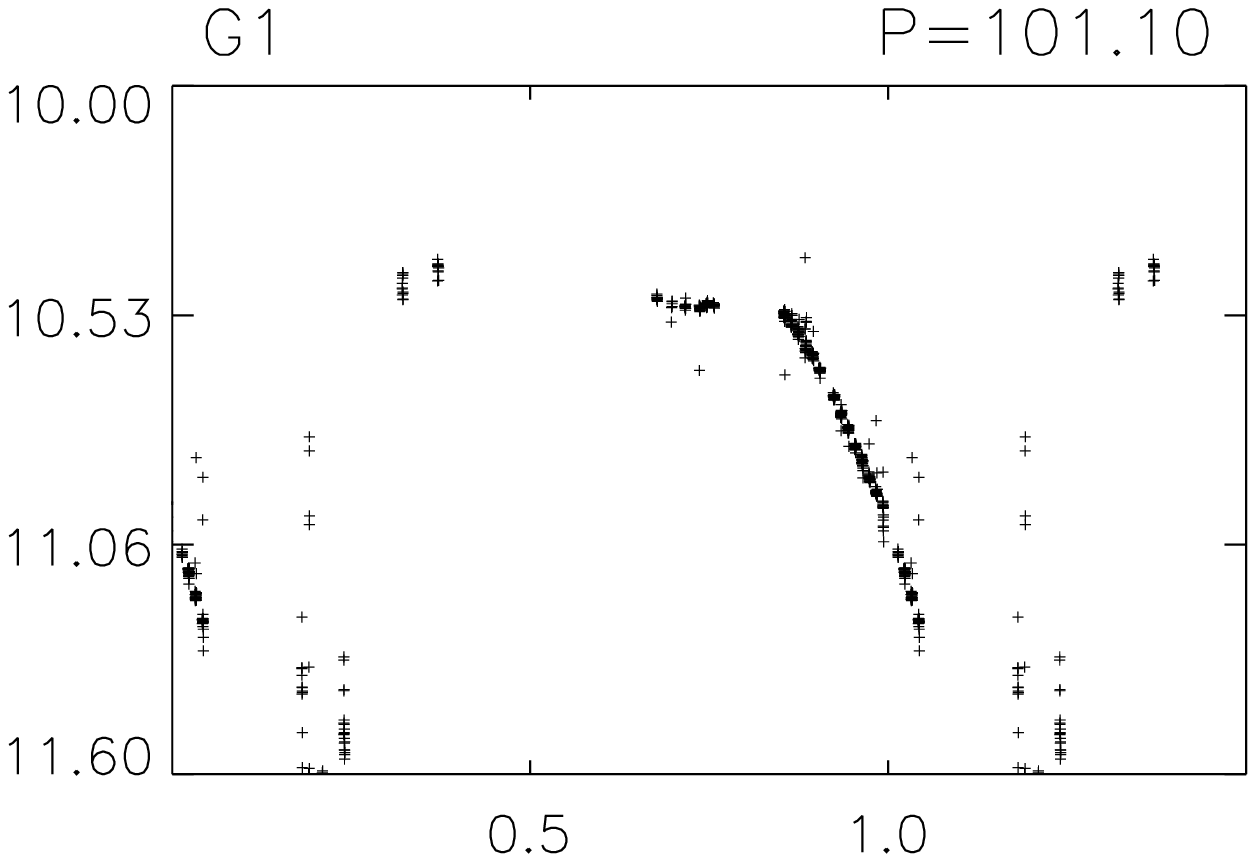}
   \hspace{-0.7cm}\includegraphics[width=4.65cm]{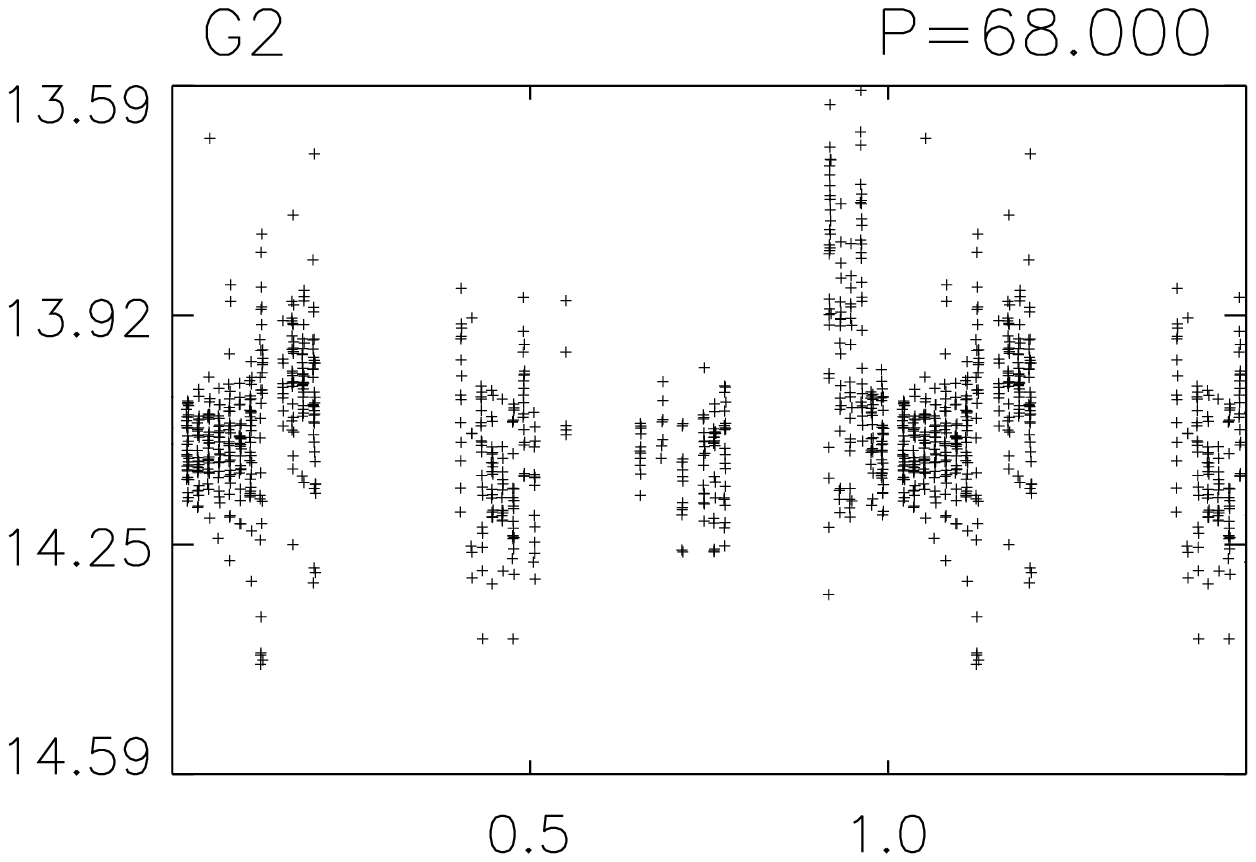}
   \hspace{-0.7cm}\includegraphics[width=4.65cm]{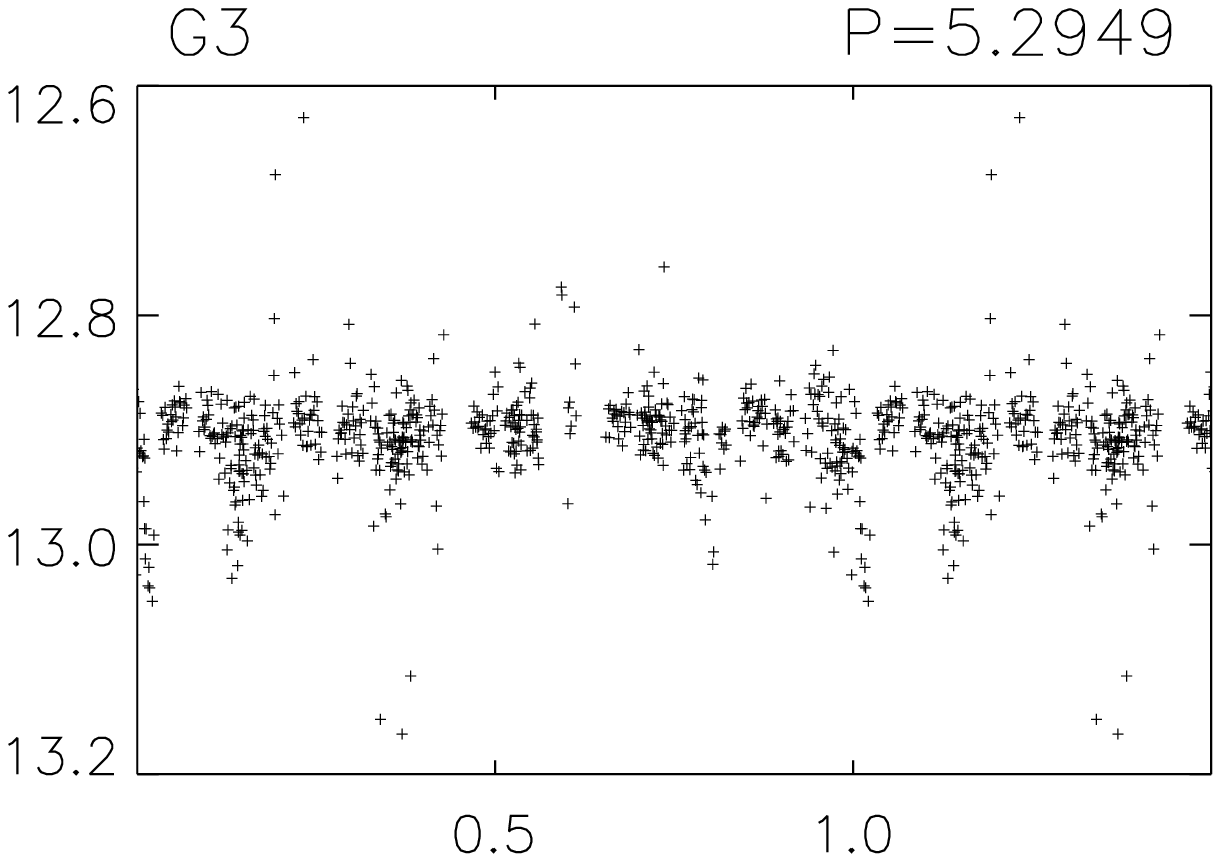}
   \hspace{-0.7cm}\includegraphics[width=4.65cm]{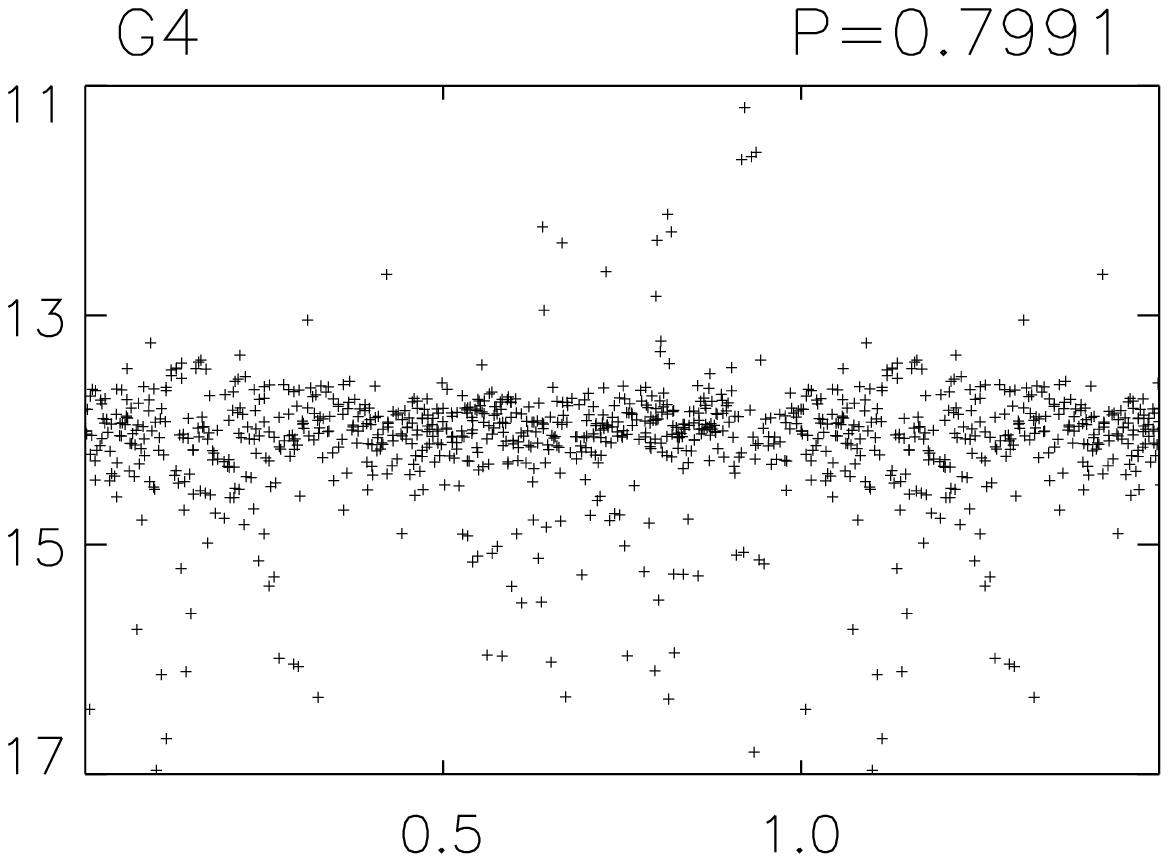}}
\vspace{-0.3cm}
   \hbox{\includegraphics[width=4.65cm]{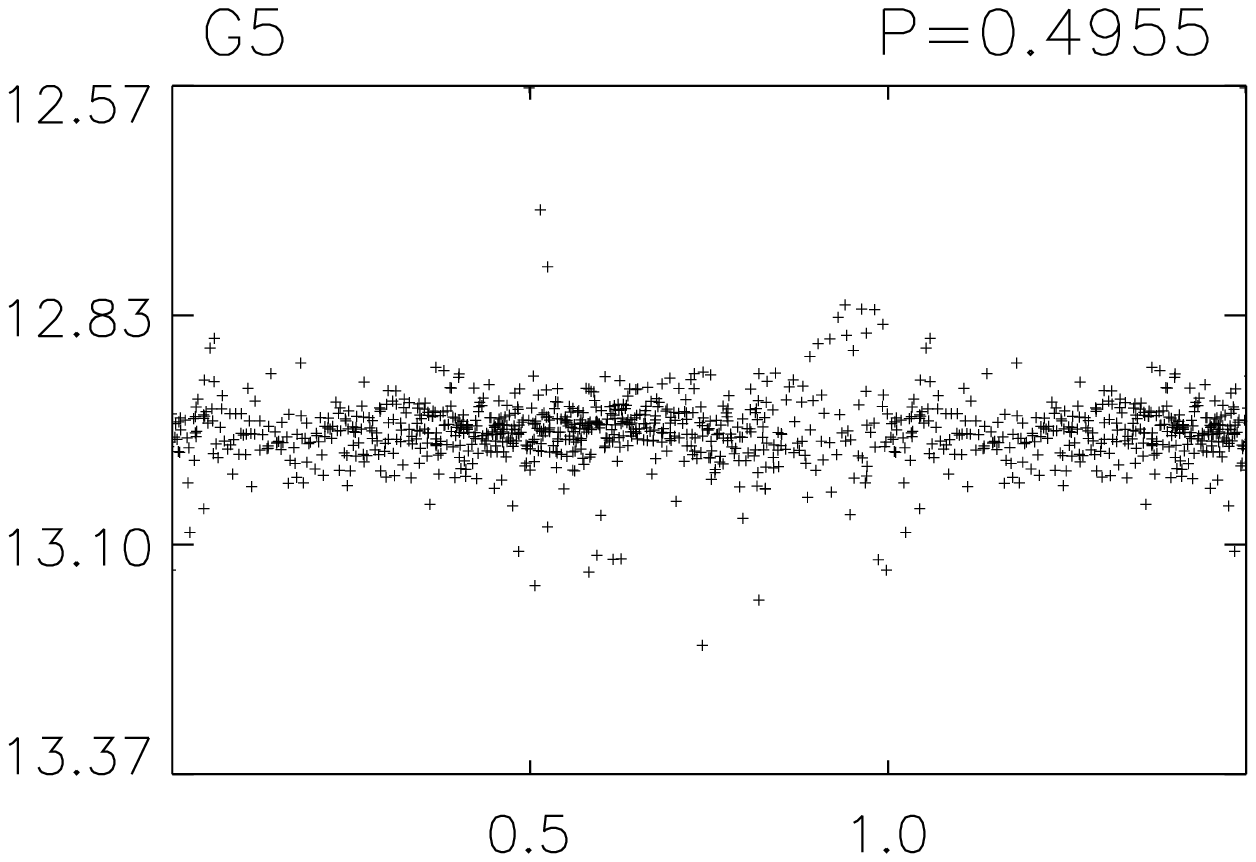}
   \hspace{-0.7cm}\includegraphics[width=4.65cm]{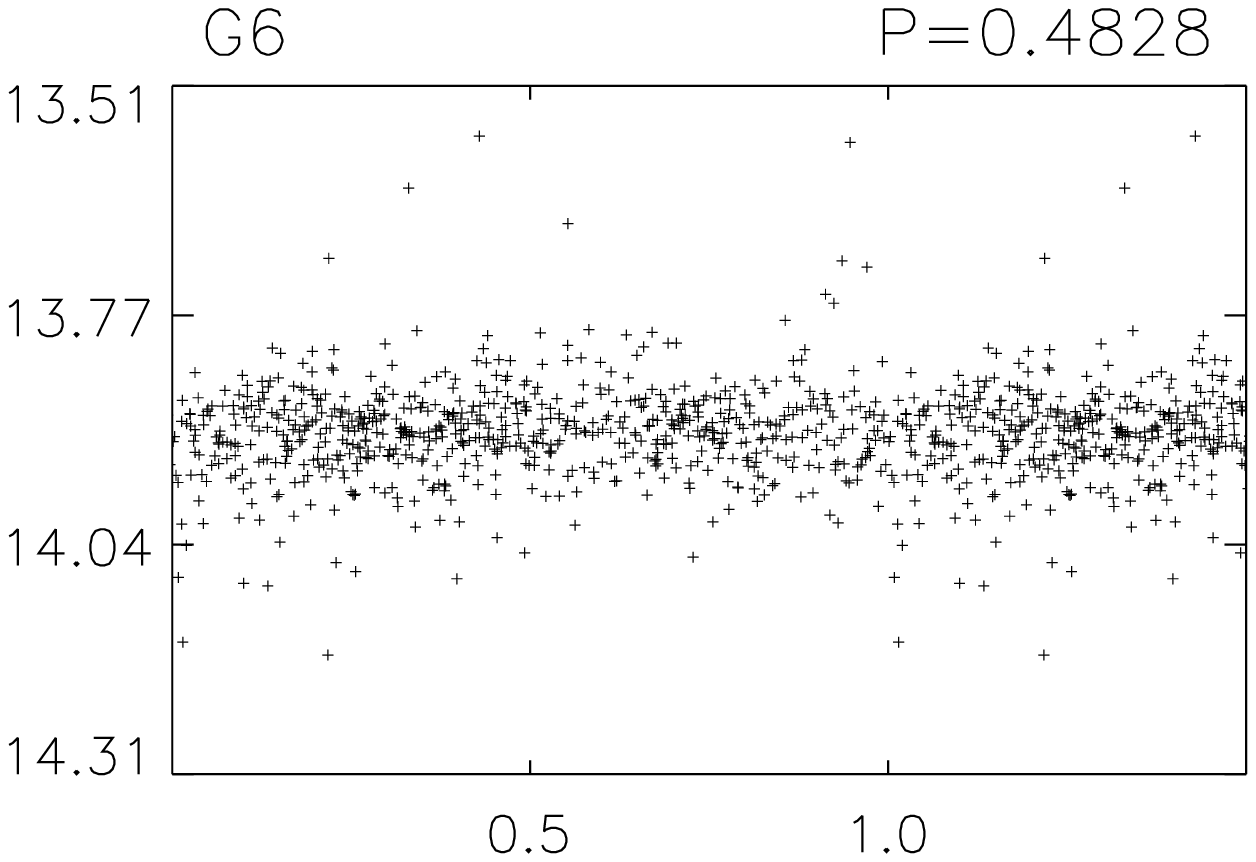}
   \hspace{-0.7cm}\includegraphics[width=4.65cm]{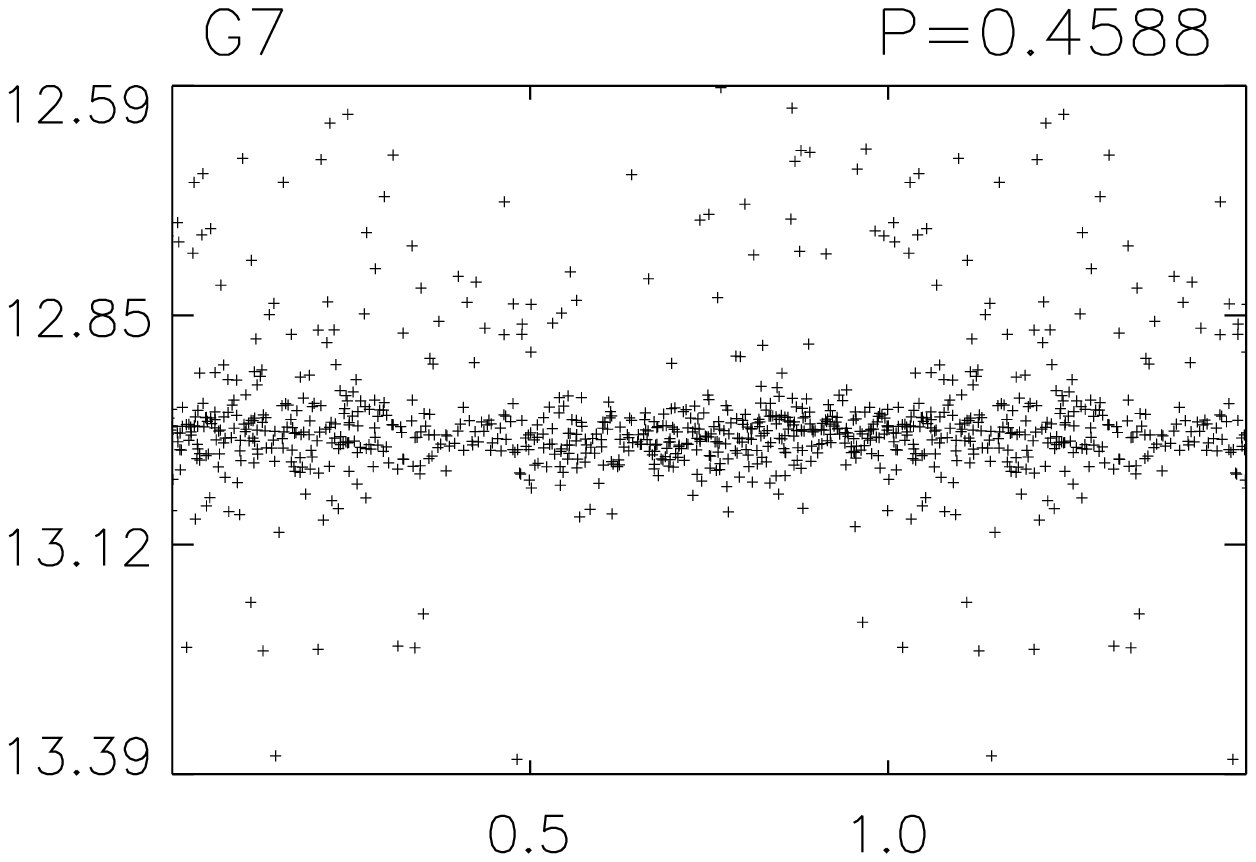}
   \hspace{-0.7cm}\includegraphics[width=4.65cm]{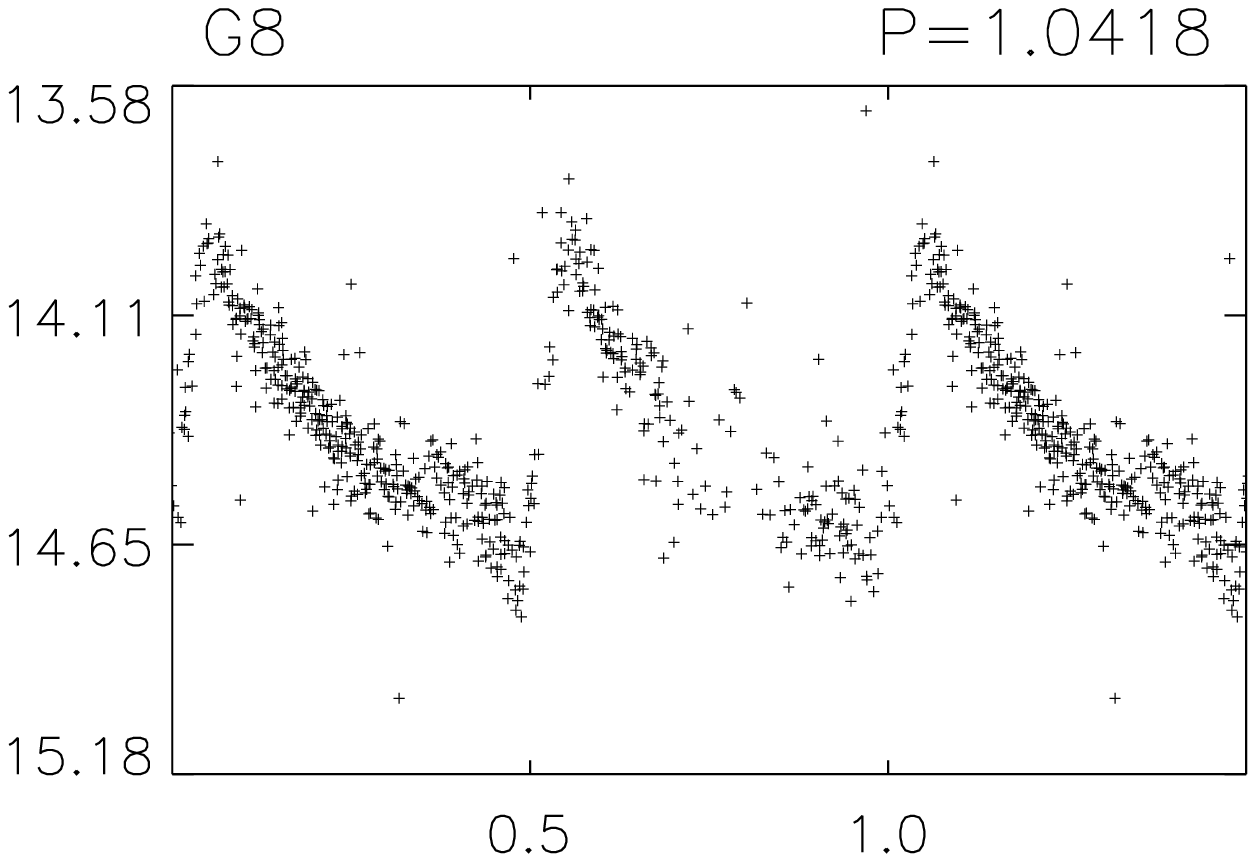}}
\vspace{-0.3cm}
   \hbox{\includegraphics[width=4.65cm]{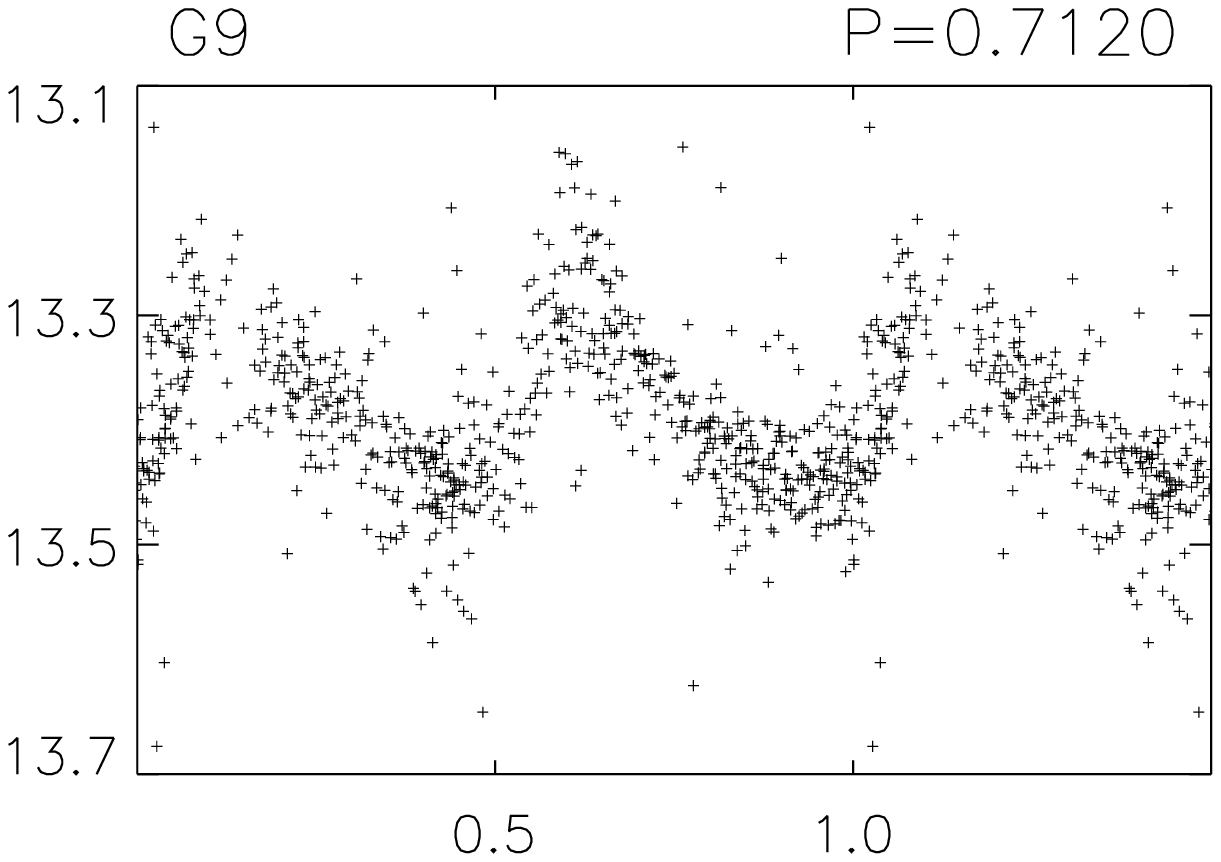}
   \hspace{-0.7cm}\includegraphics[width=4.65cm]{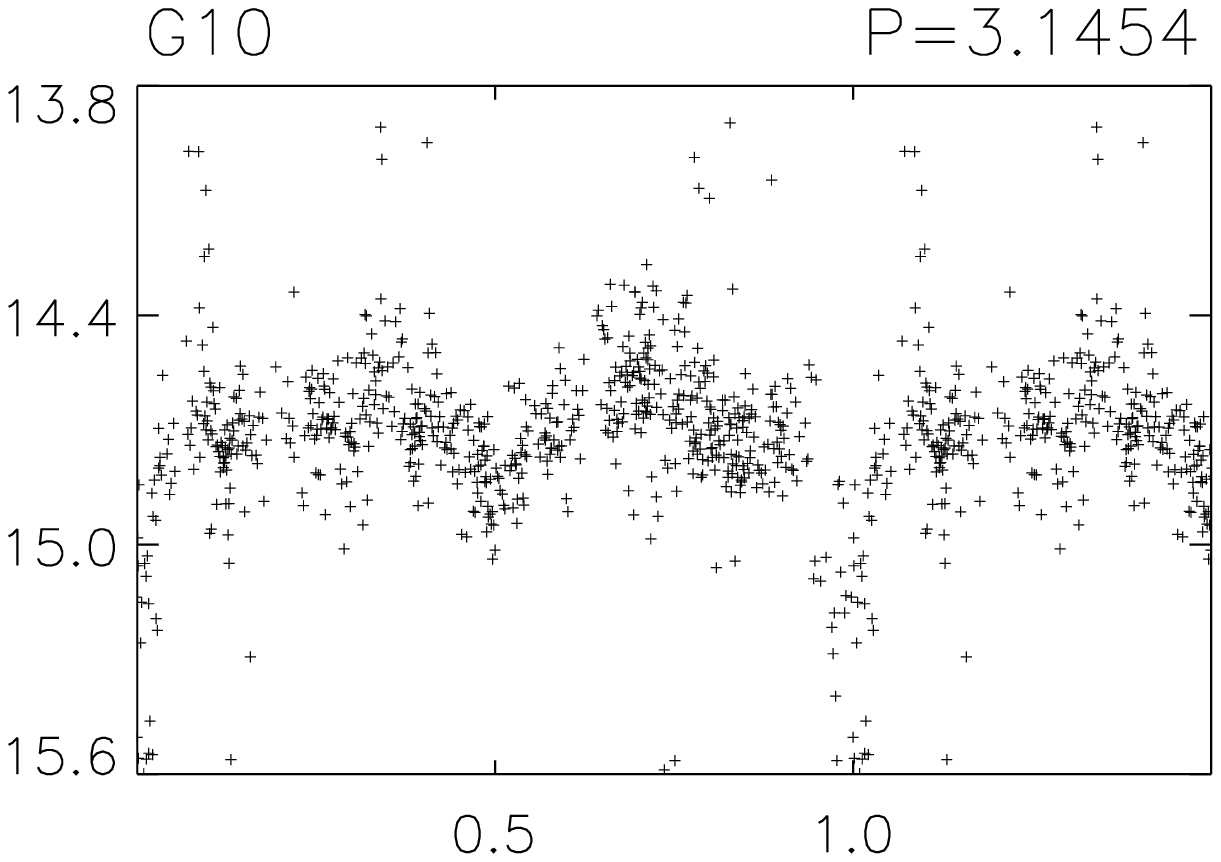}
   \hspace{-0.7cm}\includegraphics[width=4.65cm]{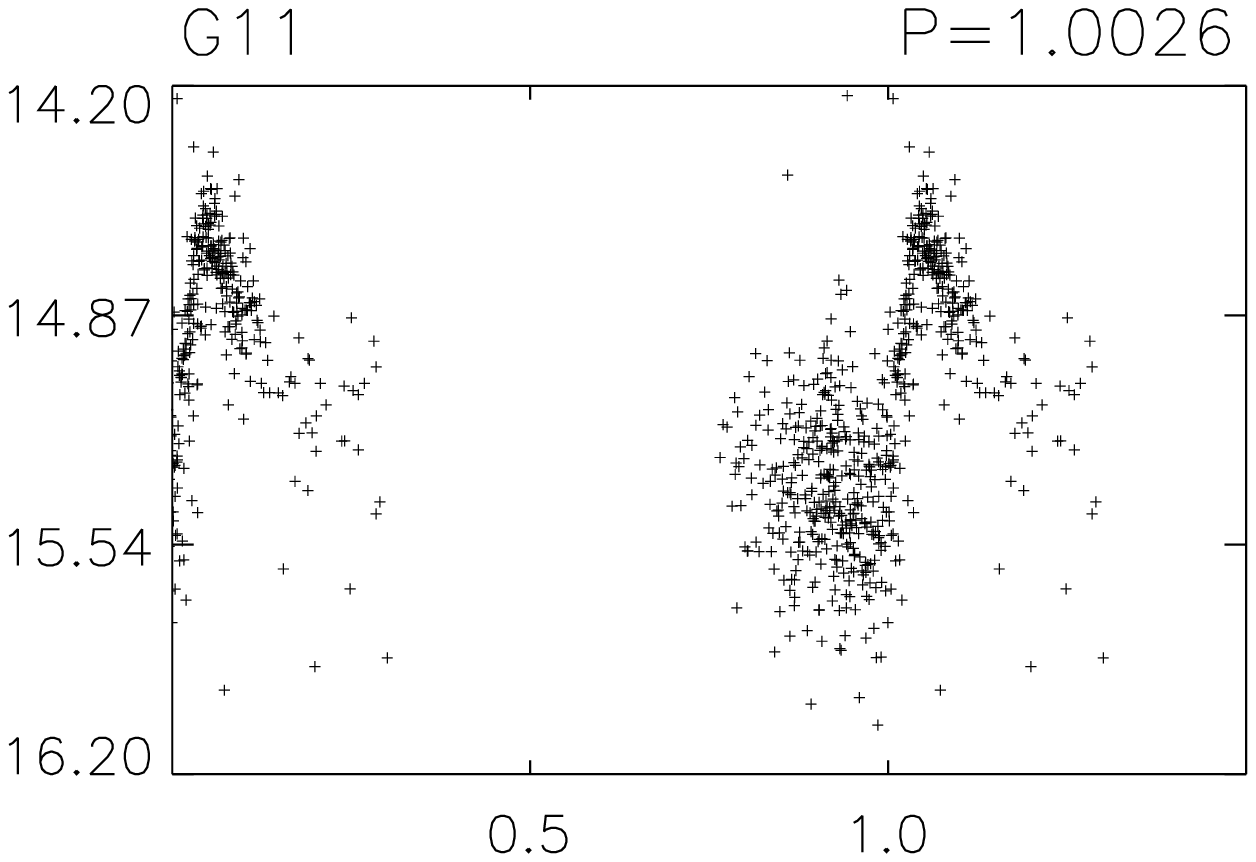}}
\caption{BEST light curves of variable stars known from the General Catalogue of Variable
Stars. See caption in Fig. \ref{fig:lcv1}.} \label{fig:glc}
\end{figure}

\begin{figure}
\centering
\includegraphics[width=\columnwidth]{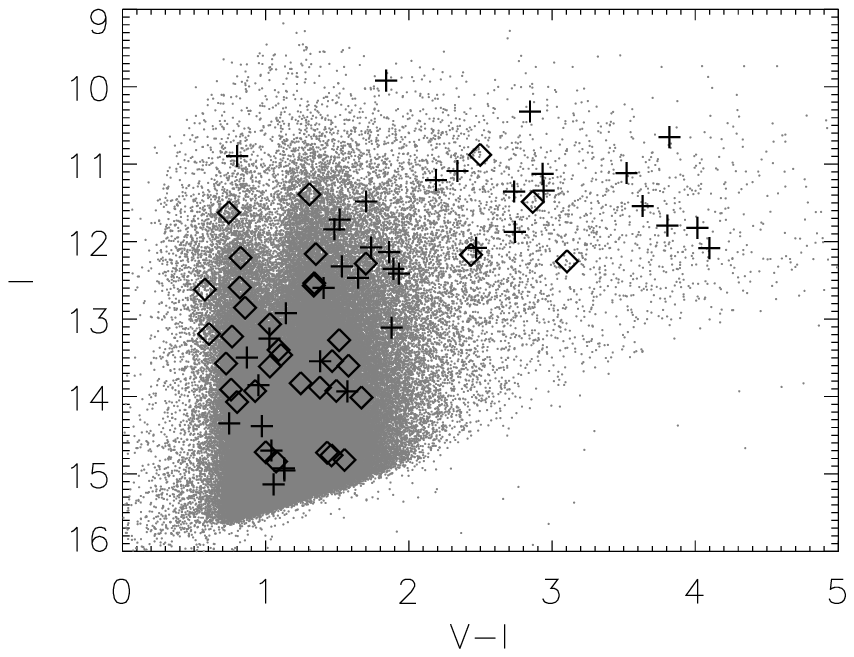}
\caption{Color-magnitude diagram for stars in the \it COROT \rm center
  filed. The gray dots are stars from the ExoDat catalog, the crosses
  are 49 pulsating periodic variable stars and the diamonds are 43
  eclipsing binaries found by BEST and identified in the ExoDat catalog.}
\label{fig:exo}
\end{figure}

\begin{figure}
\centering
\includegraphics[width=\columnwidth]{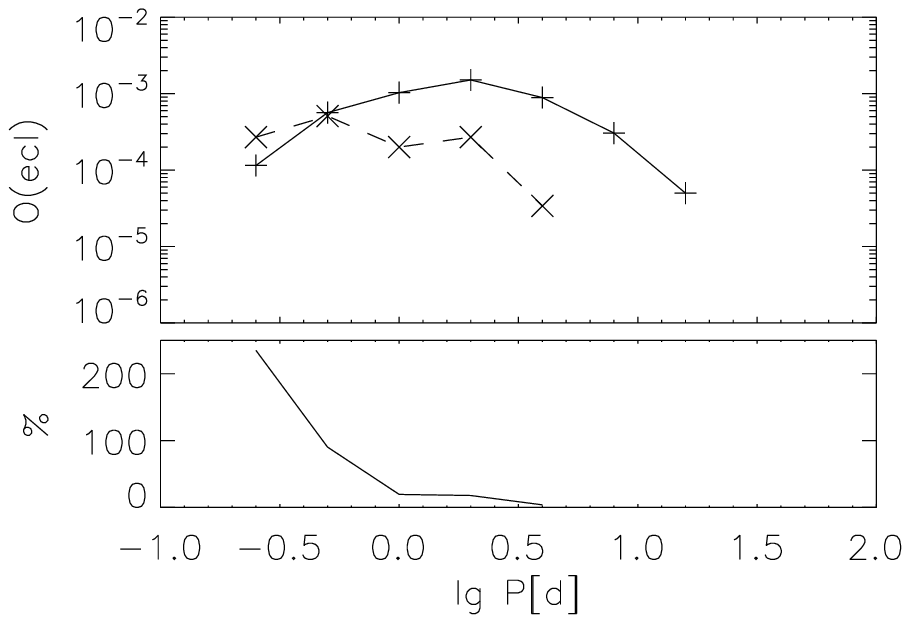}
\caption{Top: Relative number of detected eclipsing binaries as a function of
  period. The solid line shows relative number of eclipsing binaries in the 
  \it HIPPARCOS \rm data -- relative to the total number of stars observed by \it HIPPARCOS \rm
  (from \citet{2005A&A...442.1003S}) and the dashed line shows
  relative number of detected eclipsing binaries in the BEST observations 
  -- relative to the total number of stars observed in the field by BEST. Bottom: The relative difference between the relative number of detected
  eclipsing binaries in the BEST data set and in the \it HIPPARCOS \rm data set
  -- i.e. the relative difference between the two lines in the
  upper plot. The HIPPARCOS data were obtained as the mean value of Fig. 14
  and Fig. 4.15 in \citet{2005A&A...442.1003S}.}
\label{fig:ebs}
\end{figure}

\begin{table}
\begin{center}
\caption{Parameters used in ISIS}
\begin{tabular}{lcclc}
&&&&\\
\hline\hline
sub$\_$x: & 10 & & Sub$\_$y & 10\\
deg$\_$bg: & 3 & & deg$\_$spatial: & 2\\
ngause: & 3 & & &\\
sigma$\_$gauss1: & 0.7 & & deg$\_$gauss1: & 6\\
sigma$\_$gauss2: & 2.0 & & deg$\_$gauss2: &4\\
sigma$\_$gauss3: & 4.0 & & deg$\_$gauss3:  &3\\
nstamps$\_$x: & 10 & & nstamps$\_$y: & 10\\
half$\_$mesh$\_$size: & 9 & & half$\_$stamp$\_$size: & 15\\
\hline
\end{tabular}
\end{center}
\end{table}

\begin{table}
\begin{center}
\caption{Periodic variable stars in the field. $R_{mean}$ is the mean
  magnitude obtained with BEST, $A_r$ is the amplitude of the
  periodic variability in the stars (corresponding to the deeper dip for eclipsing binaries) and the variability type is given as either periodic pulsating variables (PPV) or eclipsing binaries (EB).\vspace{0.5cm}}
\tiny{
\begin{tabular}{lcccccccc}
\hline
ID   & USNO     			     & $\alpha$(J2000) 				   & $\delta$(J2000) 						& $P$(days)  		     & $R_{\rm{mean}}$ & $A_{R}$  & Type\\
\hline\hline
V1   & U0825$\_$15631497 & 19$^{\rm{h}}$23$^{\rm{m}}$50$\fs$63 & -00$^{\circ}$34$^{\prime}$53$\farcs$9        & 1.908   & 12.21           & 0.05    &PPV\\
V2   & U0825$\_$16062830 & 19$^{\rm{h}}$30$^{\rm{m}}$46$\fs$56 & -00$^{\circ}$26$^{\prime}$26$\farcs$0        & 11.932  & 13.40           & 0.16    &PPV\\
V3   & U0825$\_$15906821 & 19$^{\rm{h}}$28$^{\rm{m}}$09$\fs$70 & -00$^{\circ}$28$^{\prime}$21$\farcs$7        & 0.487   & 13.70           & 0.15    &EB\\
V4   & U0825$\_$15803107 & 19$^{\rm{h}}$26$^{\rm{m}}$29$\fs$85 & -00$^{\circ}$28$^{\prime}$59$\farcs$0        & 2.113   & 11.54           & 0.04    &EB\\
V5   & U0825$\_$15951762 & 19$^{\rm{h}}$28$^{\rm{m}}$54$\fs$77 & -00$^{\circ}$26$^{\prime}$13$\farcs$9        & 79.478  & 12.66           & 0.09    &PPV\\
V6   & U0825$\_$15828621 & 19$^{\rm{h}}$26$^{\rm{m}}$54$\fs$41 & -00$^{\circ}$25$^{\prime}$32$\farcs$8        & 0.331   & 13.07           & 0.09    &EB\\
V7   & U0825$\_$15479782 & 19$^{\rm{h}}$21$^{\rm{m}}$36$\fs$79 & -00$^{\circ}$27$^{\prime}$18$\farcs$2        & 1.967   & 11.17           & 0.06    &PPV\\
V8   & U0825$\_$16122557 & 19$^{\rm{h}}$31$^{\rm{m}}$43$\fs$22 & -00$^{\circ}$17$^{\prime}$50$\farcs$6        & 67.898  & 10.30           & 0.12    &PPV\\
V9   & U0825$\_$15871361 & 19$^{\rm{h}}$27$^{\rm{m}}$35$\fs$18 & -00$^{\circ}$17$^{\prime}$49$\farcs$5        & 48.124  & 12.50           & 0.07    &PPV\\
V10  & U0825$\_$16065804 & 19$^{\rm{h}}$30$^{\rm{m}}$49$\fs$50 & -00$^{\circ}$12$^{\prime}$59$\farcs$4        & 6.555   & 11.26           & 0.02    &PPV\\
V11  & U0825$\_$15987353 & 19$^{\rm{h}}$29$^{\rm{m}}$31$\fs$05 & -00$^{\circ}$11$^{\prime}$33$\farcs$6        & 96.016  & 10.76           & 0.06    &PPV\\
V12  & U0825$\_$15757192 & 19$^{\rm{h}}$25$^{\rm{m}}$46$\fs$21 & -00$^{\circ}$11$^{\prime}$33$\farcs$5        & 90.120  & 12.80           & 0.15    &PPV\\
V13  & ...               & 19$^{\rm{h}}$27$^{\rm{m}}$19$\fs$88 & -00$^{\circ}$00$^{\prime}$52$\farcs$6        & 59.590  & 11.87           & 0.15    &PPV\\
V14  & U0900$\_$15449170 & 19$^{\rm{h}}$31$^{\rm{m}}$14$\fs$35 & 00$^{\circ}$02$^{\prime}$33$\farcs$5        & 102.780 & 11.37           & 0.10     &PPV\\
V15  & U0900$\_$15370565 & 19$^{\rm{h}}$30$^{\rm{m}}$15$\fs$97 & 00$^{\circ}$05$^{\prime}$37$\farcs$6        & 2.273   & 12.51           & 0.03     &EB\\
V16  & U0900$\_$14944293 & 19$^{\rm{h}}$24$^{\rm{m}}$36$\fs$78 & 00$^{\circ}$01$^{\prime}$16$\farcs$5       & 32.344  & 11.01           & 0.29      &PPV\\
V17  & U0900$\_$15533956 & 19$^{\rm{h}}$32$^{\rm{m}}$17$\fs$06 & 00$^{\circ}$08$^{\prime}$47$\farcs$7        & 1.902   & 12.22           & 0.05     &EB\\
V18  & U0900$\_$14773005 & 19$^{\rm{h}}$22$^{\rm{m}}$02$\fs$23 & 00$^{\circ}$00$^{\prime}$30$\farcs$4        & 1.135   & 12.22           & 0.13     &EB\\
V19  & U0900$\_$14801088 & 19$^{\rm{h}}$22$^{\rm{m}}$27$\fs$37 & 00$^{\circ}$06$^{\prime}$59$\farcs$9        & 1.870   & 13.15           & 0.05     &EB\\
V20  & U0900$\_$14803632 & 19$^{\rm{h}}$22$^{\rm{m}}$29$\fs$64 & 00$^{\circ}$09$^{\prime}$57$\farcs$4        & 0.975   & 13.16           & 0.18     &EB\\
V21  & U0900$\_$14963651 & 19$^{\rm{h}}$24$^{\rm{m}}$52$\fs$18 & 00$^{\circ}$20$^{\prime}$50$\farcs$3        & 85.208  & 12.94           & 0.09     &PPV\\
V22  & ...               & 19$^{\rm{h}}$21$^{\rm{m}}$18$\fs$72 & 00$^{\circ}$17$^{\prime}$50$\farcs$8        & 93.828  & 12.51           & 0.12     &PPV\\
V23  & U0900$\_$15317860 & 19$^{\rm{h}}$29$^{\rm{m}}$34$\fs$44 & 00$^{\circ}$32$^{\prime}$54$\farcs$9        & 15.250  & 11.77           & 0.08     &PPV\\
V24  & U0900$\_$14875071 & 19$^{\rm{h}}$23$^{\rm{m}}$36$\fs$07 & 00$^{\circ}$31$^{\prime}$39$\farcs$4        & 0.678   & 11.91           & 0.04     &EB\\
V25  & U0900$\_$15102073 & 19$^{\rm{h}}$26$^{\rm{m}}$47$\fs$78 & 00$^{\circ}$34$^{\prime}$47$\farcs$2        & 78.260  & 12.52           & 0.07     &PPV\\
V26  & U0900$\_$15105221 & 19$^{\rm{h}}$26$^{\rm{m}}$50$\fs$33 & 00$^{\circ}$34$^{\prime}$53$\farcs$4        & 77.540  & 12.09           & 0.11     &PPV\\
V27  & U0900$\_$15227975 & 19$^{\rm{h}}$28$^{\rm{m}}$26$\fs$95 & 00$^{\circ}$43$^{\prime}$19$\farcs$9        & 116.850 & 10.64           & 0.12     &PPV\\
V28  & U0900$\_$14801012 & 19$^{\rm{h}}$22$^{\rm{m}}$27$\fs$34 & 00$^{\circ}$38$^{\prime}$02$\farcs$4        & 0.670   & 12.46           & 0.02     &EB\\
V29  & U0900$\_$14785145 & 19$^{\rm{h}}$22$^{\rm{m}}$13$\fs$39 & 00$^{\circ}$40$^{\prime}$37$\farcs$7        & 43.135  & 10.71           & 0.04     &PPV \\
V30  & U0900$\_$15459522 & 19$^{\rm{h}}$31$^{\rm{m}}$21$\fs$44 & 00$^{\circ}$53$^{\prime}$04$\farcs$3        & 55.864  & 11.70           & 0.07     &PPV \\
V31  & U0900$\_$15338939 & 19$^{\rm{h}}$29$^{\rm{m}}$50$\fs$48 & 00$^{\circ}$55$^{\prime}$01$\farcs$6        & 4.420   & 12.28           & 0.04     &EB\\
V32  & U0900$\_$15147997 & 19$^{\rm{h}}$27$^{\rm{m}}$23$\fs$94 & 00$^{\circ}$53$^{\prime}$07$\farcs$7        & 0.454   & 13.38           & 0.08     &EB\\
V33  & U0900$\_$14857685 & 19$^{\rm{h}}$23$^{\rm{m}}$20$\fs$96 & 01$^{\circ}$50$^{\prime}$48$\farcs$6        & 0.324   & 14.64           & 0.19     &EB\\
V34  & U0900$\_$14842306 & 19$^{\rm{h}}$23$^{\rm{m}}$06$\fs$29 & 01$^{\circ}$06$^{\prime}$49$\farcs$3        & 0.549   & 12.71           & 0.06     &EB\\
V35  & ...               & 19$^{\rm{h}}$31$^{\rm{m}}$48$\fs$52 & 01$^{\circ}$14$^{\prime}$41$\farcs$7        & 0.306   & 13.72           & 0.13     &EB\\
V36  & U0900$\_$15489834 & 19$^{\rm{h}}$31$^{\rm{m}}$42$\fs$44 & 01$^{\circ}$14$^{\prime}$44$\farcs$2        & 67.996  & 12.38           & 0.07     &PPV\\
V37  & U0900$\_$14842441 & 19$^{\rm{h}}$23$^{\rm{m}}$06$\fs$42 & 01$^{\circ}$07$^{\prime}$06$\farcs$8        & 1.174   & 12.82           & 0.09     &EB\\
V38  & U0900$\_$15096269 & 19$^{\rm{h}}$26$^{\rm{m}}$42$\fs$91 & 01$^{\circ}$11$^{\prime}$48$\farcs$7        & 0.452   & 12.20           & 0.07     &EB\\
V39  & U0900$\_$15088039 & 19$^{\rm{h}}$26$^{\rm{m}}$36$\fs$27 & 01$^{\circ}$13$^{\prime}$15$\farcs$5        & 0.942   & 13.53           & 0.10     &PPV\\
V40  & U0900$\_$15245627 & 19$^{\rm{h}}$26$^{\rm{m}}$37$\fs$33 & 01$^{\circ}$13$^{\prime}$33$\farcs$9        & 1.783   & 14.65           & 0.10     &EB\\
V41  & U0900$\_$14903885 & 19$^{\rm{h}}$24$^{\rm{m}}$02$\fs$05 & 01$^{\circ}$21$^{\prime}$50$\farcs$2        & 39.304  & 12.18           & 0.20     &PPV\\
V42  & U0900$\_$15033451 & 19$^{\rm{h}}$25$^{\rm{m}}$51$\fs$33 & 01$^{\circ}$29$^{\prime}$12$\farcs$5        & 0.431   & 14.02           & 0.18     &EB\\
V43  & U0900$\_$15172322 & 19$^{\rm{h}}$27$^{\rm{m}}$43$\fs$57 & 01$^{\circ}$32$^{\prime}$23$\farcs$8        & 63.682  & 11.62           & 0.03     &PPV\\
V44  & U0900$\_$15110355 & 19$^{\rm{h}}$27$^{\rm{m}}$42$\fs$76 & 01$^{\circ}$32$^{\prime}$50$\farcs$7        & 1.942   & 11.37           & 0.03     &EB\\
V45  & U0900$\_$15171348 & 19$^{\rm{h}}$26$^{\rm{m}}$54$\fs$39 & 01$^{\circ}$32$^{\prime}$32$\farcs$1        & 1.110   & 13.23           & 0.10     &PPV\\
V46  & U0900$\_$15245627 & 19$^{\rm{h}}$28$^{\rm{m}}$40$\fs$54 & 01$^{\circ}$34$^{\prime}$50$\farcs$4        & 1.966   & 11.53           & 0.06     &EB\\
V47  & U0900$\_$15190277 & 19$^{\rm{h}}$27$^{\rm{m}}$58$\fs$26 & 01$^{\circ}$35$^{\prime}$04$\farcs$1        & 1.439   & 11.61           & 0.07     &EB\\
V48  & U0900$\_$15295393 & 19$^{\rm{h}}$29$^{\rm{m}}$18$\fs$23 & 01$^{\circ}$37$^{\prime}$29$\farcs$9        & 55.496  & 12.99           & 0.10     &PPV\\
V49  & U0900$\_$14925806 & 19$^{\rm{h}}$24$^{\rm{m}}$21$\fs$19 & 01$^{\circ}$33$^{\prime}$31$\farcs$7        & 0.334   & 12.98           & 0.04     &EB\\
V50  & U0900$\_$14877836 & 19$^{\rm{h}}$23$^{\rm{m}}$38$\fs$36 & 01$^{\circ}$34$^{\prime}$44$\farcs$0        & 1.383   & 12.35           & 0.06     &EB\\
V51  & U0900$\_$15318544 & 19$^{\rm{h}}$29$^{\rm{m}}$34$\fs$91 & 01$^{\circ}$42$^{\prime}$55$\farcs$4        & 52.763  & 11.17           & 0.05     &PPV\\
V52  & U0900$\_$14757657 & 19$^{\rm{h}}$21$^{\rm{m}}$47$\fs$43 & 01$^{\circ}$37$^{\prime}$15$\farcs$3        & 0.414   & 13.91           & 0.18     &EB\\
V53  & U0900$\_$14975844 & 19$^{\rm{h}}$25$^{\rm{m}}$02$\fs$35 & 01$^{\circ}$41$^{\prime}$43$\farcs$0        & 0.355   & 11.83           & 0.04     &EB\\
V54  & ...               & 19$^{\rm{h}}$23$^{\rm{m}}$59$\fs$33 & 01$^{\circ}$41$^{\prime}$30$\farcs$1        & 1.110   & 13.85           & 0.18     &PPV\\
V55  & U0900$\_$15323691 & 19$^{\rm{h}}$29$^{\rm{m}}$38$\fs$78 & 01$^{\circ}$46$^{\prime}$35$\farcs$4        & 94.240  & 11.99           & 0.19    &PPV\\
V56  & U0900$\_$14715405 & 19$^{\rm{h}}$21$^{\rm{m}}$05$\fs$86 & 01$^{\circ}$46$^{\prime}$41$\farcs$9        & 89.746  & 11.80           & 0.06    &PPV\\
V57  & U0900$\_$15337670 & 19$^{\rm{h}}$29$^{\rm{m}}$49$\fs$43 & 01$^{\circ}$54$^{\prime}$27$\farcs$4        & 38.362  & 12.10           & 0.02    &PPV\\
V58  & U0900$\_$14845601 & 19$^{\rm{h}}$23$^{\rm{m}}$09$\fs$52 & 01$^{\circ}$49$^{\prime}$30$\farcs$9        & 47.360  & 10.07           & 0.10    &PPV\\
V59  & ...               & 19$^{\rm{h}}$26$^{\rm{m}}$37$\fs$45 & 01$^{\circ}$52$^{\prime}$57$\farcs$2        & 0.286   & 13.11           & 0.08    &EB\\
V60  & U0900$\_$15090051 & 19$^{\rm{h}}$26$^{\rm{m}}$37$\fs$87 & 01$^{\circ}$56$^{\prime}$02$\farcs$9        & 90.288  & 12.02           & 0.10    &PPV\\
\hline
\end{tabular}
}
\end{center}
\end{table}
\addtocounter{table}{-1}
\begin{table}
\begin{center}
\caption{Continued.}
 \vspace{0.5cm}
\tiny{
\begin{tabular}{lcccccccc}
\hline
ID   & USNO     			     & $\alpha$(J2000) 				   & $\delta$(J2000) 						& $P$(days)  		     & $R_{\rm{mean}}$ & $A_{R}$  & Type\\
\hline\hline
V61  & U0900$\_$15089049 & 19$^{\rm{h}}$26$^{\rm{m}}$37$\fs$11 & 01$^{\circ}$56$^{\prime}$17$\farcs$0        & 2.045   & 12.02           & 0.06    &EB\\
V62  & U0900$\_$14770477 & 19$^{\rm{h}}$22$^{\rm{m}}$02$\fs$70 & 01$^{\circ}$53$^{\prime}$01$\farcs$5        & 74.072  & 12.34           & 0.11    &PPV\\
V63  & U0900$\_$15513261 & 19$^{\rm{h}}$32$^{\rm{m}}$00$\fs$07 & 02$^{\circ}$04$^{\prime}$19$\farcs$9        & 51.242  & 11.60           & 0.06    &PPV\\
V64  & U0900$\_$15286457 & 19$^{\rm{h}}$29$^{\rm{m}}$11$\fs$40 & 02$^{\circ}$02$^{\prime}$07$\farcs$5        & 75.340  & 11.92           & 0.06    &PPV\\
V65  & U0900$\_$15129182 & 19$^{\rm{h}}$27$^{\rm{m}}$08$\fs$90 & 02$^{\circ}$04$^{\prime}$04$\farcs$2        & 0.349   & 12.59           & 0.05    &EB\\
V66  & U0900$\_$14931322 & 19$^{\rm{h}}$24$^{\rm{m}}$25$\fs$96 & 02$^{\circ}$01$^{\prime}$51$\farcs$4        & 94.194  & 11.70           & 0.08    &PPV\\
V67  & U0900$\_$14809956 & 19$^{\rm{h}}$22$^{\rm{m}}$35$\fs$19 & 02$^{\circ}$02$^{\prime}$04$\farcs$5        & 1.362   & 12.87           & 0.09    &EB\\
V68  & U0900$\_$15529595 & 19$^{\rm{h}}$32$^{\rm{m}}$13$\fs$30 & 02$^{\circ}$12$^{\prime}$33$\farcs$4        & 73.852  & 10.69           & 0.08    &PPV\\
V69  & U0900$\_$15083048 & 19$^{\rm{h}}$26$^{\rm{m}}$32$\fs$09 & 02$^{\circ}$07$^{\prime}$48$\farcs$6        & 76.042  & 11.40           & 0.03    &PPV\\
V70  & U0900$\_$14819436 & 19$^{\rm{h}}$22$^{\rm{m}}$44$\fs$20 & 02$^{\circ}$05$^{\prime}$24$\farcs$1        & 93.784  & 11.02           & 0.10    &PPV\\
V71  & U0900$\_$15313364 & 19$^{\rm{h}}$29$^{\rm{m}}$31$\fs$49 & 02$^{\circ}$12$^{\prime}$54$\farcs$6        & 0.655   & 13.92           & 0.05     &EB\\
V72  & U0900$\_$15131911 & 19$^{\rm{h}}$27$^{\rm{m}}$11$\fs$01 & 02$^{\circ}$11$^{\prime}$09$\farcs$9        & 0.357   & 13.10           & 0.14     &EB\\
V73  & U0900$\_$14936196 & 19$^{\rm{h}}$24$^{\rm{m}}$30$\fs$18 & 02$^{\circ}$12$^{\prime}$53$\farcs$3        & 0.456   & 13.26           & 0.18     &EB\\
V74  & U0900$\_$15290990 & 19$^{\rm{h}}$29$^{\rm{m}}$14$\fs$99 & 02$^{\circ}$18$^{\prime}$20$\farcs$5        & 0.588   & 13.77           & 0.10     &EB\\
V75  & U0900$\_$14927833 & 19$^{\rm{h}}$24$^{\rm{m}}$22$\fs$78 & 02$^{\circ}$15$^{\prime}$02$\farcs$3        & 1.028   & 11.62           & 0.01     &EB\\
V76  & U0900$\_$15076715 & 19$^{\rm{h}}$26$^{\rm{m}}$26$\fs$80 & 02$^{\circ}$18$^{\prime}$43$\farcs$4        & 1.793   & 10.07           & 0.02     &EB\\
V77  & U0900$\_$15286206 & 19$^{\rm{h}}$29$^{\rm{m}}$11$\fs$29 & 02$^{\circ}$21$^{\prime}$55$\farcs$2        & 0.392   & 11.99           & 0.06     &EB\\
V78  & U0825$\_$15906838 & 19$^{\rm{h}}$28$^{\rm{m}}$09$\fs$62 & -00$^{\circ}$21$^{\prime}$58$\farcs$2       & 0.495   & 14.21           & 0.4       &EB\\
V79  & U0900$\_$14772878 & 19$^{\rm{h}}$22$^{\rm{m}}$02$\fs$07 & 00$^{\circ}$01$^{\prime}$07$\farcs$5        & 0.723   & 13.39           & 0.2       &EB\\
V80  & U0900$\_$14716806 & 19$^{\rm{h}}$21$^{\rm{m}}$07$\fs$48 & 00$^{\circ}$33$^{\prime}$09$\farcs$9        & 1.188   & 13.25           & 0.07     &PPV\\
V81  & U0900$\_$15357089 & 19$^{\rm{h}}$30$^{\rm{m}}$05$\fs$04 & 00$^{\circ}$46$^{\prime}$28$\farcs$7        & 0.307   & 13.40           & 0.1       &EB\\
V82  & ...              & 19$^{\rm{h}}$23$^{\rm{m}}$28$\fs$00 & 00$^{\circ}$41$^{\prime}$29$\farcs$1        & 0.395   & 13.04           & 0.05     &EB\\
V83  & U0900$\_$14966595 & 19$^{\rm{h}}$24$^{\rm{m}}$54$\fs$54 & 01$^{\circ}$15$^{\prime}$03$\farcs$3        & 5.872   & 12.20           & 0.04     &PPV\\
V84  & U0900$\_$15033451 & 19$^{\rm{h}}$25$^{\rm{m}}$49$\fs$73 & 01$^{\circ}$28$^{\prime}$43$\farcs$2        & 0.549   & 13.95           & 0.12     &EB\\
V85  & ...             & 19$^{\rm{h}}$25$^{\rm{m}}$50$\fs$12 & 01$^{\circ}$29$^{\prime}$43$\farcs$5        & 0.354  & 13.95           & 0.2     &EB\\
V86  & U0900$\_$15116292 & 19$^{\rm{h}}$26$^{\rm{m}}$55$\fs$97 & 01$^{\circ}$31$^{\prime}$47$\farcs$0        & 2.504   & 13.11           & 0.6     &PPV \\
\hline
\end{tabular}
}
\end{center}
\end{table}

\begin{table}[!t]
\begin{center}
\caption{Periodic variable stars in the field from GCVS.}
 \vspace{0.5cm}
\tiny{
\begin{tabular}{lcccccccc}
\hline
ID 	& GCVS 				& $\alpha$(J2000) 				& $\delta$(J2000) 						& Type & $P$(days) & $V_{\rm{max}}$  & $V_{\rm{min}}$  & $R_{\rm{mean}}$\\
\hline\hline	
G1   & ES Aql                & 19$^{\rm{h}}$32$^{\rm{m}}$21$\fs$50 & -00$^{\circ}$11$^{\prime}$31$\farcs$8        & SR      & 101.1      & 13.2            & 15.1         & 10.80   \\
G2   & V0362 Aql         & 19$^{\rm{h}}$24$^{\rm{m}}$55$\fs$67 & 01$^{\circ}$05$^{\prime}$00$\farcs$6        & RV      & 68.        & 13.1            & 14.7         & 14.09   \\
G3   & V0381 Aql         & 19$^{\rm{h}}$31$^{\rm{m}}$13$\fs$68 & 02$^{\circ}$07$^{\prime}$03$\farcs$4        & RV      & 54.8       & 12.7            & 15.4         & 12.90   \\
G4   & V0919 Aql         & 19$^{\rm{h}}$22$^{\rm{m}}$55$\fs$75 & 01$^{\circ}$02$^{\prime}$39$\farcs$7        & EB/KE   & 0.797102   & 13.0            & 14.0         & 13.98   \\
G5   & V0920 Aql         & 19$^{\rm{h}}$31$^{\rm{m}}$30$\fs$75 & 01$^{\circ}$05$^{\prime}$53$\farcs$9        & RRAB    & 0.4955033  & 13.3            & 14.3         & 12.97   \\
G6   & V0291 Aql         & 19$^{\rm{h}}$30$^{\rm{m}}$19$\fs$93 & 01$^{\circ}$35$^{\prime}$47$\farcs$1        & RRAB    & 0.482889   & 13.5            & 15.0         & 13.91   \\
G7   & V0922 Aql         & 19$^{\rm{h}}$30$^{\rm{m}}$28$\fs$35 & 01$^{\circ}$13$^{\prime}$50$\farcs$9        & RRAB    & 0.458874   & 16.3            & 17.5         & 12.99   \\
G8   & V0978 Aql         & 19$^{\rm{h}}$31$^{\rm{m}}$31$\fs$57 & 02$^{\circ}$12$^{\prime}$57$\farcs$3        & RRAB    & 0.520935   & 14.6            & 15.9         & 14.37   \\
G9   & V1127 Aql         & 19$^{\rm{h}}$24$^{\rm{m}}$00$\fs$04 & 01$^{\circ}$41$^{\prime}$46$\farcs$1        & RRAB    & 0.356005   & 14.8            & 16.0         & 13.42   \\
G10  & V1135 Aql        & 19$^{\rm{h}}$31$^{\rm{m}}$04$\fs$27 & -00$^{\circ}$18$^{\prime}$42$\farcs$6        & EA/SD   & 3.14546    & 15.5            & 16.6         & 14.69   \\
G11  & V1215 Aql        & 19$^{\rm{h}}$22$^{\rm{m}}$24$\fs$69 & 01$^{\circ}$27$^{\prime}$55$\farcs$4        & RR      & 0.5013     & 15.3            & 16.0         & 15.20   \\
\hline
\end{tabular}
}
\end{center}
\end{table}


\begin{thebibliography}{}
\bibitem[Alard(2000)]{2000A&AS..144..363A} Alard, C.\ 2000, A\&AS, 144, 363
\bibitem[Baglin et al.(2002)]{2002sshp.conf...17B} Baglin, A., Auvergne,
M., Barge, P., Buey, J.-T., Catala, C., Michel, E., Weiss, W., \& COROT
Team 2002, ESA SP-485: Stellar Structure and Habitable Planet Finding,
17
\bibitem[Bertin \& Arnouts(1996)]{1996A&AS..117..393B} Bertin, E., \&
Arnouts, S.\ 1996, A\&AS, 117, 393
\bibitem[Bord{\'e} et al.(2003)]{2003A&A...405.1137B} Bord{\'e}, P., Rouan, 
D., \& L{\'e}ger, A.\ 2003, \aap, 405, 1137
\bibitem[Brown(2003)]{2003ApJ...593L.125B} Brown, T.~M.\ 2003, ApJ, 593,
L125
\bibitem[Deleuile et al.(2006)]{exodat} Deleuil, M., Moutou, C., Deeg, H., Meunier, C., Surace, C., Guterman, P., Almenara, J.~M., Alonso, R., Barge, P., Bouchy, F., Erikson, A., Leger, A., Loeillet, B., Ollivier, M., Pont, F., Rauer, H., Rouan, D., \& Queloz, D. 2006, ESA SP-1306: The CoRoT 
Mission, 341 
\bibitem[Hartman et al.(2004)]{2004AJ....128.1761H} Hartman, J.~D., Bakos,
G., Stanek, K.~Z., \& Noyes, R.~W.\ 2004, AJ, 128, 1761
\bibitem[Hartman et al.(2005)]{2005AJ....130.2241H} Hartman, J.~D., Stanek,
K.~Z., Gaudi, B.~S., Holman, M.~J., \& McLeod, B.~A.\ 2005, AJ, 130, 2241
\bibitem[Kane et al.(2004)]{2004MNRAS.353..689K} Kane, S.~R., Collier 
Cameron, A., Horne, K., James, D., Lister, T.~A., Pollacco, D.~L., Street, 
R.~A., \& Tsapras, Y.\ 2004, \mnras, 353, 689 
\bibitem[Kane et al.(2005a)]{2005MNRAS.362..117K} Kane, S.~R., Lister, 
T.~A., Collier Cameron, A., Horne, K., James, D., Pollacco, D.~L., Street, 
R.~A., \& Tsapras, Y.\ 2005, \mnras, 362, 117 
\bibitem[Kane et al.(2005b)]{2005MNRAS.364.1091K} Kane, S.~R., Collier 
Cameron, A., Horne, K., James, D., Lister, T.~A., Pollacco, D.~L., Street, 
R.~A., \& Tsapras, Y.\ 2005, \mnras, 364, 1091 
\bibitem[Karoff et al.(2006)]{2006ASP} Karoff, C., Rauer, H.,
Erikson, A., \& Voss, H.\ 2006, Astronomical Society of the Pacific
Conference Series, 349, 261
\bibitem[Kholopov et al.(1998)]{1998GCVS4.C......0K} Kholopov, P.~N., et
al.\ 1998, in Combined General Catalogue of Variable Stars, 4.1 Ed
(II/214A)
\bibitem[Kjeldsen \& Frandsen(1992)]{1992PASP..104..413K} Kjeldsen, H., \&
Frandsen, S.\ 1992, PASP, 104, 413
\bibitem[Mochejska et al.(2002)]{2002AJ....123.3460M} Mochejska, B.~J.,
Stanek, K.~Z., Sasselov, D.~D., \& Szentgyorgyi, A.~H.\ 2002, AJ, 123,
3460
\bibitem[Monet et al.(1998)]{1998USNO2.C......0M} Monet, D., et al.\
  1998, The PMM USNO-A2.0 Catalog.~(1998)
\bibitem[Rauer et al.(2004)]{2004PASP..116...38R} Rauer, H., Eisl{\"o}ffel,
J., Erikson, A., Guenther, E., Hatzes, A.~P., Michaelis, H., \& Voss, H.\
2004, PASP, 116, 38
\bibitem[Schwarzenberg-Czerny(1996)]{1996ApJ...460L.107S}
Schwarzenberg-Czerny, A.\ 1996, APJ, 460, L107
\bibitem[Stetson(1996)]{1996PASP..108..851S} Stetson, P.~B.\ 1996, PASP,
108, 851
\bibitem[S{\"o}derhjelm \& Dischler(2005)]{2005A&A...442.1003S}
S{\"o}derhjelm, S., \& Dischler, J.\ 2005, A\&A, 442, 1003
\bibitem[Valdes et al.(1995)]{1995PASP..107.1119V} Valdes, F.~G.,
Campusano, L.~E., Velasquez, J.~D., \& Stetson, P.~B.\ 1995, PASP, 107,
1119
\bibitem[Tamuz et al.(2005)]{2005MNRAS.356.1466T} Tamuz, O., Mazeh, T., \& 
Zucker, S.\ 2005, \mnras, 356, 1466 
\end{thebibliography}
\end{document}